\documentclass[12pt, reqno]{article}
\usepackage{amsmath}%
\usepackage{amsfonts}%
\usepackage{amssymb}%
\usepackage{graphicx}
\usepackage{lscape}
\usepackage{pdflscape}
\usepackage{setspace}
\usepackage{booktabs}
\usepackage[table,xcdraw]{xcolor}
\usepackage{chngcntr}
\usepackage{subfigure}
\usepackage{hyperref}
\usepackage{csquotes}

\usepackage{hyperref}
\usepackage{amsmath}

\usepackage{sectsty}
\usepackage{epsfig}
\usepackage{ctable}
\usepackage{threeparttable}
\usepackage{caption} 
\usepackage{comment}

\usepackage{float}
\usepackage{array}
\usepackage{verbatim}
\usepackage{amstext}
\usepackage{enumitem}
\usepackage{amssymb}
\usepackage{graphicx}
\usepackage{setspace}
\usepackage{gensymb}

\let\oldFootnote\footnote
\newcommand\nextToken\relax

\renewcommand\footnote[1]{%
	\oldFootnote{#1}\futurelet\nextToken\isFootnote}

\newcommand\isFootnote{%
	\ifx\footnote\nextToken\textsuperscript{,}\fi}

\usepackage[square,sort,comma,numbers]{natbib}
\bibpunct{(}{)}{;}{a}{,}{,}       

\usepackage{xcolor}
\hypersetup{
	colorlinks,
	linkcolor={black!100!black},
	citecolor={black!100!black},
	urlcolor={black!100!black}
}

\makeatletter

\renewcommand\tagform@[1]{\maketag@@@{\ignorespaces#1\unskip\@@italiccorr}}
\renewcommand\theequation{(\oldtheequation)}
\makeatother

\newcolumntype{x}[1]{>{\centering\arraybackslash\hspace{0pt}}p{#1}}

\makeatletter
\renewcommand\@biblabel[1]{}
\makeatother

\makeatletter

\DeclareRobustCommand\citepos
{\begingroup
	\let\NAT@nmfmt\NAT@posfmt
	\NAT@swafalse\let\NAT@ctype\z@\NAT@partrue
	\@ifstar{\NAT@fulltrue\NAT@citetp}{\NAT@fullfalse\NAT@citetp}}

\let\NAT@orig@nmfmt\NAT@nmfmt
\def\NAT@posfmt#1{\NAT@orig@nmfmt{#1's}}

\makeatother

\usepackage[parfill]{parskip}

\usepackage{etoolbox}
\appto\TPTnoteSettings{\scriptsize}

\renewcommand{\footnotesize}{\scriptsize}
\newcommand{\double}{\renewcommand{\baselinestretch}{1.5} \tiny \normalsize}

\usepackage{geometry}
\newcommand{\sym}[1]{\rlap{#1}}

\usepackage{tgpagella}

\usepackage{threeparttable}

\begin{document}
	\title{Structural Transformation and Environmental Externalities}

\author{Teevrat Garg, Maulik Jagnani, and Hemant K. Pullabhotla\footnote{Garg: School of Global Policy and Strategy, University of California at San Diego, La Jolla, CA 92093 (email: \href{teevrat@ucsd.edu}{teevrat@ucsd.edu}); Jagnani: Department of Economics, University of Colorado Denver, Denver, CO 80204 (email: \href{maulik.jagnani@ucdenver.edu}{maulik.jagnani@ucdenver.edu}); Pullabhotla: Center on Food Security and the Environment, Stanford University, Stanford, CA 94305 (email: \href{hemantpk@stanford.edu}{hemantpk@stanford.edu}). We thank Sam Asher, Chris Barrett, Patrick Behrer, Jim Berry, Marshall Burke, C. Austin Davis, Paul J. Ferraro, Sam Heft-Neal, Paul Novosad, Tom Vogl, and numerous seminar audiences for valuable feedback. An earlier version of this paper was circulated as ``Agricultural Labor Exits Increase Crop Fires".} }

\date{This draft: September 2022 \\ \smallskip
	First draft: May 2020}	
	\maketitle
		\thispagestyle{empty}

	\begin{abstract}
\noindent 

Even as policymakers seek to encourage economic development by addressing misallocation due to frictions in labor markets, the associated production externalities -- such as air pollution -- remain unexplored. Using a regression discontinuity design, we show access to rural roads increases agricultural fires and particulate emissions. Farm labor exits are a likely mechanism responsible for the increase in agricultural fires: rural roads cause movement of workers out of agriculture and induce farmers to use fire -- a labor-saving but polluting technology -- to clear agricultural residue or to make harvesting less labor-intensive. Overall, the adoption of fires due to rural roads increases infant mortality rate by 5.5\% in downwind locations.

	\end{abstract}

	\setlength{\parindent}{2em}	
	\setlength{\parskip}{0em}
	\double
	\newpage
	\setcounter{page}{1}
	
\section{Introduction}

The persistence of the agricultural productivity gap -- a stylized fact that marginal product of labor is substantially lower in agriculture than in other sectors, especially in low- and middle-income countries -- suggests that labor is greatly misallocated across sectors \citep{gollin2014agricultural,Lewis1954,Kuznets1955}. Understanding the causes of this labor misallocation and identifying policies that enable structural transformation away from low productivity agriculture has been a central focus of development economics \citep{restuccia2017causes,Banerjee1998,bryan2014underinvestment}. However, we know little about production externalities associated with policies that enable labor reallocation away from agriculture. If production externalities exist, by focusing exclusively on private productivity, these policies may over- or under-correct the extent of labor misallocation. In this paper, we provide the first evidence that policies that address misallocation due to frictions in labor markets can generate environmental externalities.

Using a regression discontinuity design that leverages a sharp increase in the likelihood of road construction at population thresholds under the Pradhan Mantri Gram Sadak Yojana (PMGSY) in India, we show rural roads increase agricultural fires and particulate emissions. A new road results in a 60\% increase in the annual number of fires (or 1.6 additional fires) and a 1.1\% (0.5 $\mu$g/$m^3$) increase in annual satellite-measured PM2.5 levels in the village. The effect of rural roads on particulate pollution is driven almost entirely by increase in emissions from biomass burning, and concentrated in the winter harvest and post-harvest months. This is consistent with the explanation that agricultural fires (as opposed to, for example, vehicular emissions or economic activity) are the primary mechanism linking rural roads to increased local PM2.5 levels. 

To explore mechanisms we look to the literature on the impacts of rural roads constructed under PMGSY in India. Using the same regression discontinuity design described above, \cite{Asher2020} find rural roads singularly increased labor exits from agriculture while leaving most other economic outcomes unaffected.\footnote{\cite{Asher2020} find rural roads increased the availability of transportation services, and led to a 10 percentage point (20\%) decrease in share of workers in agriculture and a quantitatively similar increase in share of workers in the non-agricultural sector. They find no evidence for increases in assets or income. Farmers do not own more agricultural equipment, move out of subsistence crops, or increase agricultural production. We replicate their results in our sample.} Consistent with farm labor exits, we find descriptive evidence that rural roads increased village agricultural wages. Higher agricultural wages can cause farmers to substitute labor with fires to harvest sugarcane or clear rice harvest residue. Indeed, we show that at baseline there exists a robust positive correlation between relative farm wage rate and agricultural fires as well as a robust negative correlation between farm labor share and agricultural fires. 

Sub-sample analyses lend further support for farm labor exits as a likely mechanism responsible for the increase in agricultural fires. First, rural roads facilitate movement of workers out of agriculture, and increase agricultural fires and particulate emissions in districts where relative agricultural wage is lower (below median) at baseline, with comparatively modest effects in districts where relative agricultural wage is higher (above median). Second, the effect of rural roads on agricultural fires and particulate emissions is concentrated in districts with a higher (above median) production of crops that benefit from use of fires in the face of agricultural labor exits at baseline, either to clear harvest residue off fields within a narrow time window before planting in the next season (rice) or to make harvest less labor-intensive (sugarcane). Together, precisely as one might expect, we find the increase in agricultural fires and particulate emissions is driven by districts with a higher production of rice or sugarcane \textit{and} where relative agricultural wage is lower at baseline, with small and statistically insignificant effects in other districts. 

Lastly, we fail to find evidence for alternative explanations linking rural roads to agricultural fires, like changes in planting dates, and adoption of mechanized harvesting.

In the final part of the paper, we test whether rural roads, which have significant effects on agricultural fires and air pollution, were associated with adverse health impacts. Using the same regression discontinuity design described above, we find villages located within a 50 km distance downwind from treated villages observe a 2.3\% (1 $\mu$g/m$^3$) increase in annual PM2.5 levels. However, as one might expect, we fail to find an increase in PM2.5 levels in villages located within a 50 km distance in all other (non-downwind) directions from treated villages. Correspondingly, we find rural roads increase the infant mortality rate by roughly 5.5\% (0.3 percentage points) in villages located within a 50 km distance downwind from treated villages, with small and statistically insignificant effects in villages located within a 50 km distance in non-downwind directions from treated villages. Because, unlike air pollution, other socioeconomic effects of rural roads are not correlated with wind direction, the increase in infant mortality in downwind locations is likely solely driven by the increase in air pollution from crop fires.

Our results contribute to a broad literature examining trade-offs and synergies between economic development and environmental quality (see \citealp{Jayachandran2022}, for a review). Within this literature, the most closely related are a set of papers that examine the effects of anti-poverty programs on environmental quality (e.g., \citealp{Alix-Garcia2013,ferraro2020conditional,Asher2020a,Behrer2020}).  \cite{Alix-Garcia2013} and \cite{ferraro2020conditional} examine how cash transfer programs that raise consumption and income affect deforestation, but find opposite results. \cite{Asher2020a} study rural road construction under PMGSY in India to examine the deforestation effect of transportation infrastructure, but find rural roads do \textit{not} affect deforestation.\footnote{They also study the effects of highways and find that those lead to large increases in deforestation through increased demand for commercial timber.} In a concurrent paper, \cite{Behrer2020} finds that a rural workfare program in India increased incomes and induced farmers to invest in labor-saving capital technologies like combine harvesters that left more crop residue and increased use of agricultural fires.\footnote{Similarly, a recent working paper, \cite{davis2017did}, shows that increased labor demand motivated sugarcane growers to adopt mechanized harvesting in Brazil. Because sugarcane fields are burned in preparation for manual harvesting but not mechanical harvesting, adoption of mechanized harvesting may have decreased crop fires and improved environmental quality. However, that paper does not examine effects on crop fires or air pollution.} That paper, like ours, is thus an exception in that its focus is on air pollution -- one of the leading causes of mortality worldwide -- and not deforestation. Overall, as these empirical investigations suggest, the relationship between economic
development and the environment is nuanced, and different elements of development can accelerate or slow
down environmental degradation. Therefore, in the absence of a general theoretical result, it is crucial to examine how and in what circumstances
economic development helps or hurts different dimensions of environmental quality \citep{Jayachandran2022}.

We make two key contributions to this literature. One, we provide the first empirical evidence for the \textit{direct} substitution between environmental quality and factor inputs like labor -- a link that has long been part of the concept of sustainability in environmental economics \citep{Solow1993}. Indeed, while environmental economists have had a keen interest in studying the implications of environmental policies for sectoral labor reallocation  \citep{walker2013transitional}, environmental externalities associated with policies that facilitate sectoral labor reallocation remain entirely unexplored.  The issue is particularly salient in many low- and middle-income countries that are undergoing rapid structural transformation while simultaneously experiencing dangerously high levels of air pollution \citep{Greenstone2015, Barrett2020}.

Two, we are the first to causally relate human health to the environmental quality impacts of anti-poverty programs, or more broadly, economic development, for that matter.  In order to calculate the environmental costs or benefits of anti-poverty programs,
accurate estimates are needed that link environmental quality effects to observable
health outcomes. However, previous work does not examine any downstream impacts, health or otherwise, likely because the environmental quality effects of anti-poverty programs are correlated with the socioeconomic effects of anti-poverty programs. We overcome this issue by exploiting
pollution variation from wind direction  (e.g., \citealp{Pullabhotla2022,Rangel2019, Schlenker2016, Deryugina2019}), that is unlikely to be correlated with the socioeconomic effects of rural roads.

\section{Background}
Air pollution remains one of the leading causes of mortality, accounting for 9 million premature deaths annually or roughly 16\% of all deaths worldwide and a staggering 268 million disability-adjusted-life-years \citep{landrigan2018pollution}. Nowhere is the problem more pronounced than in India, which is home to 14 of the 20 most polluted cities in the world. In fact, if the city of New Delhi, the capital of India, were to meet World Health Organization air quality standards, average life expectancy would increase by 10 years \citep{AQI2019}, roughly equivalent to the gains in life expectancy made by the country on average in the 21st century \citep{owidlifeexpectancy}. Of course, a number of factors -- both moderate but perpetual and seasonal but acute -- contribute to the poor air quality in India. In this section, we discuss (a) use of agricultural fires -- a point source that contributes to as much as half of the particulate pollution in many parts of the country during winter months \citep{Cusworth2018,Bikkina2019,Shyamsundar2019}, and (b) how rural roads may affect agricultural fires.

\subsection{Agricultural Fires in India}
\begin{figure}[H]
\begin{center}
	\caption{Spatial distribution of fire counts across Indian districts}
	\label{fig:distfires}
	\includegraphics[width=\linewidth]{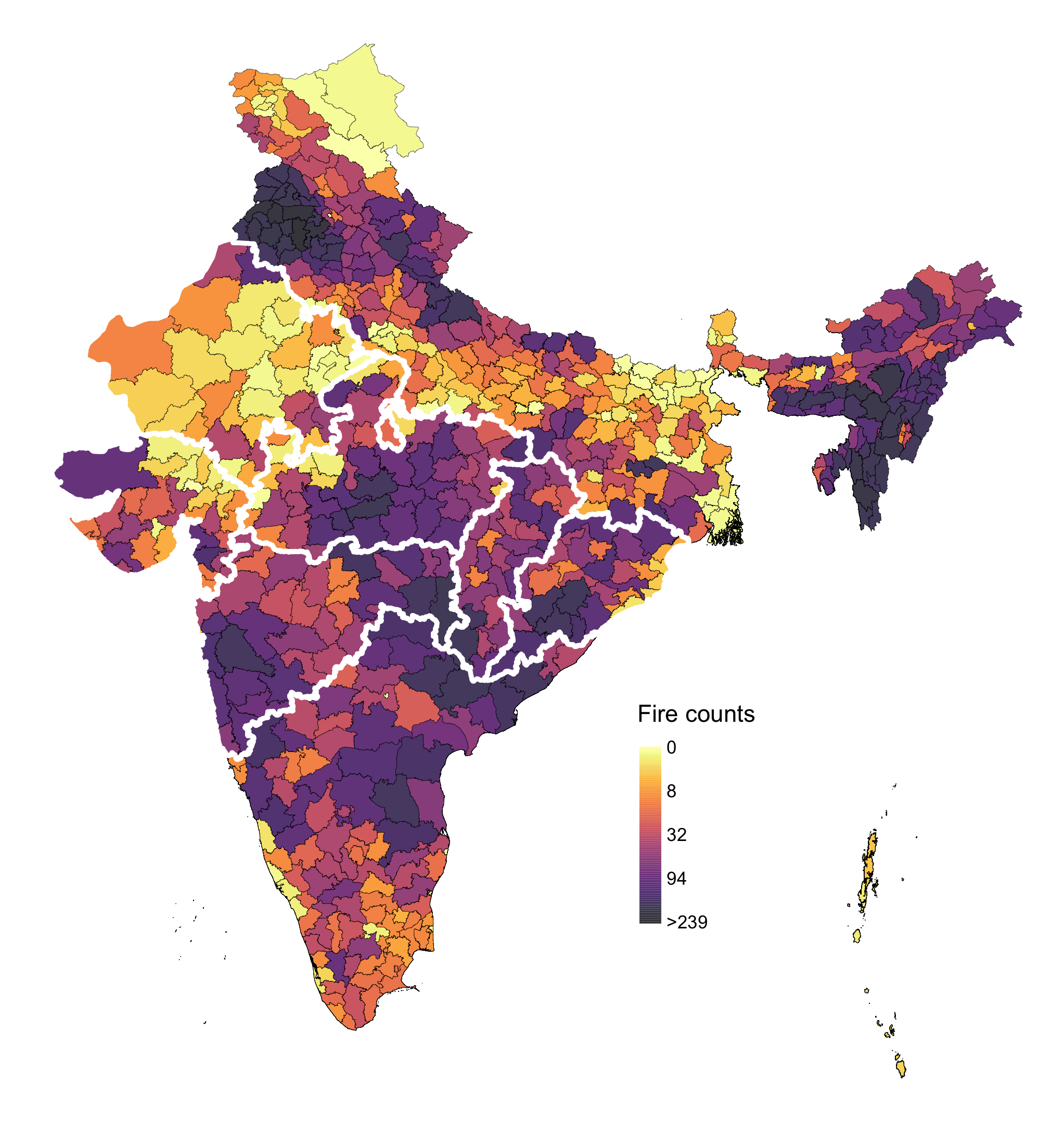} 
\end{center}
	\scriptsize Notes: This figure shows the mean annual number of fire pixels detected at the district level across India between 2001 and 2013. The states included in our main analysis sample are highlighted via white borders. The average fire counts at the district level range from a minimum of 0 to a maximum of 1810, with mean of 95.7. The legend shows the values at the  $25^{th}$, $50^{th}$, $75^{th}$ and $90^{th}$ percentiles of the average annual fire counts. 
\end{figure}

\paragraph{Why do farmers use fire?} Agricultural fires serve many purposes, including (i) clearing harvest residue off fields in preparation for planting in the next season, (ii) making sugarcane harvesting less labor-intensive, and (iii) clearing undergrowth on fields left fallow between cropping seasons. Figure \ref{fig:distfires} shows wide spatial distribution of agricultural fires across districts in India.\footnote{In Figure \ref{fig:statefires} we report state-wise annual average number of fires between 2001 and 2013.} Figure \ref{fig:annfires} shows the increasing trend in the number of fires annually between 2001 and 2013, the period of study. In 2003, there were roughly 55,000 fires whereas in 2012 there were over 90,000 fires.

The use of fire is particularly prevalent in a coupled rice-wheat cropping arrangement -- a system of agriculture widespread across India \citep{Jain2014,prasad1999}. In this system farmers grow rice during the monsoon season (\emph{kharif}) from June to November, and wheat immediately following rice harvest during the winter season (\emph{rabi}) from January to May. A narrow window of time between the harvest of rice (in October-November) and the planting of wheat (in December-January) requires large-scale and quick removal of crop residue, and setting fire to crop residue is particularly helpful in this process.\footnote{Farmers have an average of only 13 days between the harvesting of rice and the sowing of wheat. On the other hand, in the case of the wheat harvest, farmers report a window of 46-48 days between wheat harvest and planting of rice. Consistent with time availability being a significant factor, nearly 90\% of the farmers report that they engage in burning after the rice harvest. In contrast, with a longer interval available after the wheat harvest, only around 11\% of the farmers use burning to clear residue after the wheat harvest \citep{Kumar2015}.} 



Fire also plays a role in the production process for sugarcane - an important crop across the country \citep{FLA2012}. Farmers light sugarcane fields to remove the outer leaves around the cane stalk before harvesting the cane to make the process easier and require less manual labor \citep{Jain2014}. 

Finally, the use of fire is widespread alongside forest lands in central and north-east India that follow shifting cultivation, where fields are left fallow for more than a year, and farmers switch between alternate plots of land \citep{Venkataraman2006,Ramakrishnan1992}. Fallow fields are often overtaken by undergrowth, which needs to be cleared before subsequent season's planting.

\paragraph{The private costs of agricultural fires.} Although fires offer an easy and inexpensive means of clearing agricultural residue, they impose private costs on agricultural households. First, burning crop residue carries civil and criminal penalties under Section 188 of the Indian Penal Code and the Air and Pollution Control Act of 1981. While enforcement is not perfect, it is far from absent.\footnote{See for example, https://www.downtoearth.org.in/blog/agriculture/stubble-burning-a-problem-for-the-environment-agriculture-and-humans-64912} 

Second, crop residue burning decreases the productivity of agricultural land by destroying micro-nutrients in the soil, and removing valuable fertilizer including nitrogen and phosphorus \citep{Smil1999,Stan2014,Swayer2019,prasad1999}. Prior work has demonstrated that burning of rice and wheat residue can result in the loss of about 80\% of nitrogen, 25\% of phosphorus, 21\% of potassium and
4 to 60\% of sulphur from the soil \citep{Mandal2004}. 

Third, source-apportionment studies suggest pollution from agricultural fires can raise local concentrations of PM2.5 to more than 1,000\% above the WHO 24-hour guideline of 25$ug/m^{3}$ \citep{Bikkina2019,Liu2018}.\footnote{Local PM2.5 concentrations in almost all Indian villages are above the WHO 24-hour guideline of 25 $ug/m^{3}$ (Figure \ref{fig:pm25}).} Exposure to pollution from crop fires decreases birth
weight, gestational length, and in utero survival \citep{Rangel2019}, increases infant mortality \citep{Pullabhotla2022}, decreases child height for age and weight for age scores \citep{Singh2019}, decreases cognitive performance \citep{zivin2020unintended}, and increases risk of acute respiratory infections \citep{Chakrabarti2019}. Unsurprisingly, local households incur significant expenses to mitigate the consequences of burning-induced air pollution: medical records of local hospitals in the rice-wheat belt showed a 10\% increase in the number of patients within 20–25 days of the burning period
every season \citep{Kumar2015}; the same study finds inhabitants spend roughly USD 1 million every year on treatment for ailments caused by stubble burning.

Fourth, residue collected from the fields is of substantial economic value \citep{berazneva2018allocation}. Crop residue generated from the coupled rice-wheat cropping system can be used as livestock feed, which is in short supply across India, by roughly 40\% \citep{Kumar2015}. Moreover, soil treated with crop residues can hold 5 to 10 times more aerobic bacteria and 1.5 to 11 times more fungi than soil from which residues were burnt, providing higher yields \citep{Beri1992,Sidhu1995}.

\subsection{Agricultural Fires and Rural Roads}
\paragraph{Pradhan Mantri Gram Sadak Yojana (PMGSY)} \label{pmgsy}
In 2000, an estimated 330,000 of India's 825,000 rural villages lacked any all-weather road access. The Pradhan Mantri Gram Sadak Yojana (PMGSY) -- the Prime Minister's Village Road Program -- was launched in 2000 with the goal of providing all-weather road access to unconnected villages across India. Importantly, the national program guidelines prioritized larger villages according to arbitrary thresholds based on the 2001 Population Census \citep{Asher2020}. The guidelines aimed to connect all villages with populations greater than 1,000 by 2003, all villages with population greater than 500 by 2007, and villages with population over 250 after that.\footnote{The unit of targeting in the PMGSY is the habitation, defined as a cluster of population whose location does not change over time. Revenue villages, which the Economic and Population Censuses use, are comprised of one or more habitations (National Rural Roads Development Agency, 2005). In this paper, we aggregate all data to the revenue village level.} These rules were to be applied on a state-by-state basis, meaning that states that had connected all larger villages could proceed to smaller localities; for instance, states with few unconnected villages with over 1,000 people used the 500-person threshold immediately.\footnote{Some states did not comply with the threshold guidelines. Some states included several other prioritization
guidelines, and it is possible that political patronage played a role. See \cite{Asher2020} for more details. In the empirical exercise that follows we limit our analysis to states that complied with these guidelines and where we can show that there was a clear discontinuity in the probability of receiving a rural road around the relevant population thresholds.} Rural road construction under PMGSY began in 2000 but only really took off in 2002, continuing steadily thereafter through to the end of the sample period in 2013 (Figure \ref{fig:roadsyears}). 


\paragraph{How do rural roads affect agriculture?}
\cite{Asher2020} leverage the discontinuous increase at the aforementioned population thresholds for complying states using a fuzzy regression discontinuity design to show rural roads -- constructed under PMGSY -- cause a substantial increase in the availability of transportation services but do not increase assets or income. We replicate the results from \cite{Asher2020} in Appendix \ref{asher_novosad}. Farmers do not own more agricultural equipment, move out of subsistence crops, or increase agricultural production. They do find that rural roads lead to a large reallocation of workers out of agriculture. 
They find suggestive evidence that the growth in non-agricultural workers is due to greater access to jobs outside the village. Finally, they decisively rule out small changes in permanent migration, implying that their results are not the product of compositional changes to the village population.\footnote{They also show rural roads do not affect the share of adults who are either not working or who are in occupations that they are unable to classify, suggesting that changes in the workforce are not responsible for changes in the share of workers in agriculture or non-agricultural wage work.} Overall, the main effect of rural roads is to facilitate the movement of workers out of agriculture with no major changes in agricultural outcomes, income, or assets.


\paragraph{How could rural roads affect agricultural fires?} \label{pos_mech}
Labor exits from agriculture following rural road construction may raise village agricultural wages. Using the 1999 and 2006 Rural Economic and Demographic Survey (REDS), we find suggestive evidence that rural-road-induced agricultural labor exits increased agricultural wages (Appendix \ref{reds}). Higher labor costs will make farm work more expensive and can cause farmers to use fires as a labor saving technology to clear rice harvest residue off fields in preparation for the subsequent season's wheat planting or make sugarcane harvesting less labor-intensive; Indian authorities have long speculated that decreased availability of labor is one important reason for increase in agricultural fires over the last two decades \citep{goi2014}. Indeed, we show that at baseline there exists a robust positive correlation between relative agricultural wage rate and crop fires at baseline as well as a robust negative correlation between agriculture labor share and crop fires (Figure \ref{fig:corrplots}). 


Other possibilities that would link road construction to fire use include changes in planting dates and adoption of mechanized harvesting. 

In Section \ref{alternative}, we examine these mechanisms. We find evidence for farm labor exits as one mechanism responsible for the increase in agricultural fires and fail to find evidence for alternative channels.

\section{Data} \label{data}
Our primary analysis combines information on village-level rural road construction under PMGSY with (a) satellite-based measures of agricultural fires and PM2.5 collapsed at the village level and (b) village-level infant mortality rates imputed from birth history data.

\paragraph{Rural roads.} We use the Socioeconomic High-resolution Rural-Urban Geographic Dataset (SHRUG) for information on village-level rural road construction dates under PMGSY \citep{Asher2019}. We merge data on roads with village-level shapefiles \citep{Meiyappan2017,Meiyappan2018}. We follow \cite{Asher2020}, who worked closely with the National Rural Roads Development Agency to identify the state-specific thresholds that were followed, to define our sample. That is, we restrict our analysis to villages from states that adhered to the population priority criterion (in parentheses) set forth by the national government: Chhattisgarh (500, 1,000), Gujarat (500), Madhya Pradesh (500, 1,000), Maharashtra (500), Orissa (500), and Rajasthan (500). The analysis sample is restricted to the 11,151 villages that (i) did not have a paved road in 2001; (ii) were matched across all primary data sets in SHRUG; (iii) had identifiable geographic coordinates required to generate satellite-based village-level outcome measures; and (iv) had populations within the optimal bandwidth from a treatment threshold.\footnote{Our sample is slightly smaller than the analysis sample in \cite{Asher2020} because we could not identify geographic coordinates for 281 (roughly 2\%) villages.}

\paragraph{Agricultural fires.} We merge data on roads with satellite-detected fire activity data from NASA's Earth Observing System Data and Information System (EOSDIS) to capture annual agricultural fires within 10 kilometers of each village. The fire activity data are based on detections of infrared radiation that are a signature of biomass fires \citep{Giglio2006}. The underlying MODIS algorithm identifies a pixel (approximately one square kilometer area) with fire activity if at least one thermal anomaly is detected within that pixel. Unlike other remotely sensed products (for example, forest cover), fires data are considerably more precise since temperatures of fires are orders of magnitude higher than those of other non-fire pixels.  These data on thermal anomalies have been used extensively by atmospheric scientists to study the effects of agricultural fires on pollution in India and elsewhere \citep{Liu2018}. The MODIS data provides us with a daily, geocoded record of fire pixels from 2001 - 2013. We estimate annual counts of fires within a 10-kilometer radial buffer around the centroid of each village polygon.

\paragraph{Air pollution.} To generate pollution indicators at the village level, we rely on modeled PM2.5 pollution estimates from \cite{VanDonkelaar2016}. These gridded data are derived from satellite measures of aerosol density, combined with chemical transport models and calibrated to global ground-based observations of PM2.5. The \cite{VanDonkelaar2016} data provide annual average PM2.5 values at $0.1\degree \times 0.1\degree$ resolution for 1998-2013.

Furthermore, to examine seasonal impacts on air pollution, we complement these annual PM2.5 data with monthly data on black carbon and organic carbon emissions (precursors to particulate pollution) from the Modern-Era Retrospective Analysis for Research and Applications, Version 2 (MERRA-2) \citep{gelaro2017}. The MERRA-2 model apportions these emissions into those arising from biomass burning (which are driven by agricultural fires) and those arising from all other anthropogenic sources such as transportation, industrial emissions, or other combustion sources.\footnote{The MERRA-2 system produces gridded output on aerosol diagnostics on a global scale. It uses a chemical transport model (the Goddard Chemistry, Aerosol, Radiation, and Transport model (GOCART) that takes in a variety of different emission inventories as inputs to simulate aerosols arising from both natural and anthropogenic sources \citep{randles2017merra}. Emission inventories used to model aerosols from biomass burning differ from those used to capture other anthropogenic sources. Emissions from biomass burning rely on multiple inventories, including the Reanalysis of the Tropospheric Chemical Composition, the Global Fire Emissions Database, and the Quick Fire Emission Dataset. A separate emissions inventory from AeroCom Phase II \citep{diehl2012anthropogenic} is used to model anthropogenic emissions from residential, industry, power generation, and transportation sectors \citep{gelaro2017}. Validation studies in India find that  MERRA-2 simulated black carbon and organic carbon mass concentrations show a strong correlation with ground-based measures \citep{soni2021multiple}. }  This allows us to separately examine the impact of rural roads on emissions from agricultural fires and emissions from all other anthropogenic sources.   The MERRA-2 data are gridded ($0.50\degree \times 0.625\degree$) monthly estimates of emissions based on satellite and climate reanalysis measurements.  We interpolate both biomass emissions and non-biomass emissions gridded values to each village's geolocation in our sample.
 
In contrast to the fires data, our measures of air pollution suffer from two different sources of measurement error. First, air pollution data are obtained from calibrated chemical transport models that translate satellite imagery into ground level exposure of air pollution. Second, air pollution, unlike agricultural fires, can travel large distances spilling over into nearby villages. Despite these measurement error in the air pollution data, we are able to estimate precise effects of road construction on air pollution that correspond with estimates of the effects on crop fires.

\paragraph{Infant mortality.}
Infant mortality is a commonly used measure to document the external costs of air pollution \citep{currie2011traffic, Greenstone2014}. Infant mortality is highly responsive to short-run variability in air pollution since the first year of life is especially sensitive to environmental stressors. Moreover, focusing on infant mortality obviates the need to estimate life years lost and instead can be attributed to the full value of a statistical life. 

We generate infant mortality estimates using birth history data from the National Family Health Survey (NFHS-IV). The NFHS-IV, conducted in 2015-16, is the Indian wave of the Demographic and Health Surveys. The survey interviews all women aged 15-49 within the sample households and records detailed information on their birth histories. The birth histories provide us with data on the year of birth and the age at which the child died (if applicable). We use this information to assess infant mortality at the child-birth-year level from 2001 to 2013; that is, we generate an indicator variable that takes the value 1 if the  child died within 12 months of birth, and 0 otherwise. We then identify all NFHS clusters (villages) within a 50 km distance from each PMGSY village to generate estimates of infant mortality effects for children in these matched NFHS villages.\footnote{NFHS-IV sample clusters are geocoded sample enumeration areas roughly equivalent to a village in rural India.}

\section{Empirical Strategy}
Using a fuzzy regression discontinuity design, and data from 2002 to 2013, to estimate the effect of rural roads on agricultural fires, PM2.5, and infant mortality rates. Due to imperfect compliance of rural road construction with rules (population threshold) that determine award of rural roads under PMGSY, we use a two stage least squares specification with optimal bandwidth local linear regression \citep{Imbens2012,Gelman2019}:
\begin{align}\label{eq:rdspec} 
	Road_{vdst} &  = \gamma_0 + \gamma_1 \mathbf{1} (pop_{vds} \geq T) +  \gamma_2(pop_{vds} - T) + \\ \nonumber
	& + \gamma_3 (pop_{vds} - T) * \mathbf{1} (pop_{vds} \geq T) + \boldsymbol{\theta}\mathbf{X_{vds}} + \mu_{d,h} + \rho_{t} + \nu_{vdst}  \\
	Y_{vdst} & = \beta_0 + \beta_1 Road_{vdst} + \beta_2(pop_{vds} - T) + \\ \nonumber
	& + \beta_3 (pop_{vds} - T) * \mathbf{1} (pop_{vds} \geq T) +\boldsymbol{\delta } \mathbf{X_{vds}}  + \eta_{d,h} + \omega_{t} + \varepsilon_{vdst}
\end{align}

$Road_{vdst}$ takes the value 1 if village $v$ in district $d$ in state $s$ receives a PMGSY road by year $t$. $Y_{vdst}$ is outcome of interest (number of agricultural fires, PM2.5, infant mortality rate) for village $v$ in district $d$ in state $s$ in year $t$.\footnote{As we discuss in Section \ref{infant}, because the infant mortality data is available for NHFS villages and not for PMGSY villages, to estimate effects on infant mortality, in equation (2), the outcome variable is at the child-birth-year level where $Y_{invdst}$ is an indicator variable that takes the value 1 if child $i$ in NFHS village $n$, within 50 km of PMGSY village $v$ in district $d$ in state $s$, born in year $t$, died within the first year of birth, and 0 otherwise.}  The population of the village in 2001 is $pop_{vds}$, while $T$ is the treatment threshold (either 500 or 1000, depending on the state). Figure \ref{fig:runningvar} shows the density of the village population distribution is continuous across the treatment threshold; the McCrary test statistic is $-0.010$ (s.e. $0.048$) \citep{McCrary2008}. $\mu_{d,h}$ and $\eta_{d,h}$ are district-population threshold fixed effects - that is, an interaction of district dummies with an indicator variable that takes the value $1$ if village is in a state where the treatment threshold is equal to $1,000$, and $0$ otherwise. $\rho_{t}$ and $\omega_{t}$ are year fixed effects. Thus, the RD estimates compare outcomes for villages within the same district but on opposite sides of the PMGSY population threshold in year $t$. 

To ensure that our design follows closely with \citet{Asher2020}, we include a vector of baseline (2001) village characteristics, $\mathbf{X_{vds}}$, as controls, although excluding these controls does not alter our results appreciably. Specifically, we control for village amenities (primary school, medical center, electrification), agricultural characteristics including total agricultural land area, (log) share of irrigated agricultural land, and share of workers in agriculture, and village-level measures of socio-economic status and connectivity like literacy rate, share of inhabitants that belong to a scheduled caste, share of households owning agricultural land, share of households who are subsistence farmers, share of households earning over 250 INR cash per month (approximately 4 USD), and distance in km from the closest census town. We also control for count of fires or PM2.5 levels or infant mortality rate in 2001 when the outcome variable is count of fires or PM2.5 levels or infant mortality rate, respectively. We estimate equation~\ref{eq:rdspec} using an optimal bandwidth (84) with a triangular kernel that provides higher weights to observations close to the threshold \citep{calonico2014robust}. Standard errors are clustered at the village level.

\paragraph{Balance and falsification tests.} In Table \ref{tab:rdbal}, we present the mean values for various village baseline characteristics, including the set of controls that we use in all regressions, as well as mean values for our outcomes variables at baseline. Importantly, we fail to find evidence for discontinuous changes at the threshold for (i) various village baseline characteristics (balance tests) or (ii) baseline values of our outcome variables (falsification tests).

\section{Effects of Rural Roads on Crop Fires and PM2.5}

\begin{figure} [H]
	\begin{center}
		\caption{Impact of rural roads on crop fires  and PM2.5}	\label{fig:rdplots}
		\subfigure[Rural roads (0/1)]{
			\includegraphics[scale=0.25]{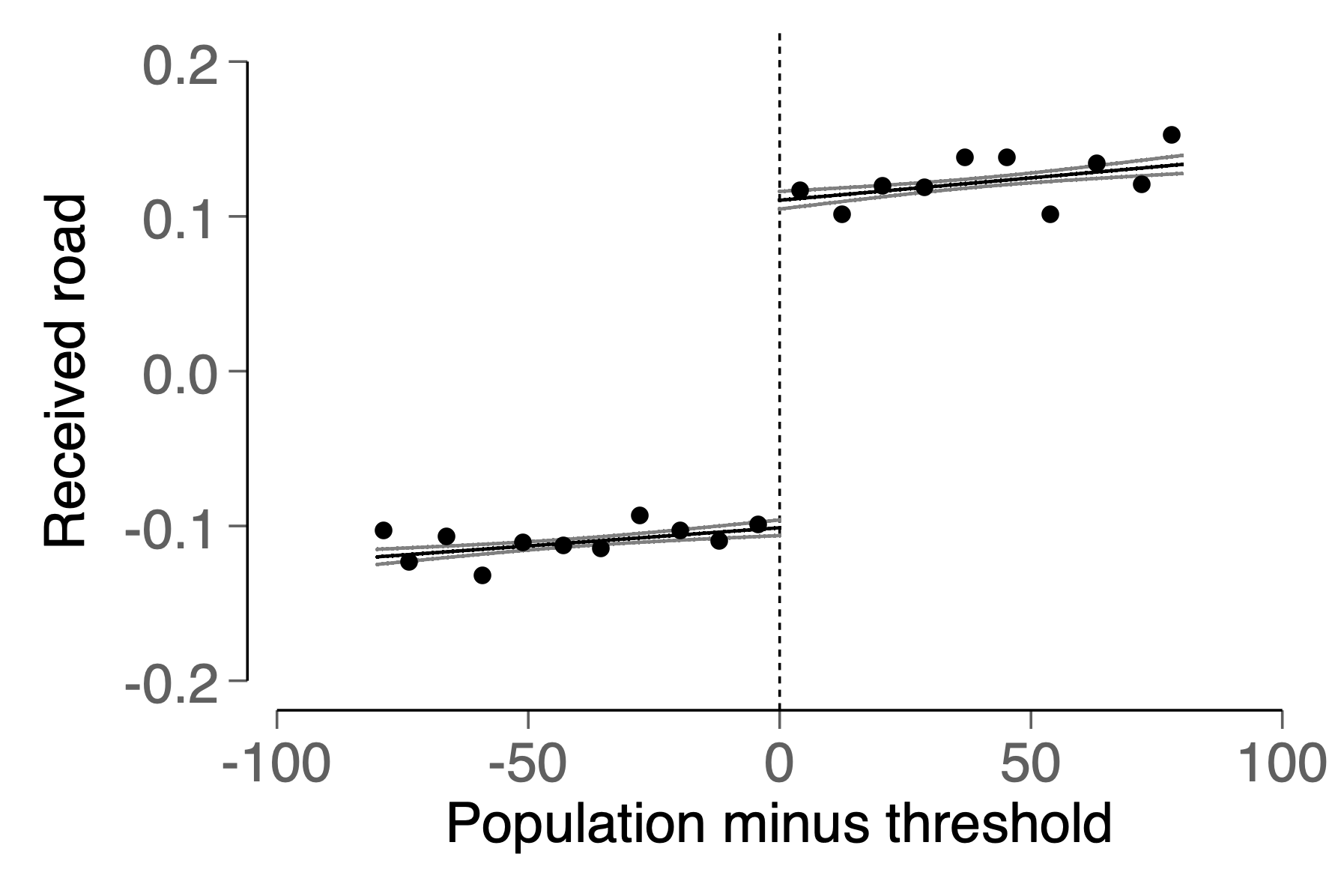}
		} 
		\subfigure[Annual fires (count)]{
			\includegraphics[scale=0.25]{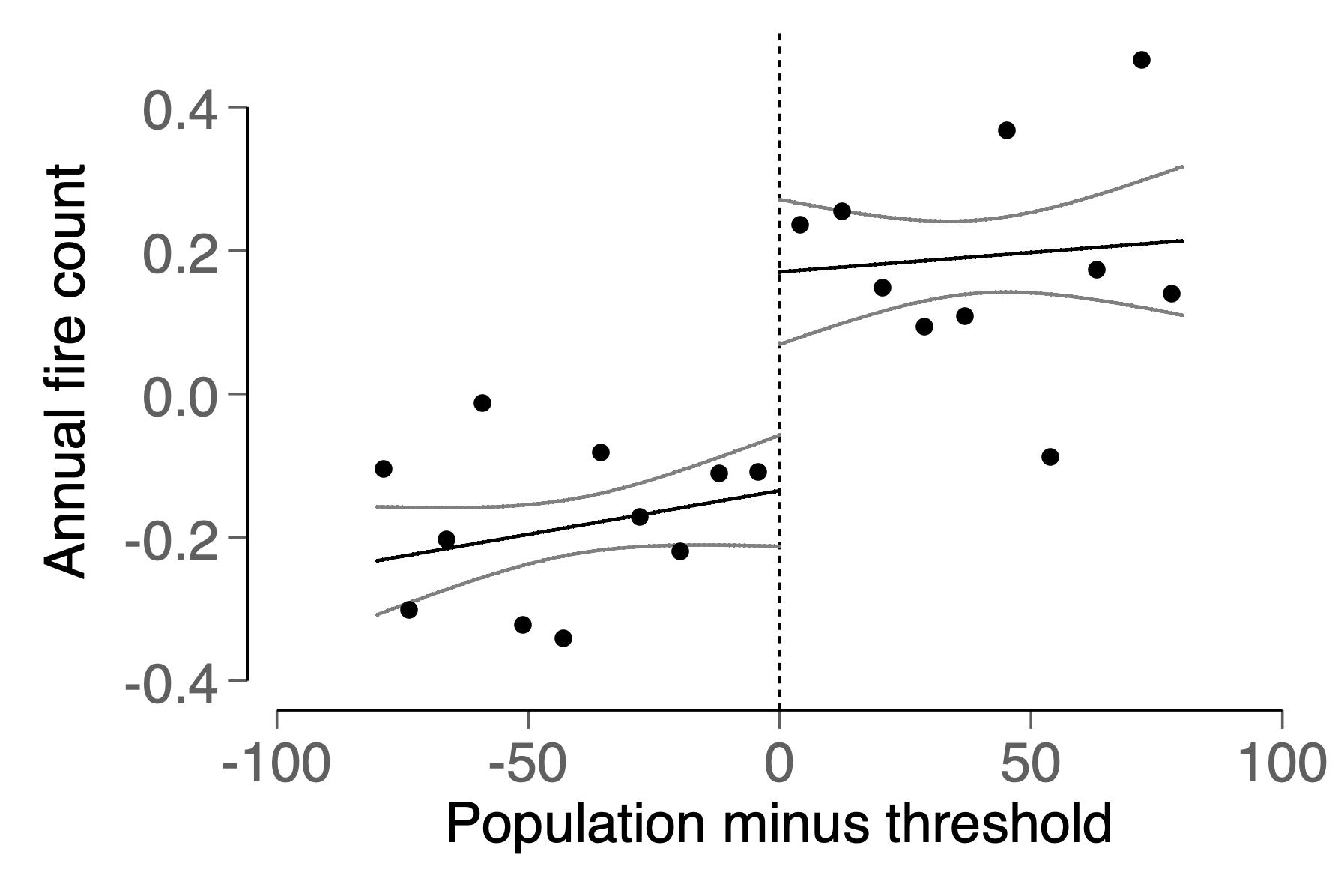} 
		}				
		\subfigure[Annual PM2.5 ($\mu g/m^3$)]{
			\includegraphics[scale=0.25]{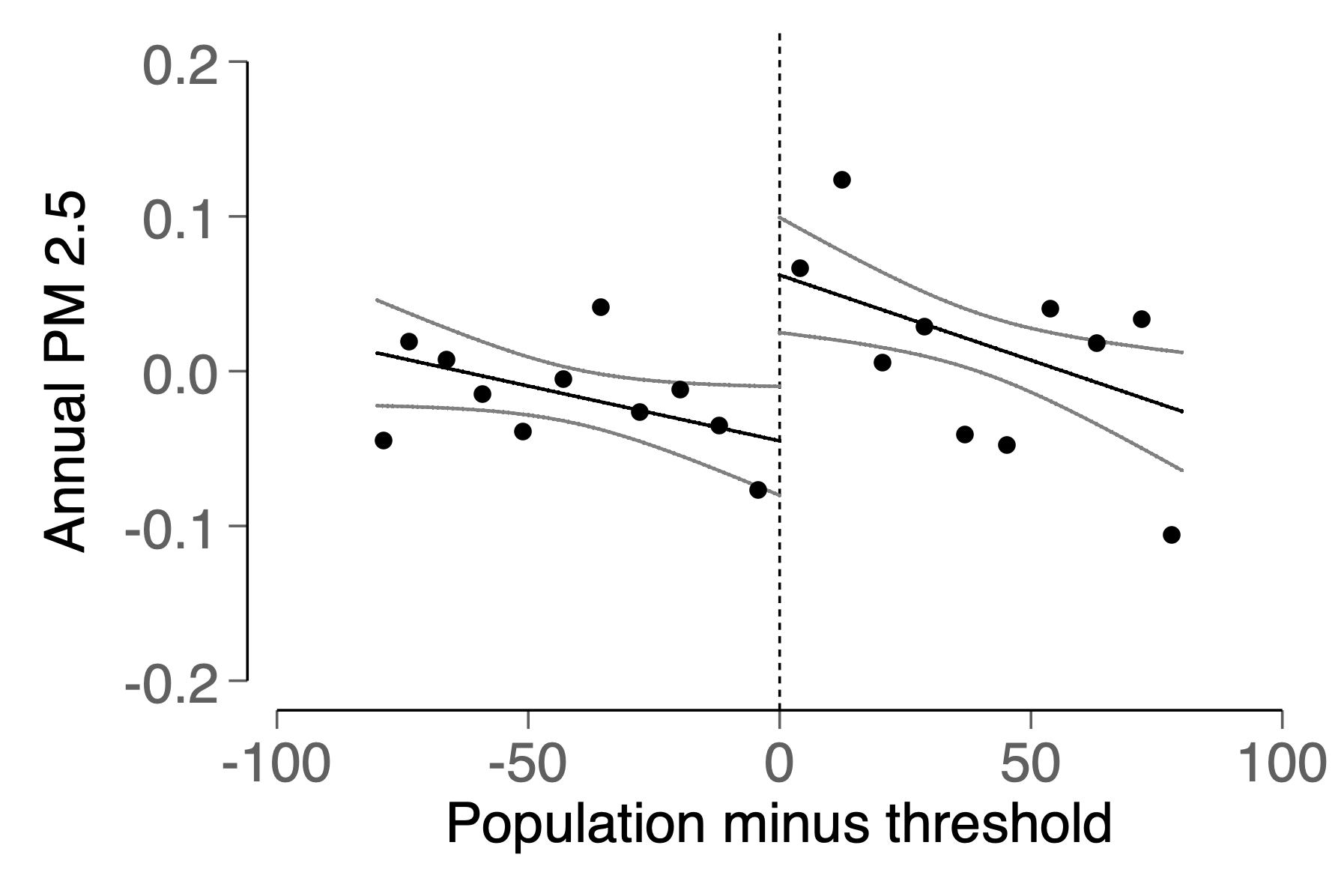} 
		}		
		
	\end{center}
	\scriptsize Notes: All panels show reduced form regression discontinuity estimates by plotting the residualized values of the outcomes (after controlling for all variables in the main specification other than population) as a function of the normalized 2001 village population relative to the threshold. Each point represents the mean of all villages in a given population bin. Figure (a) plots the residualized probability of a village receiving a new rural road between 2002 to 2013. Figure (b) plots the residualized annual number of fires between 2002 to 2013. Figure (c) plots the residualized average annual PM2.5 from 2002 to 2013. All panels control for district-threshold fixed effects, year fixed effects, and village characteristics in 2001 (baseline).  Population is centered around the state-specific threshold used for road eligibility - either 500 or 1000 depending on the state. Standard errors are clustered at the village level. \\

\end{figure}

Figure \ref{fig:rdplots} shows the graphical representation of the reduced form effect; i.e., the change in likelihood of rural road construction under PMGSY (Figure \ref{fig:rdplots}(a)) and count of agricultural fires at the (treatment) population threshold (Figure \ref{fig:rdplots}(b)). We observe a large and statistically significant increase in likelihood of rural road construction and number of fires for villages at the treatment threshold. Table \ref{tab:rdtable1} shows the corresponding point estimates. Column (1) presents the first stage result. Villages above the PMGSY population threshold observe a 23 percentage point increase in the likelihood of receiving a road. Column (2) shows the instrumental variable (IV) or local average treatment effects (LATE); we instrument access to rural roads with the PMGSY population threshold. Villages with rural road access observe a 60\% increase (1.6 additional fires) in the annual number of fires compared to villages that do not have access to a rural road.

Next, we examine the consequent effect of rural roads on local air quality. Our outcome is a satellite-based estimate of the annual average ambient PM2.5 concentrations for each village. Figure \ref{fig:rdplots}(c) shows the graphical representation of the reduced form effect of rural roads on PM2.5.  We observe a statistically significant increase in PM2.5 for villages at the population threshold that determines rural road construction under PMGSY.   Column (3) shows the IV or LATE point estimates where we instrument access to rural roads with PMGSY population threshold; villages with rural road access observe a (0.5 $\mu$g/$m^3$) 1.1\% increase in annual average PM2.5, compared to villages that do not have access to a rural road.

The 1.1\% increase in annual PM2.5 levels is comparable to the increase in particulate pollution (1.6\%) on days when air quality is not monitored in the US \citep{Zou2021}. It is roughly 25\% the size of the percentage decrease in particulates (4\%) due to subway openings worldwide \citep{gendron2022}, and about 6\% the magnitude of the percentage decrease in particulate matter (19\%) due to the most stringent air pollution regulations in India \citep{Greenstone2014}.   

\subsection{Mechanisms}
\subsubsection{Do agricultural labor exits increase crop fires?}
In this section, we examine whether farm labor exits following rural roads construction is a mechanism responsible for the increase in agricultural fires. As discussed above, consistent with farm labor exits, we find descriptive evidence that rural roads increased village agricultural wages. Higher agricultural wages can cause farmers to substitute labor with fires to harvest sugarcane or clear rice residue. Indeed, we show there exists a robust positive correlation between relative farm wage rate and crop fires at baseline.

Heterogeneity analyses lends further support for agricultural labor exits from road construction as one mechanism responsible for the increase in crop fires. First, rural roads facilitate movement of workers out of agriculture and increase agricultural fires in districts where relative agricultural wage is lower (below median) at baseline, but not in districts where relative agricultural wage is higher (above median) at baseline.\footnote{Wage data are unavailable for PMGSY villages (the unit of analysis for our regression discontinuity research design) during the period of our study. The employment and unemployment surveys conducted by the  National Sample Survey Organization (NSSO) of India collect data on the daily activities for the past seven days for all household members above four years of age. We use the $55^{th}$ round of the NSSO (1999 - 2000) and compute the average earnings per day worked in casual labor for individuals aged 18-60 residing in rural areas by dividing the weekly earnings by the number of days worked. We then calculate the district-level relative agricultural labor wage rate as the sample-weighted average of the ratio of the daily wage rate for casual labor in the agricultural sector to the non-agricultural sector. For districts in our sample, the median ratio is $0.46$. We split the sample into higher and lower relative agricultural wage rate districts based on this sample median.} The effects of rural roads on labor exits from agriculture and the increase in non-agricultural manual labor share is driven by districts with lower relative agricultural wage rates at baseline (Figure~\ref{fig:labshare_rdplots}; Table~\ref{tab:aglabexit_hetero}). Unsurprisingly, farm laborers in districts with lower relative agricultural wage  are more likely to reallocate to non-farm work outside the village due to access to rural roads.    Correspondingly, the increase in agricultural fires (Figure~\ref{fig:heteroagwage_rdplots}; Table~\ref{tab:heteroagwage}) and particulate emissions (Figure~\ref{fig:heteroagwage_rdplots}; Table~\ref{tab:heteroagwage}) is concentrated in districts with lower relative agricultural wage rates at baseline, with comparatively smaller effects in districts with higher relative agricultural wage rates at baseline.

Second, the effect of rural roads on agricultural fires is concentrated in districts with a higher (above median) baseline production of crops that benefit from use of fires in the face of agricultural labor exits, either to clear harvest residue off fields within a narrow time window before planting in the next season (rice) or to make harvesting less labor-intensive (sugarcane).\footnote{We match districts in the analysis sample to agricultural data obtained from the ICRISAT District Level Database. The ICRISAT District Level Data are available online from the following website: http://data.icrisat.org/dld/src/crops.html. We use data for the year 2001 to classify higher/lower rice or sugarcane districts based on the sample median of acreage share for the respective crops.} Figure \ref{fig:fires_crops} presents the spatial distribution of average annual fire activity and the share of rice and sugarcane grown across districts in our analysis sample. A simple eyeball test suggests fire counts are higher in districts where the share of rice or sugarcane area is above the median at baseline compared to districts where the share of rice or sugarcane area is below the median. More formally, indeed, the effects of rural roads on agricultural fires (Figure~\ref{fig:rdricesugar}; Table~\ref{tab:ricesugar}) and particulate emissions (Figure~\ref{fig:rdricesugar}; Table~\ref{tab:ricesugar}) are driven by districts with higher share of rice or sugarcane acreage at baseline, with comparatively modest effects in districts with lower share of rice and sugarcane acreage at baseline.

\begin{figure} [H]
	\begin{center}
		\caption{Impact of rural roads on crop fires (count) and PM2.5 ($\mu g/ m^{3}$) by relative agricultural wage rate and rice/sugar cropped area}	\label{fig:agwagecrop_rdplots}
		\subfigure[Low relative farm labor wage with high rice or sugar]{
			\includegraphics[scale=0.25]{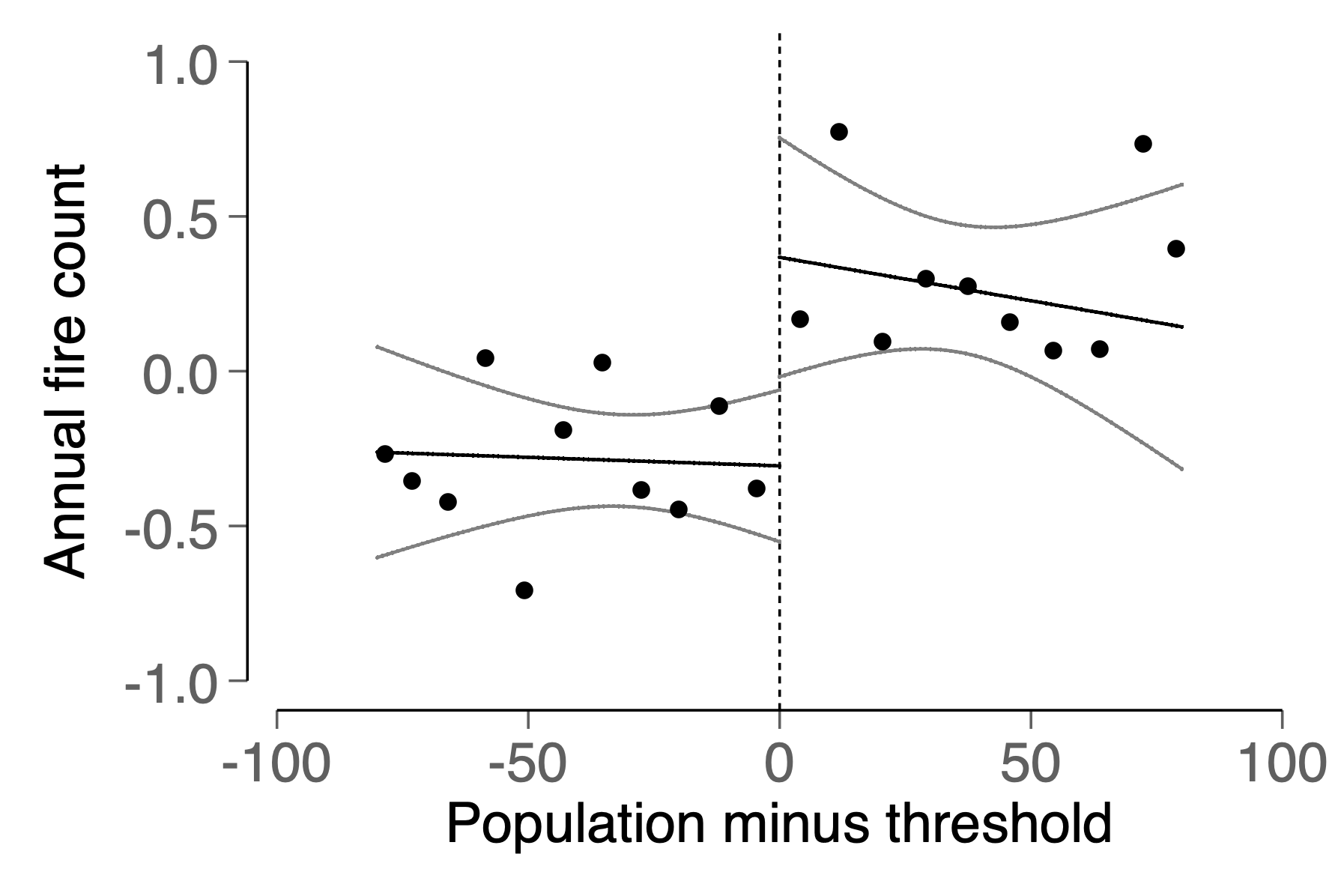}
		} 
		\subfigure[High relative farm labor wage or low relative farm labor wage with low rice \& low sugar]{
			\includegraphics[scale=0.25]{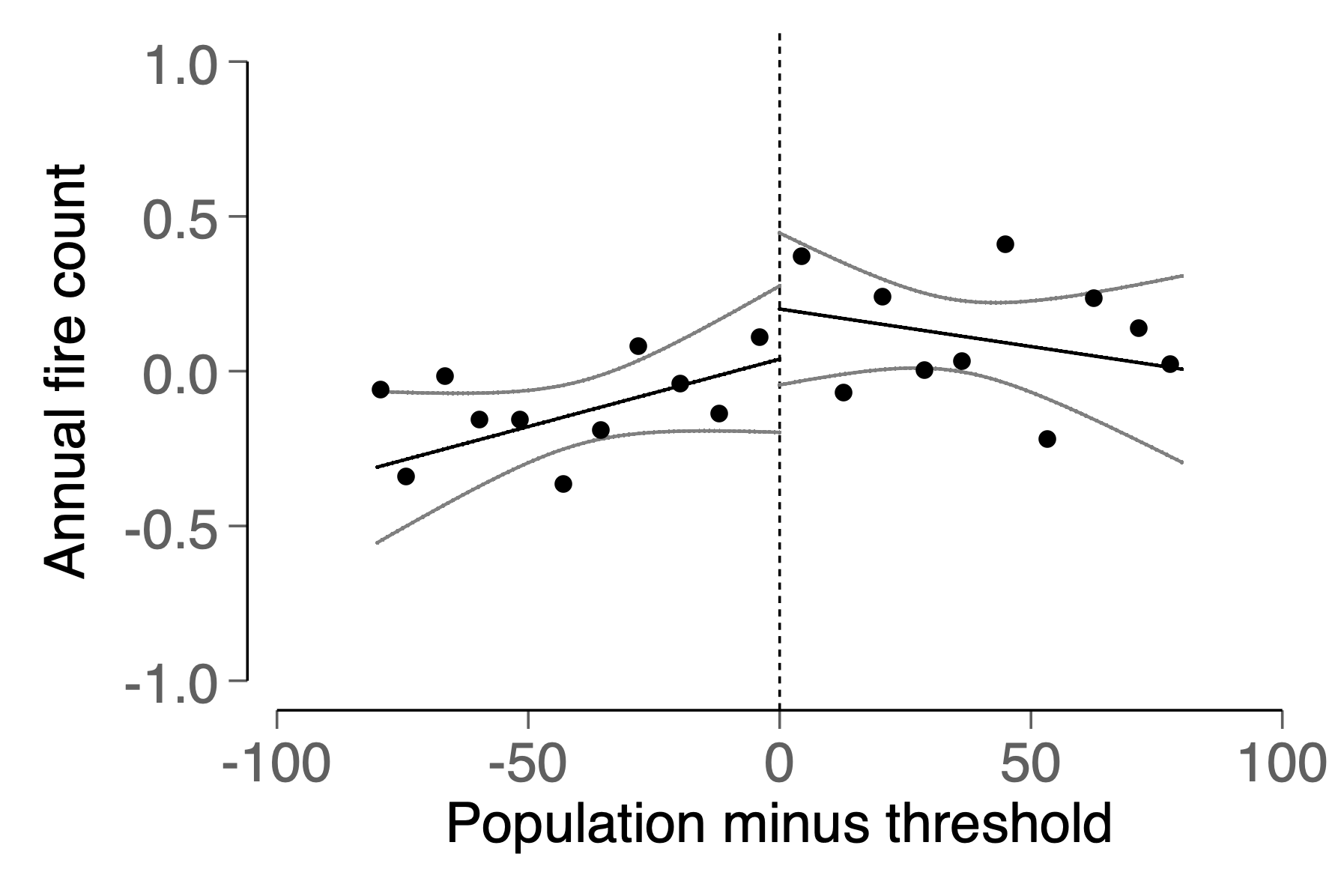} 
		}				
		\subfigure[Low relative farm labor wage with high rice or sugar]{
			\includegraphics[scale=0.25]{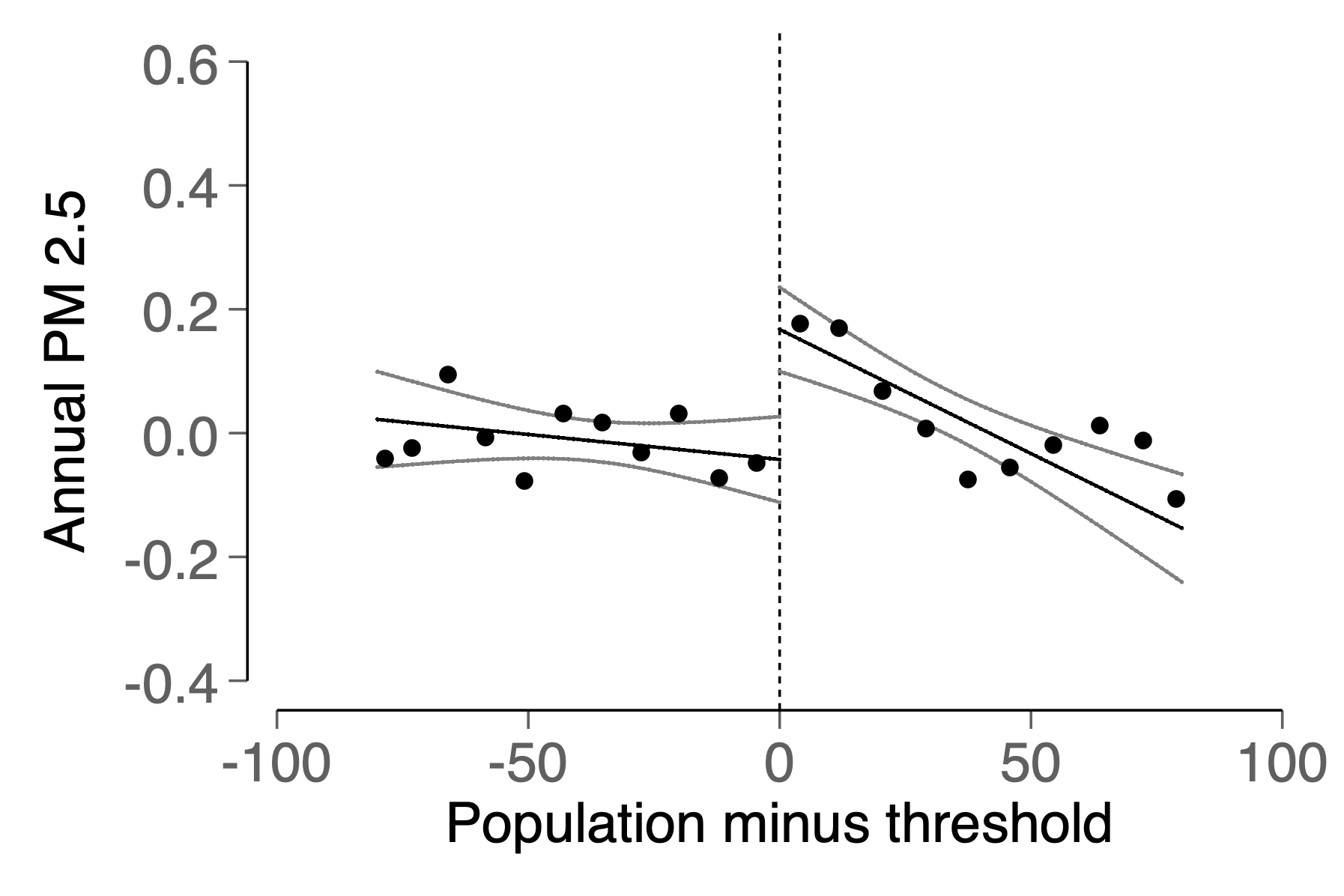}
		} 
		\subfigure[High relative farm labor wage or low relative farm labor wage with low rice \& low sugar]{
			\includegraphics[scale=0.25]{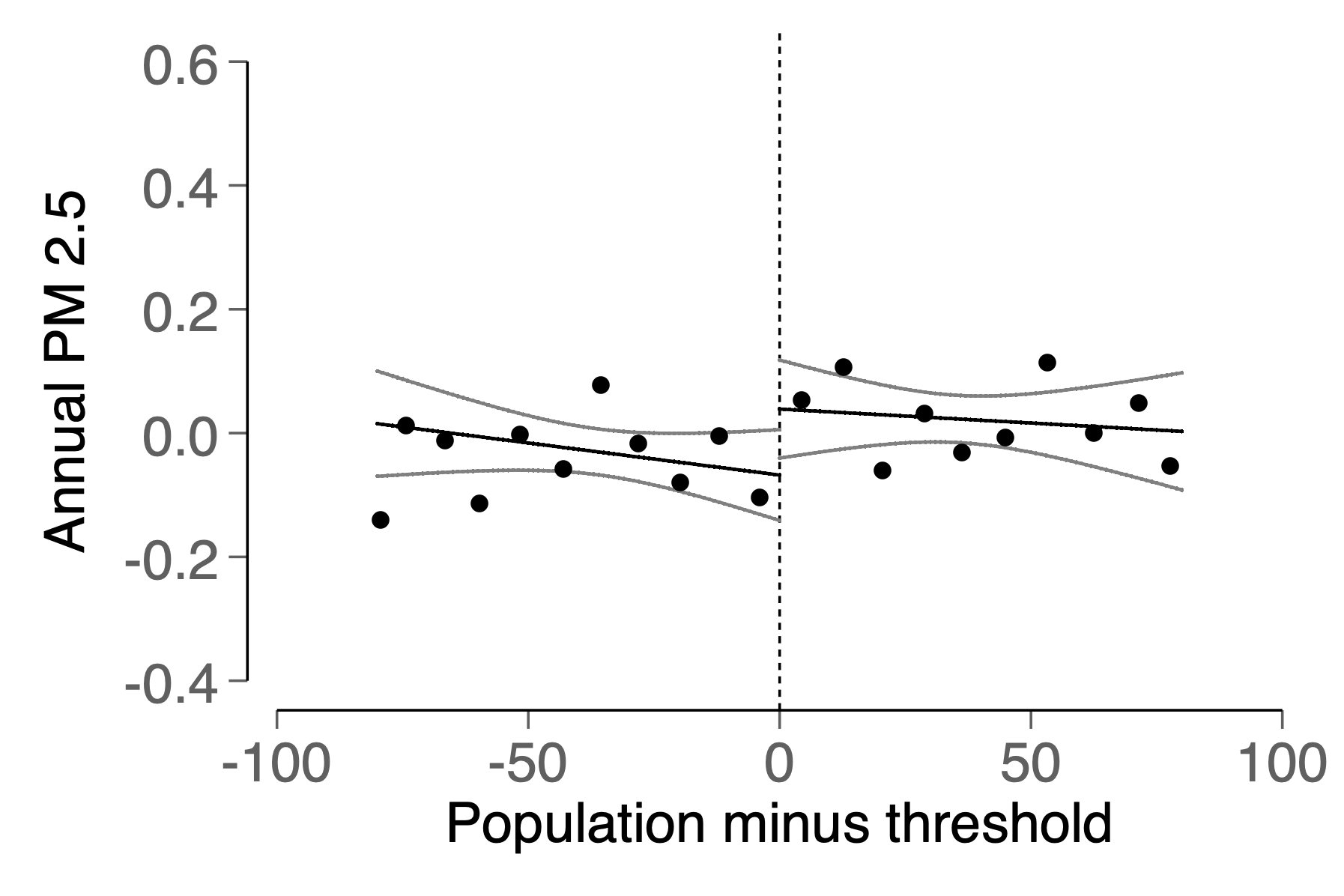} 
		}	
	\end{center}
	\scriptsize Notes: All panels show reduced form regression discontinuity estimates by plotting the residualized values of outcomes as a function of the normalized 2001 village population relative to the threshold (after controlling for fixed effects and all baseline variables in the main specification other than population). Each point represents the mean of all villages in a given population bin. Panels (a) and (b) plot the regression discontinuity relationship for  annual fire counts, while panels (c) and (d) plot the regression discontinuity relationship for annual average PM 2.5 ($\mu g/ m^{3}$). Panels (a) and (c) plots the relationship for districts which had low (below sample median) agricultural labor wages relative to non-agricultural labor wage rates in rural areas \emph{and} had high (above sample median) share of cropped area under rice or sugar at baseline. Panels (b) and (d) plots the relationship for  districts which had either (i) high (above median) relative agricultural wage rates or (ii) low relative agricultural wage rates with low rice and sugar cropped areas. Rural agricultural and non-agricultural daily labor wage rates are from the 1999 - 2000 NSSO survey data (Round 55). Rice and sugar cropped areas are from the ICRISAT district level data for 2001.  All panels control for district-threshold fixed effects, year fixed effects, and village characteristics in 2001 (baseline).  Population is centered around the state-specific threshold used for road eligibility - either 500 or 1000, depending on the state. Standard errors are clustered at the village level. \\
	
\end{figure}

Together, precisely as one might expect, we find the increase in agricultural fires (Figure \ref{fig:agwagecrop_rdplots}; Table \ref{tab:heteroagwagecrops}) and particulate emissions (Figure \ref{fig:agwagecrop_rdplots}; Table \ref{tab:heteroagwagecrops}) is concentrated in districts that observe a higher production of rice or sugarcane \textit{and} where the relative agricultural wage is lower at baseline, with small and statistically insignificant effects in other districts. Agricultural labor exits are higher in districts where relative agricultural wage rate is lower at baseline; therefore, it is likely that farmers in these districts observe a larger increase in labor costs due to rural road construction. Consequently, farmers in districts within this sub-sample where rice or sugarcane production is higher will drive the adoption of fire as a labor-saving technology: either  to clear rice residue within a short time window before planting next season's crop or to make sugarcane harvesting less labor-intensive.


\subsubsection{Alternative Explanations} \label{alternative}
Other explanations that would link road construction to air pollution include increase in economic activity or vehicular emissions. Alternative explanations that would link road construction to fire use include changes in planting dates, increase in non-farm biomass fires, adoption of mehcanized harvesting, and increase in school enrollment. Below we evaluate these alternative possibilities but fail to find evidence to support these mechanisms.


\paragraph{Is the increase in local PM2.5 levels driven by increase in economic activity or vehicular emissions?}\label{bc_oc}
\citet{Asher2019} fail to find evidence that rural roads increase local economic activity, as proxied by night lights (Appendix \ref{asher_novosad}). However, rural roads increased employment growth in manufacturing and retail (although the former is not statistically significant) as well as the availability of transportation services such as government buses. Therefore, it is plausible that economic activity or vehicular traffic and not agricultural fires relate rural roads to local particulate emissions. 

We directly test for this possibility by comparing the effects of rural roads on fires and particulate pollution in the winter harvest and post-harvest months -- months in which access to rural roads increases agricultural fires (October through March) --  with the effects of rural roads on particulate pollution during the rest of the year (April through September).\footnote{Decreased availability of farm labor or increased labor costs -- following rural road construction -- could cause farmers to use fires as a labor-saving technology to clear rice harvest residue off fields in preparation for the subsequent season’s wheat planting or to make sugarcane harvesting less labor-intensive. Crop fires associated with burning of rice residue occur between October and December \citep{Jain2014,prasad1999,Kumar2015}. Crop burning associated with the sugarcane harvest takes place between December and March \citep{doa2019}. Therefore, we examine the effects of rural roads on agricultural fires and particulate pollution between October and March and compare it to the effects of rural roads on agricultural fires and particulate pollution in the rest of the year.} If agricultural fires and not economic activity or vehicular emissions are the primary mechanism linking rural roads to increased local PM2.5 levels, one would expect to see the effects of rural roads on particulate pollution concentrated from October through March, with comparatively modest effects in the rest of the year. Indeed, we find rural roads increase both agricultural fires and particulate pollution from all sources in the winter harvest and post-harvest months of October to March, with comparatively modest effects in the rest of the year (Table \ref{tab:rdtable3}, Columns 1-3).\footnote{The PM2.5 data is available only at the annual level. Therefore, to examine seasonal effects, we use monthly emissions data from the Modern-Era Retrospective analysis for Research and Applications, Version 2 (MERRA-2) \citep{gelaro2017}. We focus on black carbon and organic carbon, the two major pollutants commonly associated with biomass burning, and which form precursor elements that lead to PM2.5 particles' formation in the atmosphere \citep{Cusworth2018}.} 

However, if the increase in economic activity or use of transportation services is concentrated in harvest months, the above test would fail to rule out economic activity or vehicular emissions as a cause for increased local PM2.5 level. Therefore, we further show the impact on particulate pollution is driven almost entirely by the increase in emissions from biomass burning with no effect on emissions from other (non-biomass burning) anthropogenic sources (Table \ref{tab:rdtable3}, Columns 4-7). These results suggest that agricultural fires are the primary mechanism linking rural roads to increased local PM2.5 levels.

\paragraph{Is the increase in crop fires driven by changes in planting dates?}
Rural roads may decrease the already short turnaround time between rice harvest and wheat planting even further in two ways, increasing fire use. One, rural-road-induced labor exits may delay rice harvest, leaving little time to clear the residue before subsequent season's wheat planting. Two, road drainage and excavation may lower the water table in surrounding areas \citep{Tsunokawa1997}. Therefore, rice planting might be delayed, and farmers may be rushed to harvest and clear the monsoon season's crop (rice) and plant winter season's crop (wheat) on time. In Appendix \ref{dates}, we examine the effect of rural roads on satellite-based measures of harvest (end-)date and planting date for the monsoon season crop (rice). We fail to find evidence that access to rural roads affects harvest or planting dates for rice.

\paragraph{Is the increase in crop fires driven by increase in adoption of mechanized harvesting?}
Farm labor exits due to rural roads may also affect number of agricultural fires by increasing adoption of mechanized harvesting. \cite{Behrer2020} uses a difference-in-differences design to show India's Mahatma Gandhi National Rural Employment Guarantee Act (NREGA), which guaranteed rural employment, increased the incidence of agricultural fires. The author's results suggest income growth associated with NREGA induced farmers to invest in labor-saving technologies like combine harvesters, which leave more crop residue, increasing use of agricultural fires. Because rural roads do not increase ownership of mechanized farm equipment (Appendix \ref{asher_novosad}), such an indirect mechanism is unlikely to be operational in our study. Nevertheless, farmers often do not own  combines but rather rent a combine \citep{Shyamsundar2019}. Therefore, we use the 1999 and 2006 REDS to examine the effect of rural roads on village-level stock of agricultural machinery (Appendix \ref{reds}). We fail to find evidence that rural roads increase the local stock of combines. We also fail to find evidence that rural roads increase household use of hired mechanized agricultural equipment (tractors, combine harvesters, threshers etc.). 

\subparagraph{High vs. low mechanization at baseline.}
As a second test for the influence of combine harvesters on our results, like \citet{Behrer2020}, we examine whether districts where farmers could feasibly respond to farm labor exits by mechanizing their harvests saw larger increases in crop fires. We compare the effect of rural roads for districts with high (above median) mechanization at baseline to districts with low (below median) mechanization at baseline.\footnote{To generate an agricultural mechanization index at baseline, we use the 2001-02 Agricultural Input Census available here:   \url{http://inputsurvey.dacnet.nic.in/districttables.aspx}. The 2001-02 census collected information on the number of mechanized equipment used in ploughing, leveling, seeding and planting, tilling, spraying, harvesting (combines), threshing, cane crushing, etc. Unfortunately, district-level mechanization data are unavailable for two states in our sample -- Gujarat and Maharashtra -- which are therefore excluded from the heterogeneity analysis.} We fail to find evidence for any significant differential effect of rural roads on crop fires (Table \ref{tab:hetero_mech}). In fact, the effect is slightly larger for low mechanization districts; in high mechanization districts, rural roads increase crop fires by 46\%, compared to 58\% in low mechanization districts. We find the same pattern of results if we restrict our sample to districts with a higher production of rice at baseline, a crop that would benefit from combine harvesters in the face of farm labor exits; rural roads increase crop fires by 110\% in high mechanization rice-growing districts, compared to 131\% increase in low mechanization rice-growing districts (Table \ref{tab:hetero_mech_rice}). We also find the same pattern of results if we split our sample by a mechanization index that only includes rice harvesting equipment (self propelled combine harvesters and tractor drawn combines); rural roads increase crop fires by 96\% in high mechanization rice-growing districts, compared to 104\% increase in low mechanization rice-growing districts (Table \ref{tab:hetero_mech_rice_v2}).

\subparagraph{High sugarcane vs. high rice at baseline.} As a final test for the influence of mechanized harvesting on our results, we leverage the fact that unlike mechanized rice harvesting, mechanized sugarcane harvesting should \textit{decrease} and not increase crop fires. Farmers light sugarcane fields to remove the outer leaves around the cane stalk before harvesting the cane manually to make the process
easier and require less labor \citep{Jain2014}. However, mechanical harvesting does not require burning \citep{davis2017did,Rangel2019}.  Therefore, if farm labor exits motivate sugarcane growers to adopt mechanized harvesting \citep{davis2017did}, one would expect to find a decrease in crop fires in districts with a high (above median) baseline production of sugarcane. 

We find the opposite result. We find that in high sugarcane districts, like in high rice districts, rural roads increase crop fires. In fact, the point estimates are almost identical: rural roads increase crop fires by 104\% in sugarcane-growing districts, compared to 103\% increase in rice-growing districts (Table \ref{tab:ricehi_sugarhi}). We find the same pattern of results if we restrict our sample to high sugarcane but low (below median) rice and high rice but low sugarcane districts at baseline: rural roads increase crop fires by 89\% in high sugarcane-growing but low rice-growing districts, compared to 83\% increase in high rice-growing but low sugarcane-growing districts (Table \ref{tab:ricehi_sugarhi_only}).

\paragraph{Is the increase in school enrollment responsible for the increase in crop fires?}

\cite{adukia2020} show rural road construction under PMGSY significantly increased middle school enrollment. If, prior to road construction, children were employed to harvest sugarcane or clear rice harvest residue, it is possible that the increase in crop fires is driven by \textit{child} farm labor exits due to increase in school enrollment. 

To test this hypothesis, we examine whether crop fires increased most (least) where enrollment increased most (least); \cite{adukia2020} show enrollment increased most where road connections are expected to raise the returns to education the most and least where roads are expected to most raise the opportunity cost of education. We proxy for the expected size of the opportunity cost effect with the district-level urban-rural wage gap. To proxy for the expected size of the returns to education effect, we identify the difference in returns to education between individuals in rural and urban areas.\footnote{The assumption underlying their heterogeneity analysis is that reductions in transportation costs will lead to factor price equalization: when a rural village receives a new road, its wages and returns to education will adjust toward the wages and returns in the broader geographic area.} We define binary indicator measures for the opportunity cost proxy and the returns to education proxy, based
on whether each proxy is above the sample median. 

We fail to find evidence that increase in school enrollment is responsible for the increase in crop fires (Table \ref{tab:hetero_educ}). First, we find the increase in crop fires is concentrated in districts where roads are expected to most raise the opportunity cost of education, with small and statistically insignificant effects in other districts, the opposite of what one would expect if increase in school enrollment were a mechanism. However, this result is exactly what one would expect if farm labor exits were a mechanism for increase in crop fires: after road construction, rural (farm) workers who faced the highest opportunity cost of (farm) work in the village were most likely to exit the local labor market in favor of (non-farm) work outside the village. Second, we fail to find evidence that fires increased most where road connections are expected to raise the returns to education the most: rural roads increased crop fires by 62\% in districts where road connections are expected to raise the returns to education the most, compared to 66\% in districts where road connections are expected to raise the returns to education the least. 

\paragraph{Is the increase in fire activity driven by non-farm biomass fires?}

As discussed in Section \ref{data}, we use data from NASA's EOSDIS to capture annual agricultural fires within 10 kilometers of each village. The fire activity data are based on detections of infrared radiation that are a signature of biomass fires \citep{Giglio2006}. Furthermore, these data have been used extensively by atmospheric scientists to study the effects of agricultural fires in India and elsewhere \citep{Liu2018}. However, the underlying algorithm does not separately identify farm fires from non-farm biomass fires (e.g., forest fires or trash fires). Therefore, if for some reason, the increase in non-farm biomass fires, like farm fires, is concentrated in the winter harvest and post-harvest months, it is possible the increase in fire activity in response to rural roads is driven by non-farm biomass fires and not farm fires. 

To evaluate this hypothesis, we examine the effect of rural roads on fire activity in cropland areas versus non-cropland areas (Table \ref{tab:fires_cropland}). If the increase in fire activity is due to non-farm biomass fires, one would expect to observe a disproportionate increase in fire activity in non-cropland areas. However, we find the opposite result. We find the increase in fires is concentrated in cropland areas with comparatively small and statistically insignificant effects in non-cropland areas. Therefore, it is unlikely that the increase in fire activity is driven by non-farm biomass fires.

\section{Did the Increase in Crop Fires Due to Rural Roads Increase Infant Mortality?} \label{infant}

In this section, we explore whether rural road construction is associated with changes in human health, as measured by changes in the infant mortality rate. Specifically, we estimate equations (1) and (2) with two key changes. 

First, because the infant mortality data is available for NFHS villages and not for PMGSY villages, we identify all NFHS villages within 50 km of each PMGSY village, and estimate infant mortality effects for these matched NFHS villages.\footnote{There exist no NFHS villages within a 50 km distance for 841 of the 11151 PMGSY villages in our main analysis sample. In Table \ref{tab:rdbaldown}, we show that unmatched PMGSY villages that had to be excluded from the infant mortality analysis are largely similar on baseline characteristics to matched PMGSY villages included in the analysis. We also show that, for the sample of matched PMGSY villages, there is no discontinuous change at the treatment threshold for (i) various village baseline characteristics (balance tests) or (ii) baseline values of our outcome variables (falsification tests).} The outcome variable in equation (2) is at the child-birth-year level where $Y_{invdst}$ takes the value 1 if child $i$ in NFHS village $n$, within 50 km of PMGSY village $v$ in district $d$ in state $s$, born in year $t$, died within the first year of birth, and 0 otherwise.

\begin{figure}[H]
   \begin{center}
    \caption{Schematic showing NFHS-IV villages located downwind versus NFHS-IV villages located in all other directions from a PMGSY village}
    \label{fig:schematic}
    \includegraphics[scale=0.9]{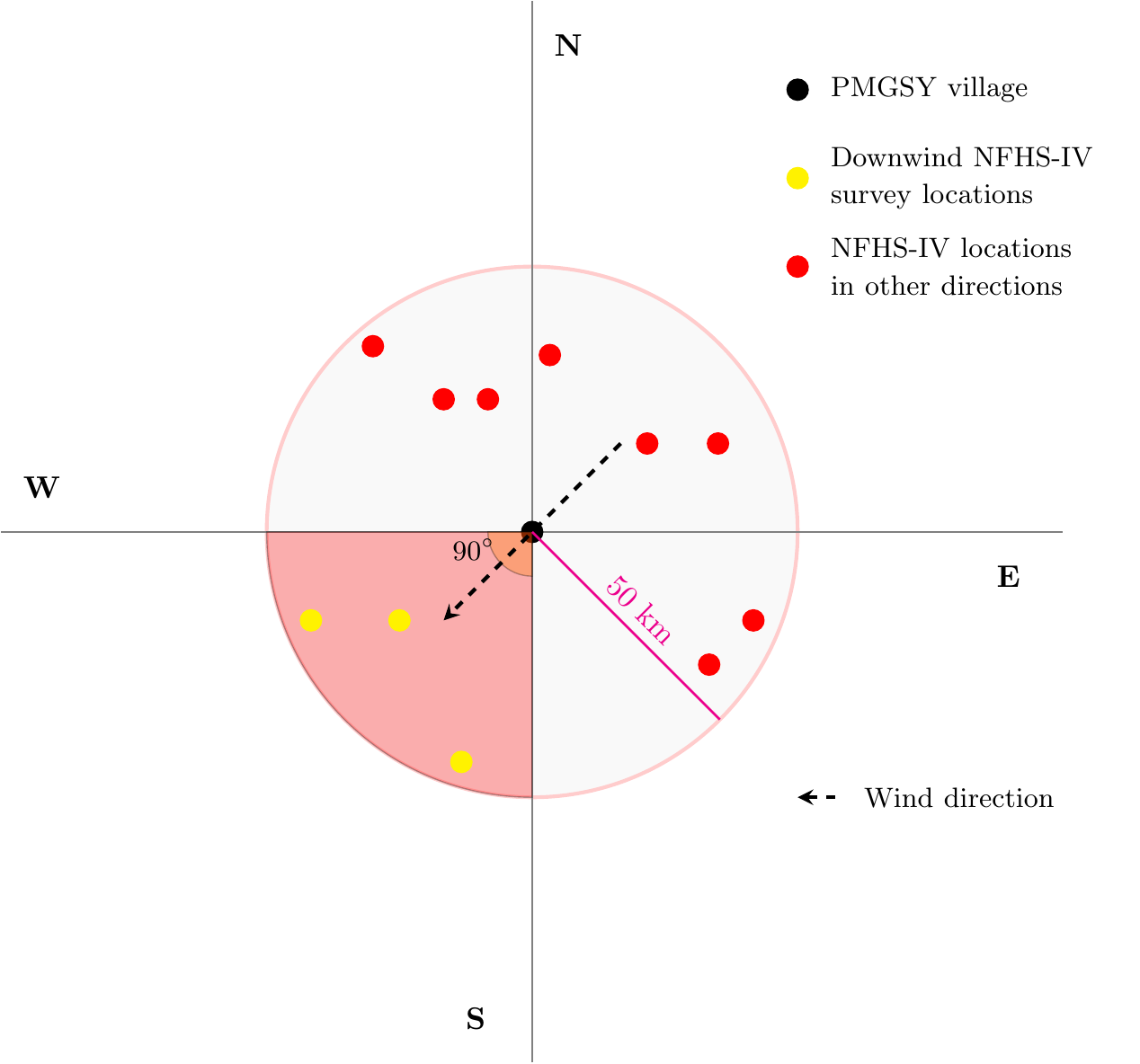}
    \end{center}
    	\scriptsize Notes: In the example depicted above, the prevailing wind direction at the PMGSY village is from the northeast to the southwest. Therefore, downwind locations are those NFHS-IV survey clusters within a 45-degree angle around the wind vector southwest of the PMGSY village. NFHS-IV survey villages outside of this cone comprise non-downwind (other directions) locations. \\
\end{figure}

Second, following a growing literature that exploits
pollution variation from wind direction (e.g., \citealp{Pullabhotla2022,Rangel2019, Schlenker2016, Deryugina2019}), we estimate effects for NFHS villages located within 50 km downwind from PMGSY villages separately from NFHS villages located within 50 km in all other directions from PMGSY villages (Figure~\ref{fig:schematic}).\footnote{We use wind direction data from a climate reanalysis dataset (ERA-5) to identify downwind locations by calculating the prevailing wind direction at each PMGSY village location. Prevailing wind direction is calculated by extracting monthly easterly ($u$) and northerly ($v$) wind vectors from ERA-5 for each PMGSY village location, averaging them to the annual-level, and calculating the direction corresponding to the annual wind vector.} 

The latter change allows us to distinguish the mechanism of air pollution associated with crop fires from other socioeconomic effects of rural roads. That is, it is likely that NFHS villages downwind from treated PMGSY villages experience disproportionate exposure to air pollution associated with crop fires compared to NFHS villages located in non-downwind directions from treated PMGSY villages. However, other socioeconomic effects of rural roads are likely not correlated with transient changes in wind direction. Therefore, comparing infant mortality estimates between downwind and non-downwind NFHS villages allows us to isolate the effect of air pollution from crop fires on infant mortality.

\begin{figure} [H]
	\begin{center}
		\caption{Impact of rural roads on PM 2.5 and infant mortality rates (IMR) for NFHS villages located downwind from PMGSY villages versus NFHS villages located in all other direction from PMGSY villages}	\label{fig:imr_rdplots}
		\subfigure[Downwind PM 2.5 ($\mu g/ m^{3}$)]{
			\includegraphics[scale=0.2]{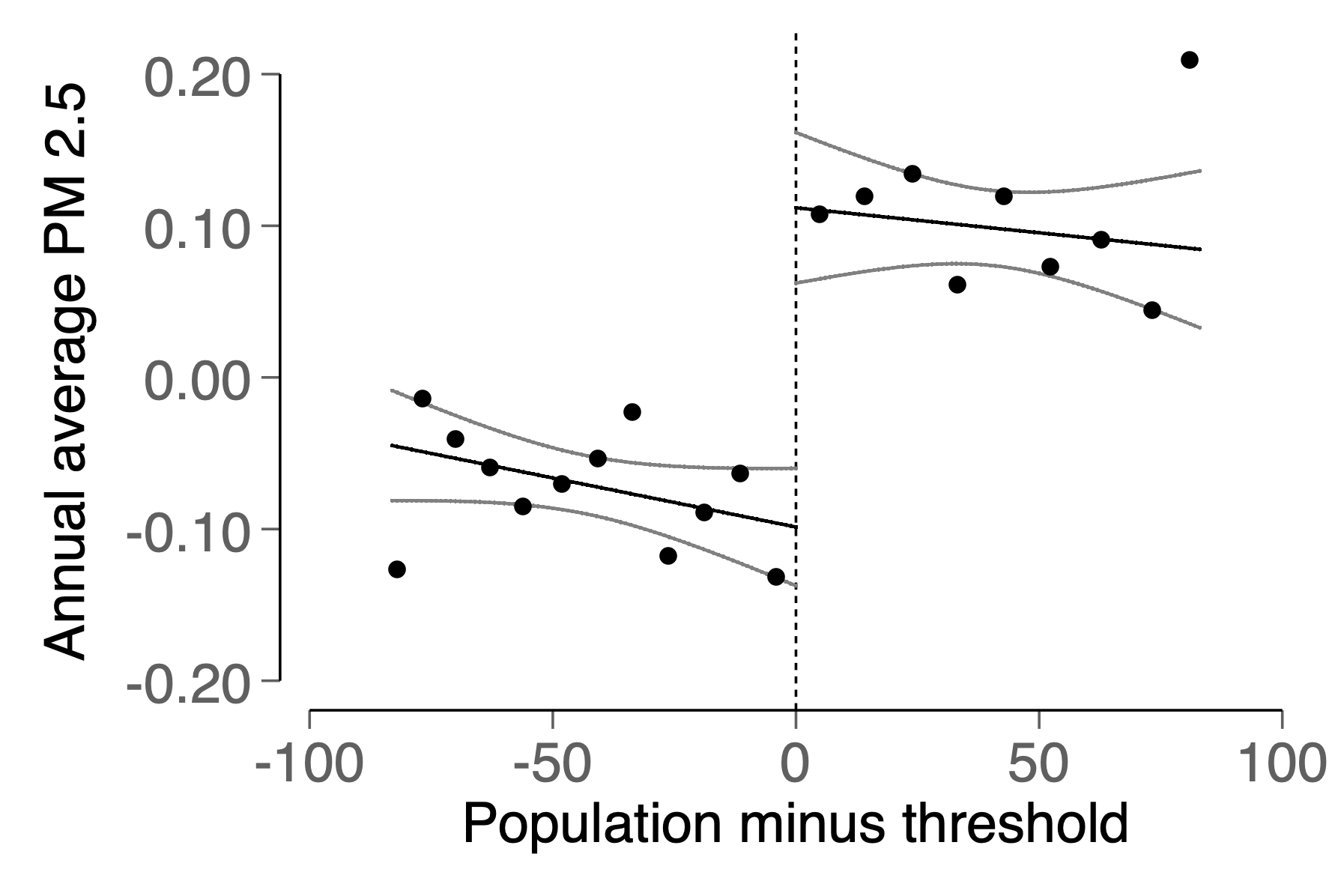} 
		}	
		\subfigure[Downwind IMR (0/1)]{
			\includegraphics[scale=0.2]{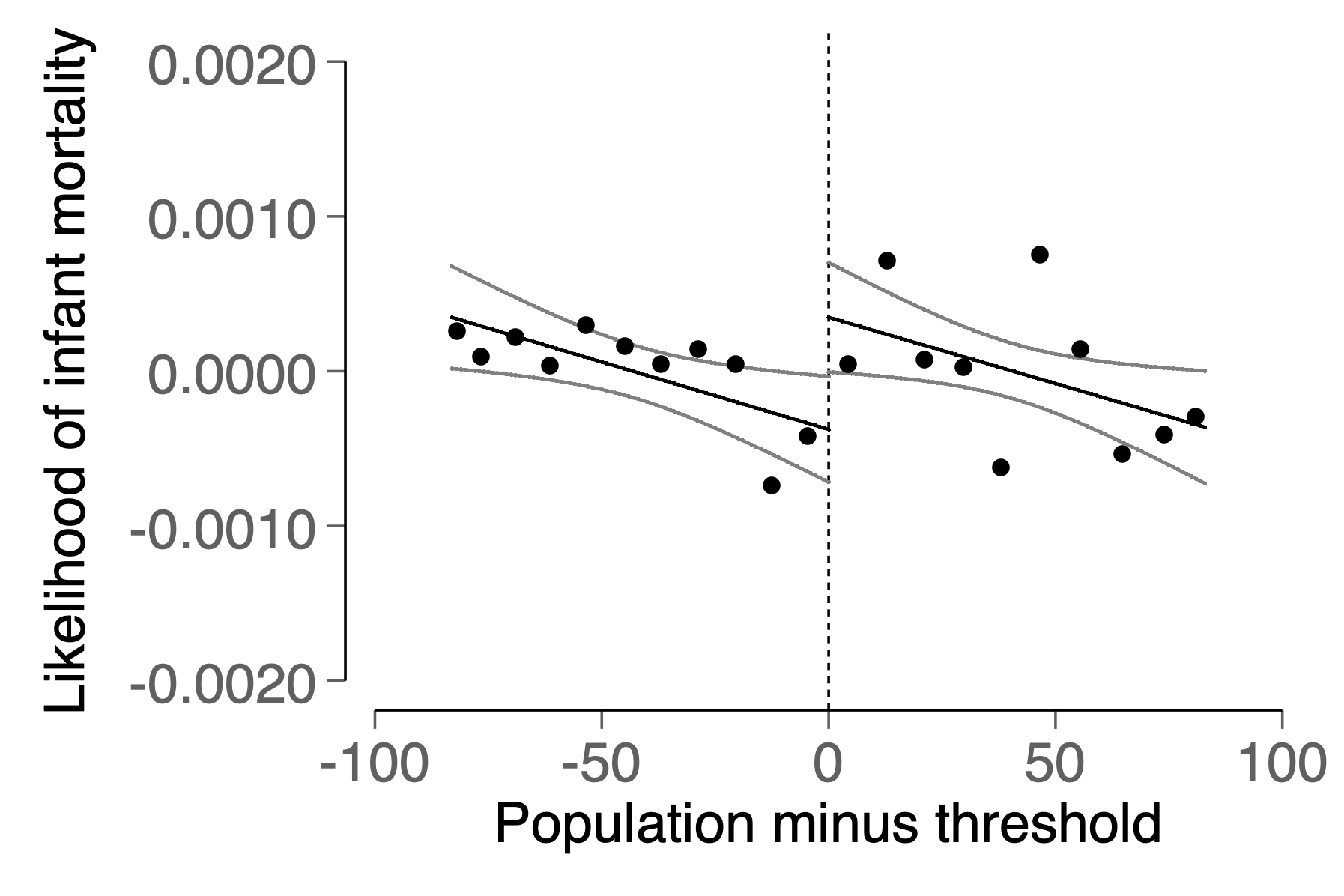}
		} 
		\subfigure[Other directions PM 2.5 ($\mu g/ m^{3}$)]{
			\includegraphics[scale=0.2]{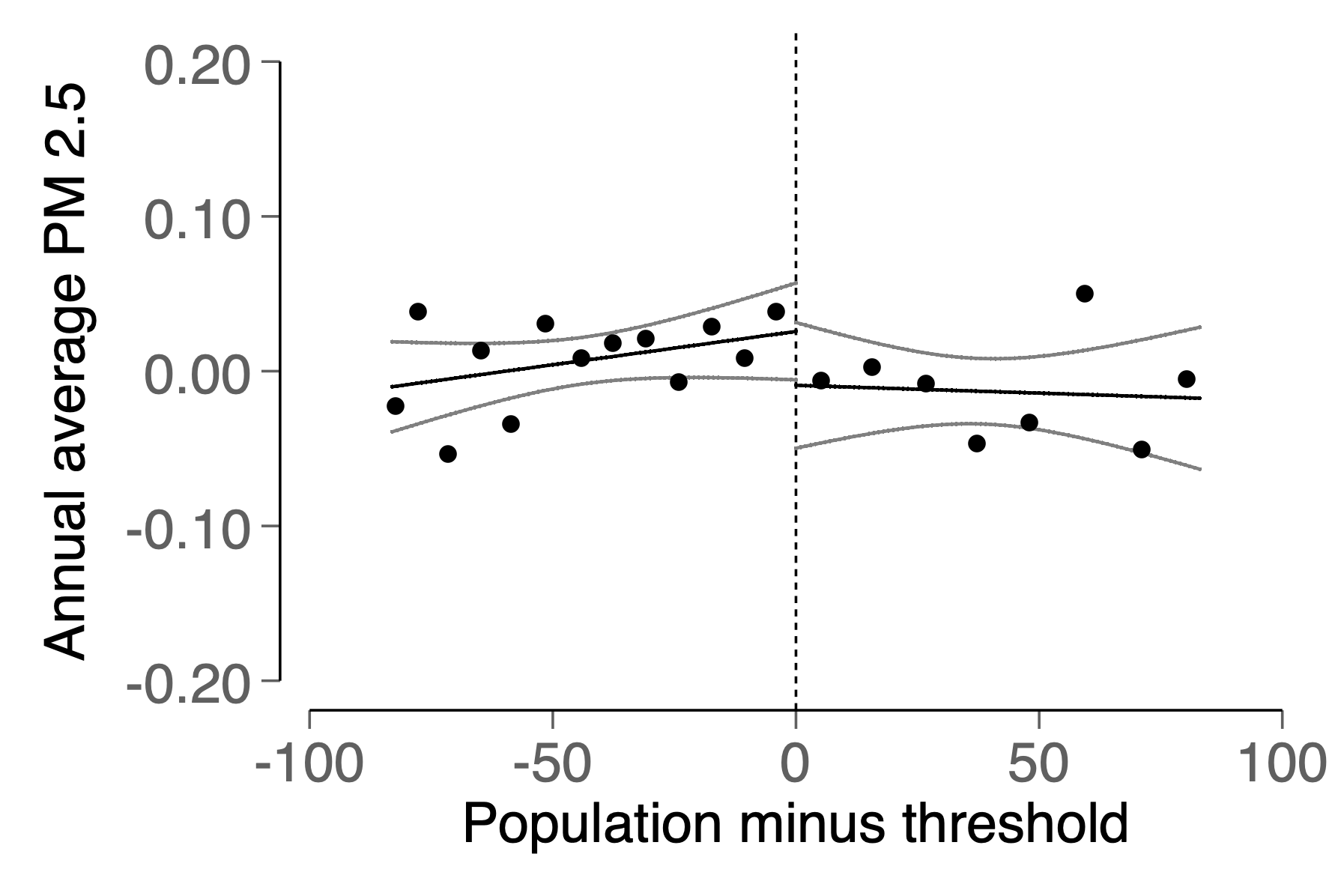} 
		}		
		\subfigure[Other directions IMR (0/1)]{
			\includegraphics[scale=0.2]{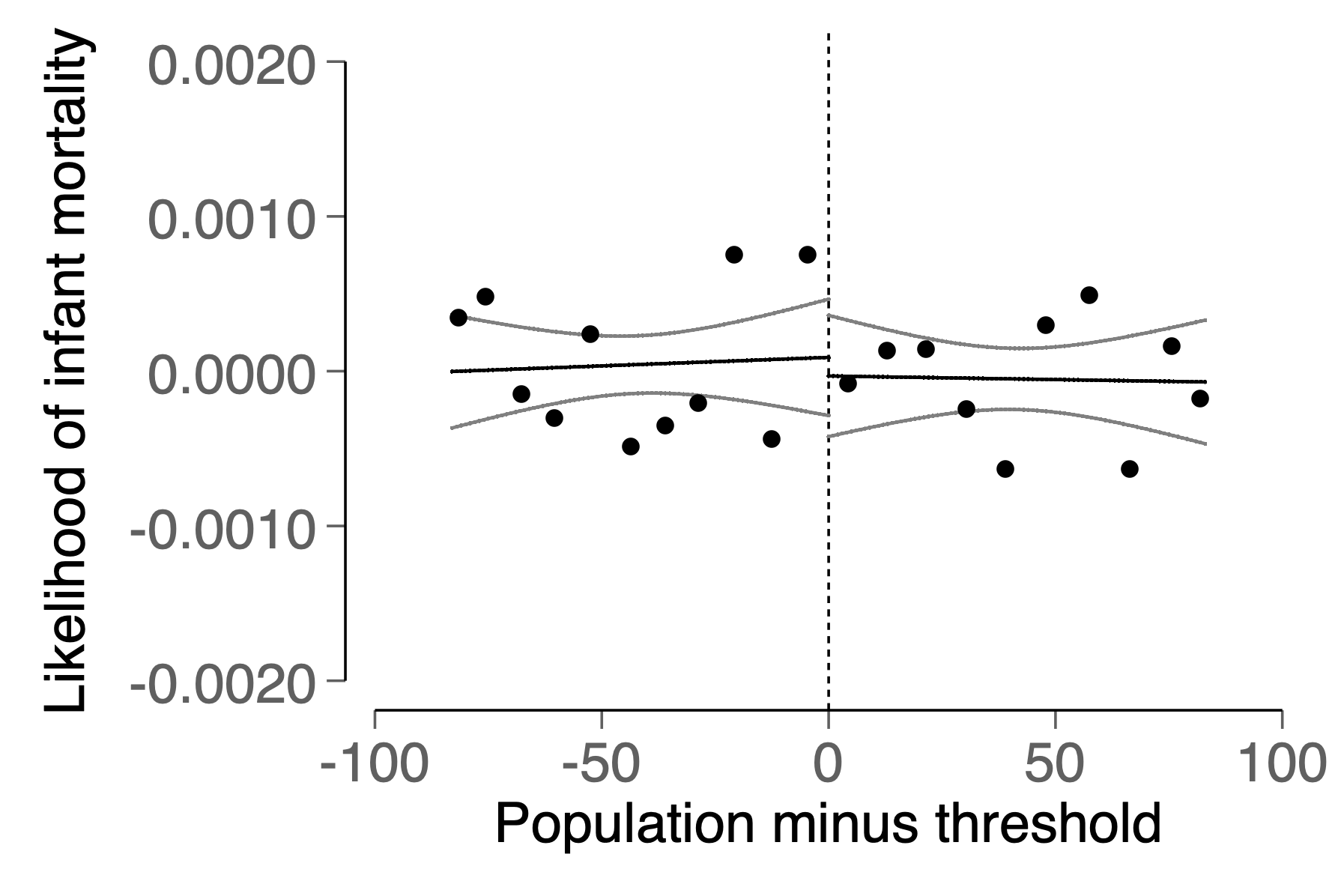}
		} 
	\end{center}
	\scriptsize Notes: All panels show reduced form regression discontinuity estimates by plotting the residualized values of outcomes as a function of the normalized 2001 village population relative to the threshold (after controlling for fixed effects and all baseline variables in the main specification other than population). Panels (a) and (b) plot the regression discontinuity relationship for annual average PM 2.5 and likelihood of infant death within 12 months of birth, respectively, in NFHS-IV villages that are downwind from PMGSY villages.  Panels (c) and (d) plots the relationship for PM 2.5 and infant mortality, respectively,  in NFHS-IV villages in all other directions from PMGSY villages. To define downwind, we calculate the prevailing vector-averaged annual wind direction at each PMGSY village. Then, as described in Figure \ref{fig:schematic}, we locate NFHS-IV villages  that are at a distance less than 50 km and lie within  45-degrees on either side of the wind direction vector for each PMGSY village. The outcomes in panels (c) and (d) are based on NFHS survey locations that are within 50 km but more than 45-degrees away from the wind direction vector. All panels control for district-threshold fixed effects, year fixed effects, and village characteristics in 2001 (baseline).  Population is centered around the state-specific threshold used for road eligibility - either 500 or 1000, depending on the state. Standard errors are clustered at the PMGSY village level. \\
	
\end{figure}

Figure \ref{fig:imr_rdplots} shows the graphical representation of the reduced form effects, while Table \ref{tab:rdimr_updn} presents the corresponding instrumental variable point estimates. Indeed, we find villages located within a 50 km distance downwind from treated villages observe a 2.3\% (1 $\mu$g/m$^3$) increase in annual average PM2.5 levels, with null effects for villages located within a 50 km distance in non-downwind directions from treated villages.\footnote{We find matched PMGSY villages at the population threshold observe a 0.72 $\mu$g/m$^3$ increase in PM2.5 levels (Table 
\ref{tab:fsiv_nfhsmatched}). The 1 $\mu$g/m$^3$ increase in PM2.5 levels for NFHS villages located within a 50 km distance downwind from treated PMGSY villages is consistent with the increase in PM2.5 levels in treated PMGSY villages; that is, because the same NFHS village may be within a 50 km distance from multiple treated PMGSY villages, the increase in PM2.5 levels for matched downwind NFHS villages should be at least as large as the increase in PM2.5 levels for PMGSY villages at the treatment threshold.} 

Correspondingly, we observe a statistically significant 5.5\% (0.3 percentage points) increase in infant mortality rate in villages located within a 50 km distance downwind from treated villages at the treatment threshold. However, as one might expect, we fail to find evidence for infant mortality effects in villages located within a 50 km distance in all other directions from treated villages. The point estimate is small and statistically insignificant. 

Overall, these estimates suggest a 1 $\mu$g/m$^3$ increase in PM2.5 levels increases infant mortality by 5.5\%. This increase in infant mortality is comparable to the effect of a 1 $\mu$g/m$^3$ increase in PM2.5 levels due to dust carried by Harmattan trade winds (5.8\%) in West Africa \citep{Adhvaryu2022}. It is smaller than the effect of a 1 $\mu$g/m$^3$ increase in PM2.5 levels due to automobile congestion (7.4\%) in the US \citep{Knittel2016}, but larger than the effect of a 1 $\mu$g/m$^3$ increase in PM2.5 levels due to expansion of natural gas infrastructure (2.8\%) in Turkey \citep{Cesur2017}, and due to the meteorological phenomenon of thermal inversions (0.7\%) in Mexico \citep{arceo2016does}. It is also larger than the effect of a 1 $\mu$g/m$^3$ decrease in PM2.5 levels due to the 1981-1982 recession (1.7\%) in the US \citep{Chay2003}.\footnote{To convert the effect of increase in PM10 or TSP levels on infant mortality across these studies to the effect of a 1 $\mu$g/m$^3$ increase in PM2.5 levels, we employ the estimated mean ratios of PM2.5/PM10 and PM2.5/TSP in the US \citep{Lall2004}.}

\section{Conclusion}

In this paper, we employ a regression discontinuity design that leverages a sharp increase in the likelihood of rural road construction at population thresholds in India to show rural roads increase agricultural fires by 60\% (1.6 additional fires) and particulate emissions by 1.1\% (0.5 $\mu g/m^3$). We find evidence that reallocation of farm labor to non-farm sectors due to rural roads is one mechanism responsible for the increase in agricultural fires. In effect, labor exits motivate the adoption of fire as a labor-saving but polluting technology to clear agricultural residue or to make harvesting less labor-intensive. Overall, the adoption of fires due to rural roads increases infant mortality by 5.5\% (0.3 percentage points) in villages located downwind of treated villages.  

The persistence of the agricultural productivity gap has generated considerable interest amongst governments and international agencies to devise policies that reduce frictions in labor reallocation across sectors. Our research does not imply such efforts are misguided or should be discouraged. Instead, our results underscore the need to complement these policies with strategies to mitigate their potential negative environmental externalities \citep{jayachandran_pes2022}. In our context, future research could investigate the design and implementation of monetary and non-monetary incentives to alter farmers' decisions to engage in using fire in agriculture in the face of soaring labor costs. 

\newpage

\singlespacing
\bibliographystyle{aea} 
\bibliography{gjp}
\newpage

\section*{Figures and Tables}

\newpage

\newpage

\begin{table}[H] 
	\centering 
	\caption{Main analysis sample: summary statistics, and balance and falsification tests}
	\label{tab:rdbal}
	\resizebox{\linewidth}{!}
		{\begin{threeparttable}
			\setlength{\linewidth}{.1cm} \begin{center} 
{\begin{tabular}{l c c c c c c c}
\hline\hline 
Variable                           & Full                              & Below                           & Over                            & Difference                      & p-value on                      & RD                                & p-value on                     \\ 
                                   & sample                            & threshold                       & threshold                       & of means                        & difference                      & estimate                          & RD estimate                    \\ 
\hline                                                               
Primary school (0/1)                     & 0.959           & 0.955           & 0.964           & 0.01           & 0.02           & -0.018           & 0.59           \\
Medical center (0/1)                     & 0.166               & 0.155               & 0.177               & 0.02               & 0.00               & -0.097               & 0.14               \\
Electrified (0/1)                        & 0.430                    & 0.414                    & 0.447                    & 0.03                    & 0.00                    & -0.014                    & 0.87                \\
Distance from nearest town (km)    & 26.490                    & 26.379                    & 26.613                    & 0.23                    & 0.58                    & -3.426                    & 0.34                    \\
Land irrigated (share)             & 0.281                & 0.276                & 0.287                & 0.01                & 0.05                & -0.025                & 0.59                \\
Ln land area                       & 5.151                  & 5.093                  & 5.215                  & 0.12                  & 0.00                  & -0.103                  & 0.33                  \\
\hline 
\end{tabular} }  
\end{center}
		\end{threeparttable}}
		\begin{minipage}{\linewidth} \footnotesize  
			\scriptsize Notes: This table presents mean values for village characteristics, measured at baseline. The first eight variables come from the 2001 Population Census, the next three (below the first line) come from the 2002 BPL Census, while the final six variables are our outcome variables measured at baseline (2001).  Columns 1-3 show the unconditional means for all villages, villages below the treatment threshold, and villages above the treatment threshold, respectively. Column 4 shows the difference of means across Columns 2 and 3, and Column 5 shows the p-value for the difference of means. Column 6 shows the regression discontinuity estimate, following the main estimating equation, of the effect of being above the treatment threshold on the baseline variable (with the outcome variable omitted from the set of controls), and Column 7 is the p-value for this estimate, using heteroskedasticity robust standard errors.
		\end{minipage}
\end{table}
	
\begin{table}[H] 
	\centering 
	\caption{First stage and IV estimates of impact of rural roads on agricultural fires (count) and PM2.5 ($\mu g/ m^{3}$)}
	\label{tab:rdtable1}
	\resizebox{0.8\linewidth}{!}{\begin{threeparttable}
			{
\def\sym#1{\ifmmode^{#1}\else\(^{#1}\)\fi}
\begin{tabular}{l*{3}{c}}
\hline\hline
                         &\multicolumn{1}{c}{New road}&\multicolumn{1}{c}{Annual fire activity}&\multicolumn{1}{c}{Annual average PM 2.5}\\\cmidrule(lr){2-2}\cmidrule(lr){3-3}\cmidrule(lr){4-4}
                         &\multicolumn{1}{c}{(1)}&\multicolumn{1}{c}{(2)}&\multicolumn{1}{c}{(3)}\\
                         &\multicolumn{1}{c}{$1^{st}$ stage}&\multicolumn{1}{c}{IV}&\multicolumn{1}{c}{IV}\\
\hline
Above threshold pop.     &       0.230***&               &               \\
                         &     (0.017)   &               &               \\
[1em]
Road built               &               &       1.590** &       0.463** \\
                         &               &     (0.667)   &     (0.199)   \\
\hline
N                        &     133,788   &     133,788   &     133,788   \\
Control group mean       &        0.10   &        2.77   &       42.59   \\
\hline\hline
\end{tabular}
}
		\end{threeparttable}}
		\begin{minipage}{\linewidth} \footnotesize \smallskip 
			\scriptsize Notes: This table shows  the first stage estimates (probability of receiving a rural road) as well as the effects of rural roads on count of agricultural fires and PM 2.5. The sample consists of the panel of villages from 2002 - 2013. ``Above threshold pop." is an indicator for the village population being above the treatment threshold. Column (1) shows the first stage, with the dependent variable (``Road built") taking the value 1 if the village received a new road during 2002-2013, 0 otherwise. Columns (2) and (3) present the IV estimates of the treatment effects of new roads on annual fire counts and annual average PM 2.5  ($\mu g/m^{3}$), respectively. All regressions include district-threshold fixed effects, year fixed effects, and baseline control variables. Standard errors in parentheses are clustered at village level. Significance at 1\%, 5\% and 10\% are indicated by $^{***}$, $^{**}$ and $^{*}$, respectively.
		\end{minipage}
		
	\end{table}
\newpage

\begin{table}[H] 
	\centering 
	\caption{Impact of rural roads on annual agricultural fires (count) and PM2.5 ($\mu g/ m^{3}$) by relative agricultural wage rate and rice or sugar production}
	\label{tab:heteroagwagecrops}
	\resizebox{0.8\linewidth}{!}{\begin{threeparttable}
			{
\def\sym#1{\ifmmode^{#1}\else\(^{#1}\)\fi}
\begin{tabular}{l*{4}{c}}
\hline\hline
                         &\multicolumn{2}{c}{\shortstack{Low rel. ag. wage\\with high rice or high sugar}}&\multicolumn{2}{c}{\shortstack{High rel ag. wage\\or low rel. ag wage with\\low rice \& low sugar}}\\\cmidrule(lr){2-3}\cmidrule(lr){4-5}
                         &\multicolumn{1}{c}{(1)}&\multicolumn{1}{c}{(2)}&\multicolumn{1}{c}{(3)}&\multicolumn{1}{c}{(4)}\\
                         &\multicolumn{1}{c}{Fires}&\multicolumn{1}{c}{PM 2.5}&\multicolumn{1}{c}{Fires}&\multicolumn{1}{c}{PM 2.5}\\
\hline
Road built               &       4.432** &       1.265***&       0.694   &       0.315   \\
                         &     (1.984)   &     (0.482)   &     (0.628)   &     (0.211)   \\
\hline
N                        &      55,020   &      55,020   &      72,636   &      72,636   \\
Control group mean       &        3.29   &       38.70   &        2.39   &       45.66   \\
\hline\hline
\end{tabular}
}
		\end{threeparttable}}
		\begin{minipage}{\linewidth} \footnotesize \smallskip 
			\scriptsize Notes: This table shows the effect of rural roads on count of village-level annual fires and village-level annual average PM 2.5 ($\mu g/ m^{3}$). The sample consists of the panel of villages for the period 2002 - 2013. ``Low rel. ag. labor wage with high rice or high sugar'' sample consists of districts which had low (below sample median) agricultural labor wages relative to non-agricultural labor wage rates in rural areas \emph{and} had high (above sample median) share of cropped area under rice \emph{or} sugarcane at baseline (2001). ``High rel ag. wage or low rel. ag wage with low rice \& low sugar'' consists of villages within districts which had either (i) high (above median) relative agricultural wage rates or (ii) low relative agricultural wage rates with low rice and sugar cropped areas. Rural agricultural and non-agricultural daily labor wage rates are based on the 1999 - 2000 NSSO survey data (Round 55). Rice and sugar cropped areas are based on ICRISAT district level data for 2001. Regressions include district-threshold fixed effects, year fixed effects and baseline control variables. Standard errors in parentheses are clustered at village level. Significance at 1\%, 5\% and 10\% are indicated by $^{***}$, $^{**}$ and $^{*}$, respectively.
		\end{minipage}
\end{table}

\begin{table}[H] 
	\centering 
	\caption{Impact of rural roads on agricultural fires (count) and particulate emissions ($ng/m^{2}/s$) by season and point sources}
	\label{tab:rdtable3}
	\resizebox{\linewidth}{!}{\begin{threeparttable}
			\begin{table}[H] \centering
\def\sym#1{\ifmmode^{#1}\else\(^{#1}\)\fi}
\begin{tabular*}{1\textwidth}{@{\extracolsep{\fill}}l*{7}{c}}
\hline\hline
\multicolumn{8}{c}{Panel A: Winter harvest and post-harvest months}\\
\midrule
                         &\multicolumn{1}{c}{Fires}&\multicolumn{2}{c}{All sources}&\multicolumn{2}{c}{Biomass burning}&\multicolumn{2}{c}{Other sources}\\\cmidrule(lr){2-2}\cmidrule(lr){3-4}\cmidrule(lr){5-6}\cmidrule(lr){7-8}
                         &\multicolumn{1}{c}{(1)}&\multicolumn{1}{c}{(2)}&\multicolumn{1}{c}{(3)}&\multicolumn{1}{c}{(4)}&\multicolumn{1}{c}{(5)}&\multicolumn{1}{c}{(6)}&\multicolumn{1}{c}{(7)}\\
                         &\multicolumn{1}{c}{ }&\multicolumn{1}{c}{BC}&\multicolumn{1}{c}{OC}&\multicolumn{1}{c}{BC}&\multicolumn{1}{c}{OC}&\multicolumn{1}{c}{BC}&\multicolumn{1}{c}{OC}\\
\hline
Road built               &       1.051***&       0.285** &       3.146*  &       0.320***&       2.791** &      -0.010   &      -0.010   \\
                         &     (0.391)   &     (0.130)   &     (1.779)   &     (0.103)   &     (1.181)   &     (0.009)   &     (0.009)   \\
\hline
N                        &     133,788   &     133,788   &     133,788   &     133,788   &     133,788   &     133,788   &     133,788   \\
Control group mean       &        1.45   &       30.47   &      120.81   &        1.17   &       13.46   &       29.30   &      107.35   \\
\end{tabular*}
\end{table}
		\end{threeparttable}}
		\begin{minipage}{\linewidth} \footnotesize \smallskip 
			\scriptsize Notes: This table shows  the effect of rural roads on  agricultural fires, and black carbon (BC) and organic carbon (OC) emissions in the winter harvest and post-harvest months (Panel A) and the rest of the year (Panel B) from 2002-2013. Winter harvest and post-harvest period comprises the months from October - March. Fire activity is measured in counts and emissions measured are in nano-gram per square meter per second ($ng/m^{2}/s$). Columns (2) and (3) show impact on emissions from all sources, columns (4) and (5) are emissions from biomass sources only, and columns (6) and (7) are emissions from other (non-biomass sources) sources. All regressions control for district-threshold FE, year FE, and baseline controls. Standard errors in parentheses are clustered at village level. Significance at 1\%, 5\% and 10\% are indicated by $^{***}$, $^{**}$ and $^{*}$, respectively. 
		\end{minipage}
		
	\end{table}

\begin{table}[H] 
	\centering 
	\caption{Impact of rural roads on PM 2.5 ($\mu g/m^{3}$) and infant mortality (0/1) in NFHS villages located downwind versus all other directions from PMGSY villages}
	\label{tab:rdimr_updn}
	\resizebox{0.75\linewidth}{!}{\begin{threeparttable}
			\begin{tabular}{lcccc}
   \toprule
    & \multicolumn{2}{c}{Downwind} & \multicolumn{2}{c}{Other directions} \\ 
                      & PM 2.5         & Infant mortality & PM 2.5   & Infant mortality\\
                      & (1)            & (2)              & (3)      & (4)\\
   \midrule 
   Road built         & 1.0072$^{***}$ & 0.0031$^{**}$    & -0.1551  & -0.0005\\
                      & (0.1979)       & (0.0015)         & (0.1423) & (0.0015)\\
   \midrule 
   Control group mean & 43.1560        & 0.0569           & 46.3760  & 0.0588\\
   Observations       & 129,405        & 9,648,574        & 104,052  & 8,191,400\\
   \bottomrule
\end{tabular}
		\end{threeparttable}}
		\begin{minipage}{\linewidth} \footnotesize \smallskip 
			\scriptsize Notes: This table shows  the effects of rural roads on  PM 2.5 and infant mortality in NFHS-IV clusters -- within 50 km of PMGSY villages -- located in downwind (columns 1 and 2) versus non-downwind (columns 3 and 4) directions. Road completion is instrumented using an indicator for baseline (2001 Census) village population above the program threshold.   Downwind locations are NFHS-IV sample locations that lie within 45-degrees on either side along the wind direction vector for each PMGSY village; locations outside of this comprise of the other directions sample. The outcome variables for columns (1) and (3) is the mean annual PM 2.5 ($\mu g/m^{3}$) averaged over all NFHS-IV locations  in downwind and non-downwind directions, respectively. The sample for columns (2) and (4) consists of all births recorded during 2002 - 2013 in the NFHS-IV locations in downwind and non-downwind directions of PMGSY villages, respectively.  The outcome variable ``infant mortality'' takes the value 1 if the child born died within the first year of birth, 0 otherwise.  All regressions  control for district-threshold fixed effects, year fixed effects, and baseline controls. Standard errors in parentheses are clustered at the PMGSY village level. Significance at 1\%, 5\% and 10\% are indicated by $^{***}$, $^{**}$ and $^{*}$, respectively. 
		\end{minipage}

\end{table}

\newpage
\appendix
	

\begin{center}	
	\textbf{\Large{Online Appendices}}
\end{center}
\setcounter{section}{0}
\pagenumbering{roman}
\setcounter{page}{1}
\renewcommand{\theHsection}{A\arabic{section}}
\renewcommand{\theHfigure}{A\arabic{figure}}

\section{Appendix Figures and Tables} 
\setcounter{table}{0}
\renewcommand{\thetable}{\Alph{section}.\arabic{table}}
\setcounter{figure}{0}
\renewcommand{\thefigure}{\Alph{section}.\arabic{figure}}
\numberwithin{table}{section}

\begin{figure}[H]
\begin{center}
	\caption{Annual average fire counts per state}
	\label{fig:statefires}
	\includegraphics[width=\linewidth]{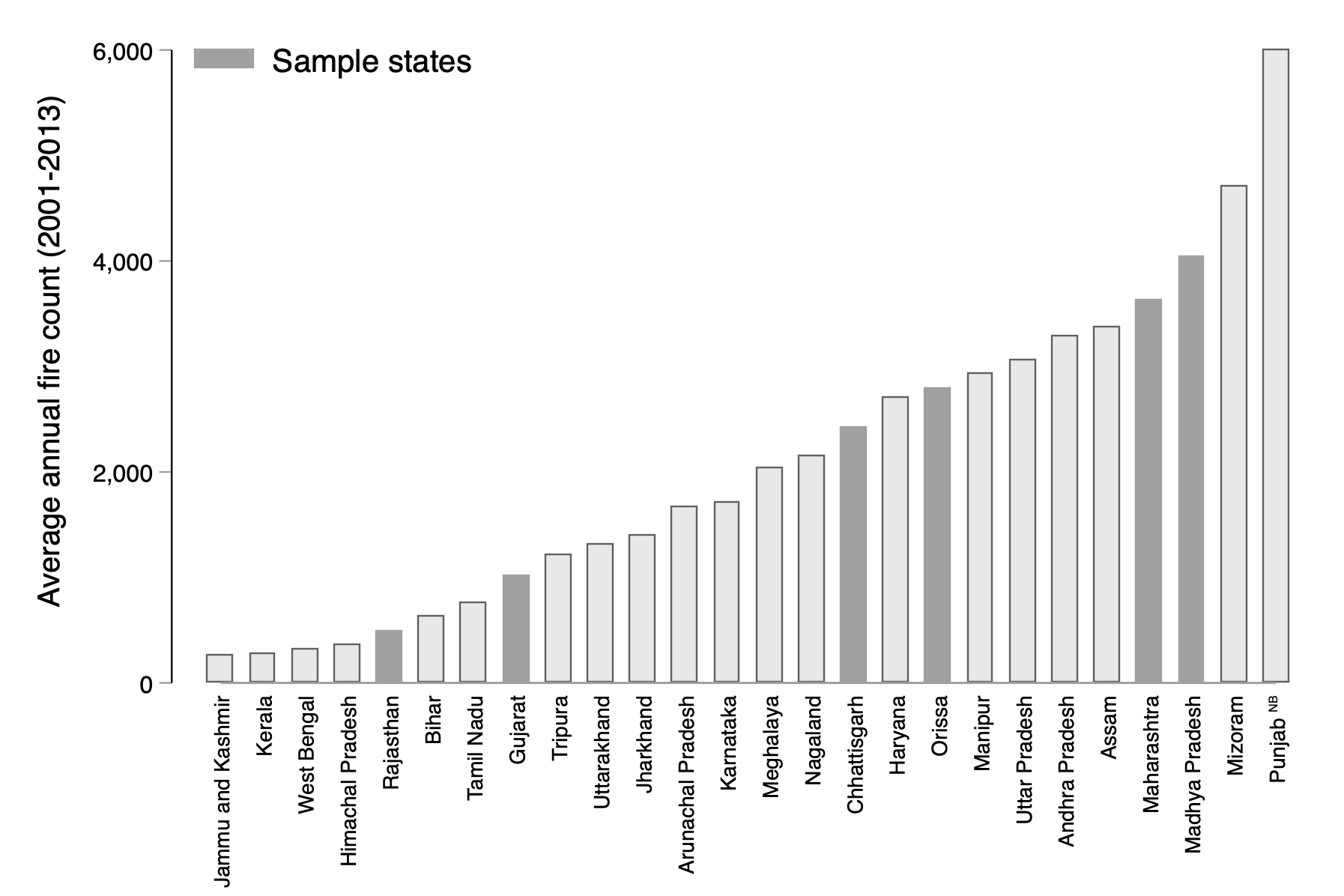}
\end{center}
	\scriptsize Notes: Figure shows the annual average fire counts for 2001 - 2013 for each state. States in the analysis sample are highlighted in darker shaded bars. Note that the value for Punjab is truncated at 6000 for ease of visual representation. The average fire counts per year for Punjab is $\approx$ 15,000 fires.
\end{figure}

\begin{figure}[H]
\begin{center}
	\caption{Annual number of fires and average PM 2.5 across India}
	\label{fig:annfires}
	\subfigure[Fire counts]{
	\includegraphics[width=0.6\linewidth]{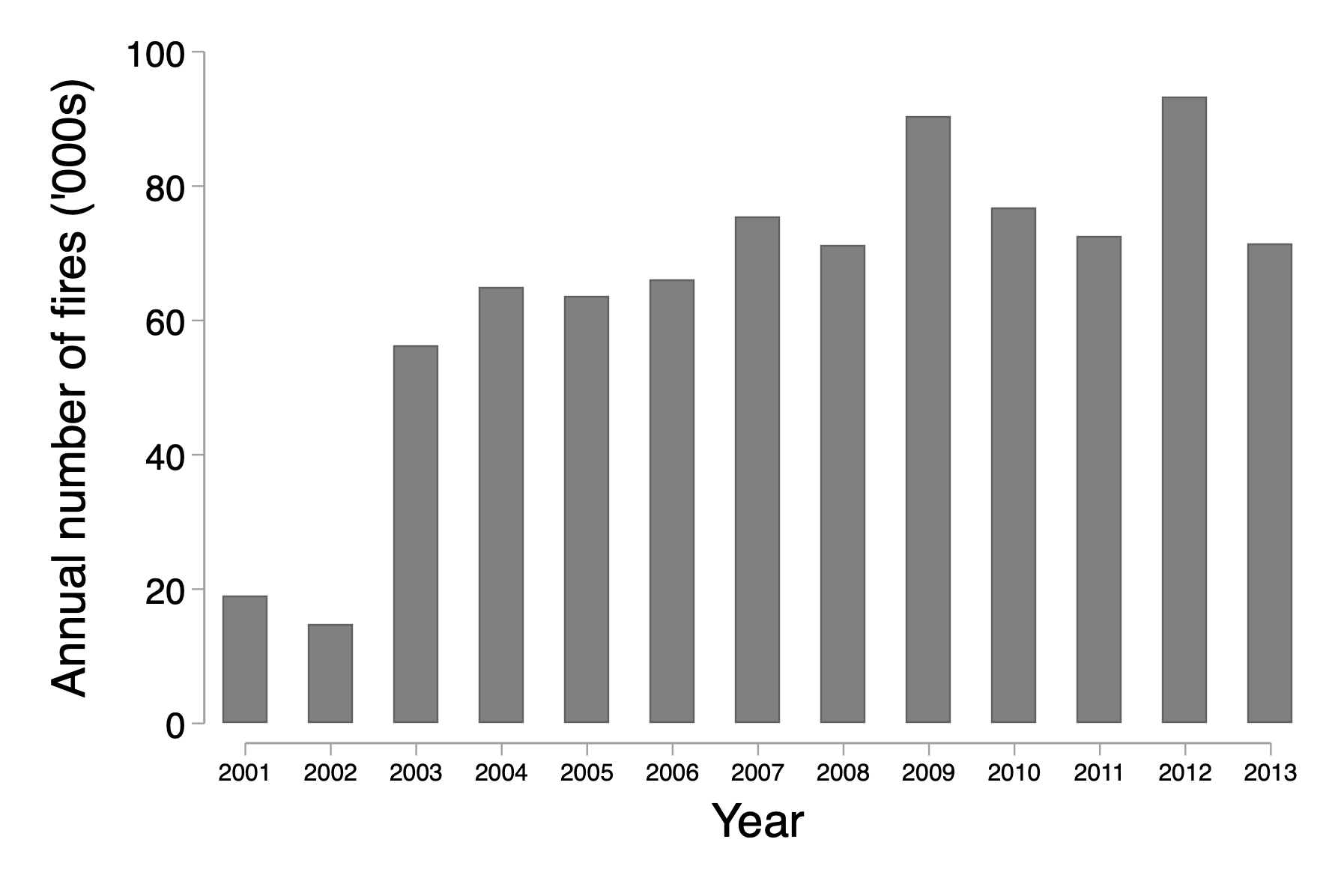}
	}
	\subfigure[PM 2.5]{
	\includegraphics[width=0.6\linewidth]{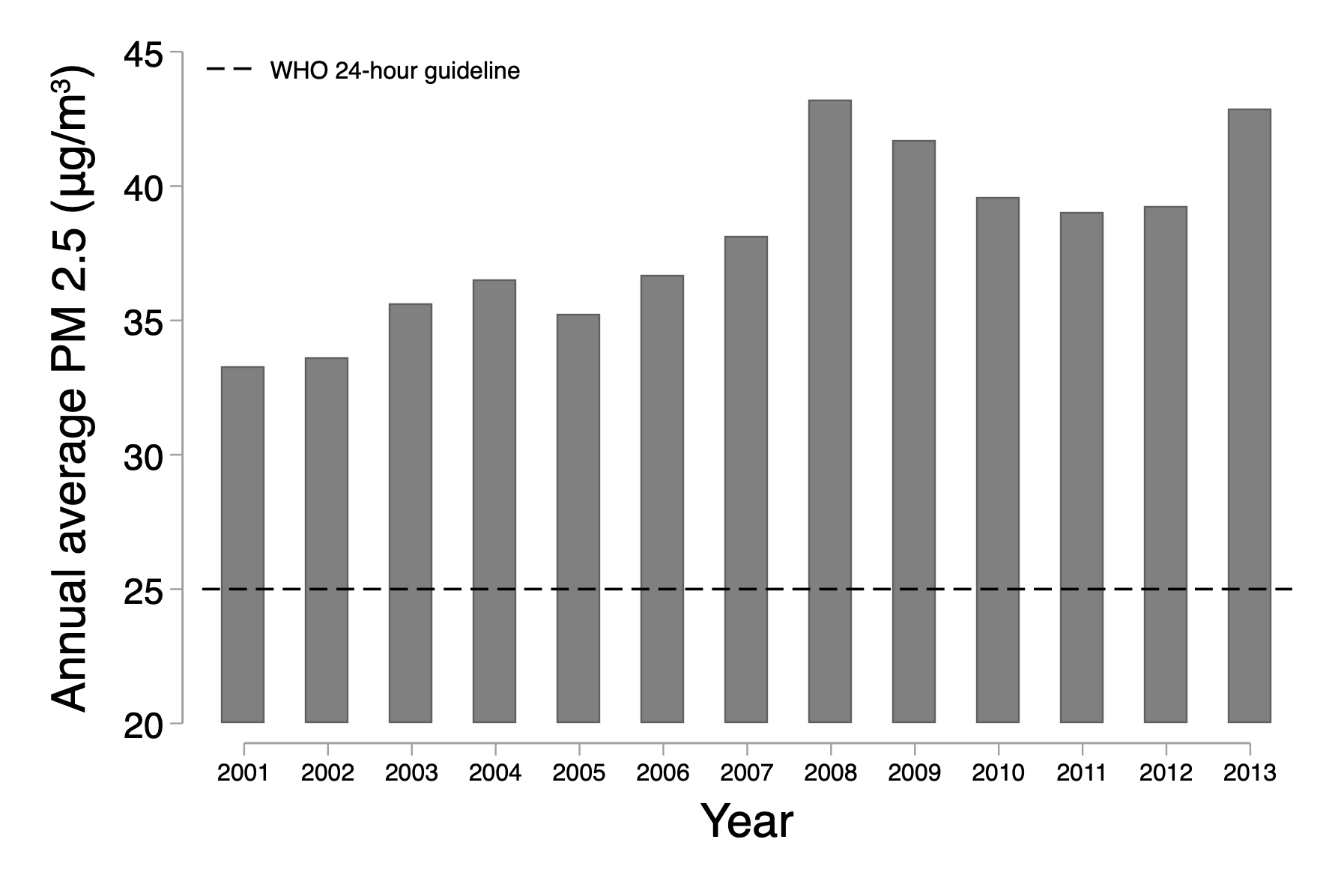}
	}
\end{center}
	\scriptsize Notes:  Panel (a) shows the number of fire pixels detected each year over India from MODIS TERRA and AQUA satellite data from 2001 to 2013. MODIS AQUA became operational only after July 2002. Therefore, the total number of fires detected for 2001 and 2002 are lower compared to later years when both TERRA and AQUA were in operation. Panel (b) shows the average annual PM 2.5 across India using data from \cite{VanDonkelaar2016}. The horizontal dashed line shows the WHO 24-hour guideline of 25 $\mu$g/$m^3$ for PM 2.5.
\end{figure}

\begin{figure}[H]
\begin{center}
	\caption{Distribution of annual average PM2.5 at the village level}
	\label{fig:pm25}
	\includegraphics[width=\linewidth]{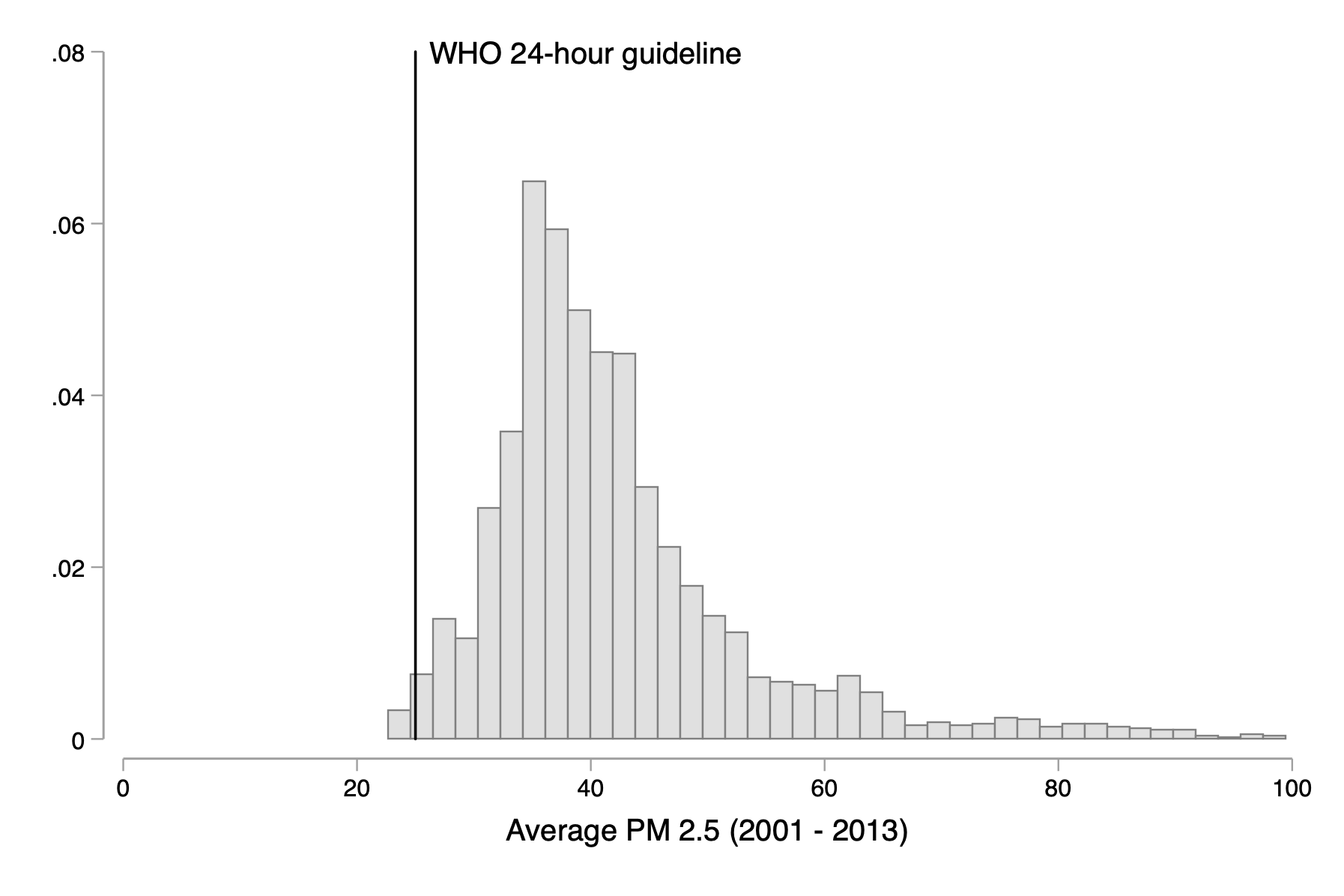}
	\end{center}
	\scriptsize Notes: Figure plots the distribution of annual average PM 2.5 ($\mu$g/$m^3$) between 2001 and 2013 in PMGSY villages. The vertical line indicates the WHO 24-hour guideline of 25 $\mu$g/$m^3$.
\end{figure}

\begin{figure}[H]
\begin{center}
	\caption{Roads constructed under the PMGSY between 2000 and 2013}
	\label{fig:roadsyears}
	\subfigure[All States]{
	    \includegraphics[width=0.8\linewidth]{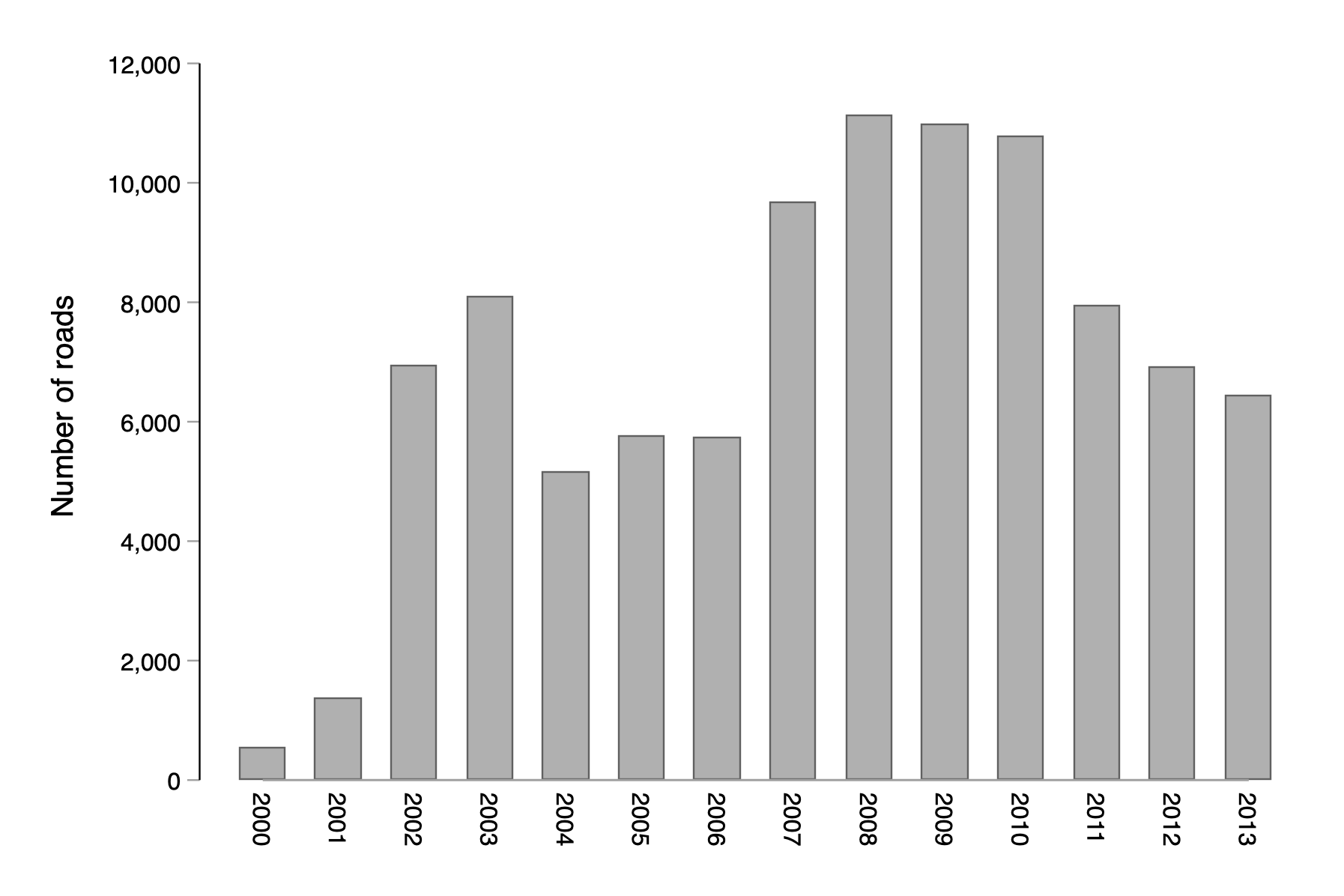}
	}
	\subfigure[Sample States]{
	    \includegraphics[width=0.8\linewidth]{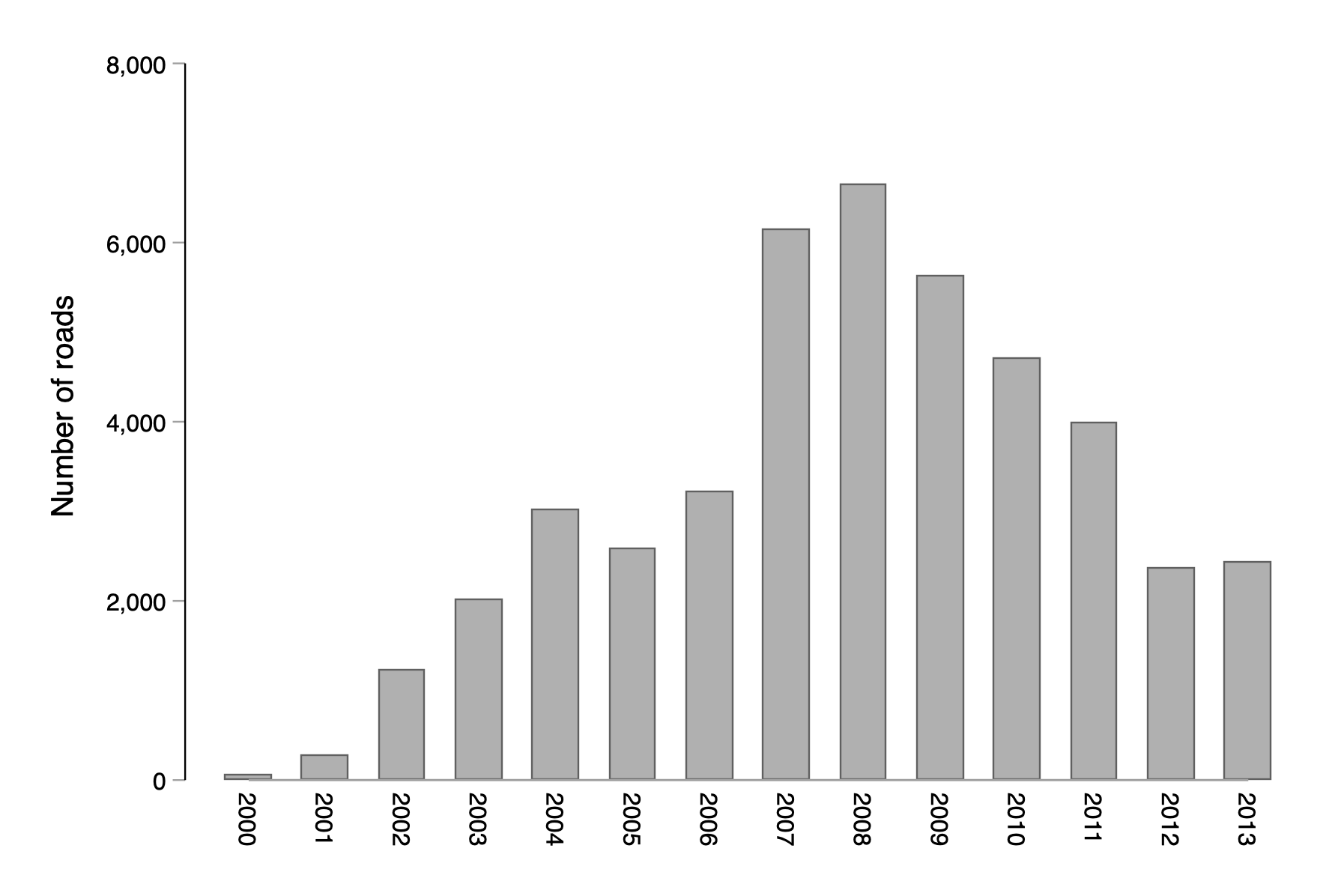}
	}
	\end{center}
	\scriptsize Notes: Panel (a) shows the total number of roads constructed per year under the PMGSY for all Indian states and Panel (b) shows total number of roads constructed per year in sample states. The road construction data is drawn from \cite{Asher2019} and \cite{Asher2020}.
\end{figure}

\begin{figure} [H]
	\begin{center}
		\caption{Correlation between farm labor share and crop fires and relative farm wage rate and crop fires at baseline}	\label{fig:corrplots}
		\subfigure[Farm labor share and crop fires (count)]{
			\includegraphics[scale=0.25]{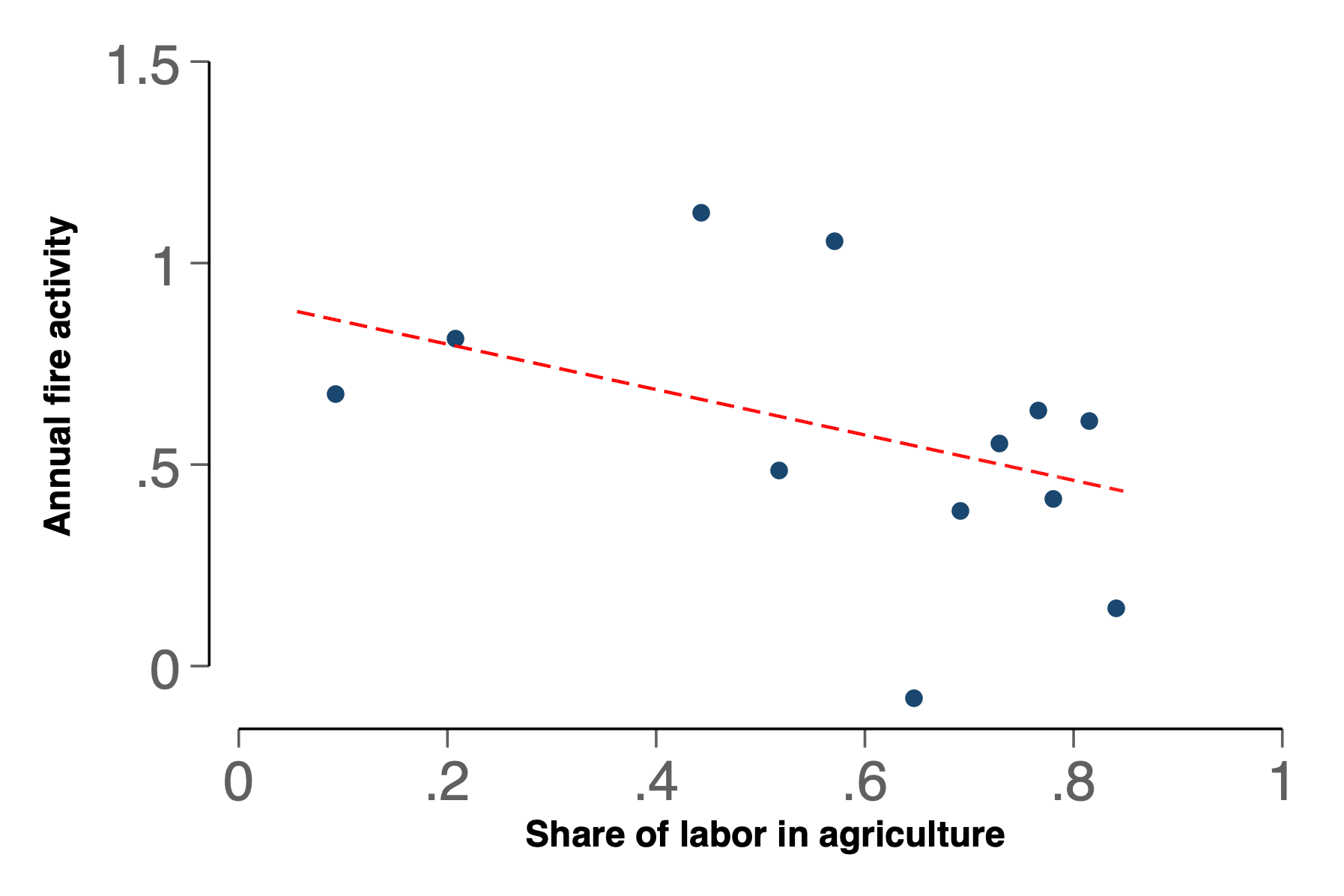}
		} 
		\subfigure[Relative farm wage rate and crop fires (count)]{
			\includegraphics[scale=0.25]{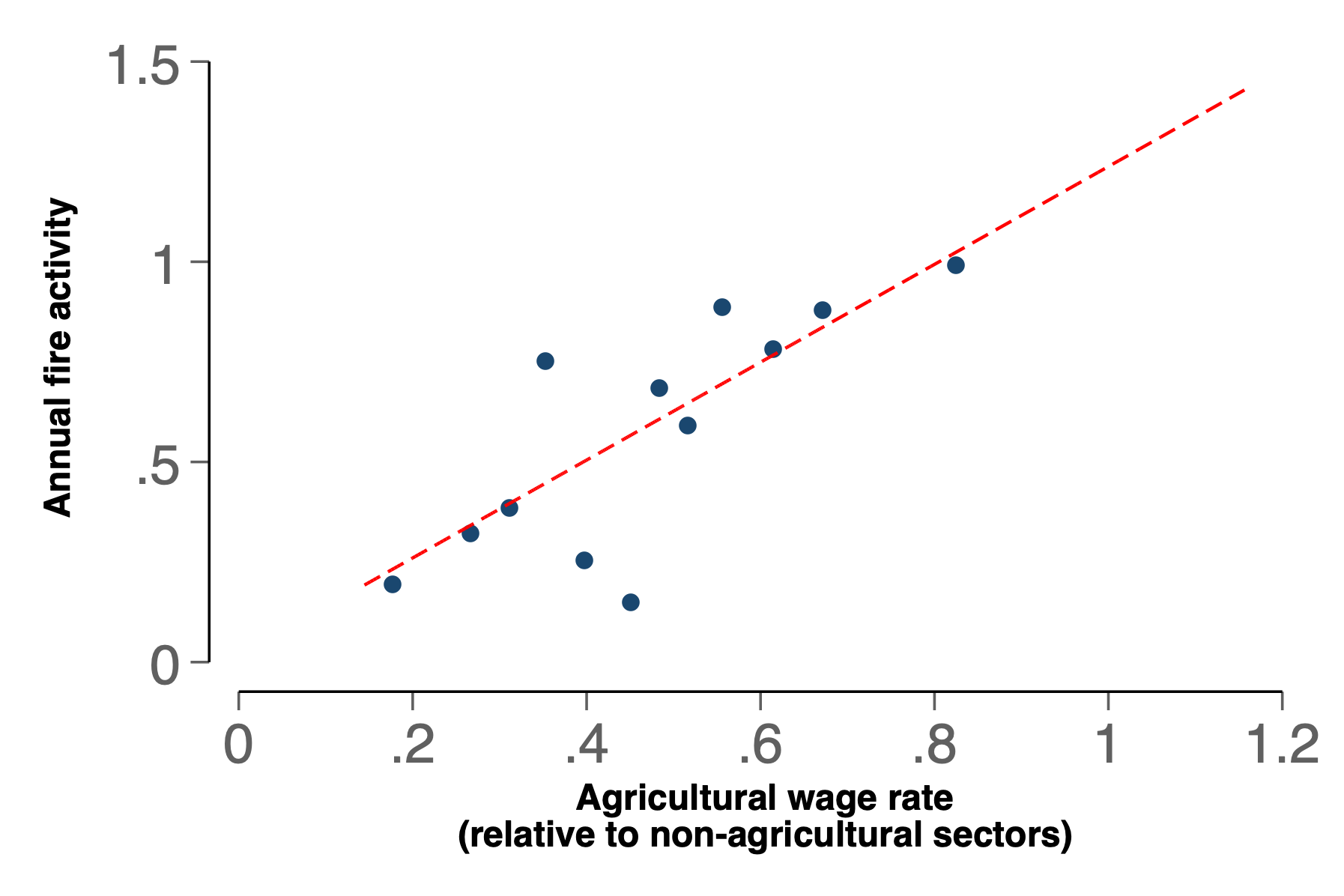} 
		}				
		
	\end{center}
	\scriptsize Notes: Both panels show a binned scatter plot and linear fit line, conditional on state fixed effects. Figure (a) shows the correlation between annual crop fires (count) and share of wage labor in agriculture in 2001. Figure (b) shows the correlation between annual crop fires (count) and relative agricultural wage rate in 2001. Rural labor shares and  agricultural and non-agricultural daily labor wage rates are district-level estimates based on data from the 1999 - 2000 NSSO survey data (Round 55). Annual crop fires is the baseline annual count of fires at the village level in 2001 within the sample of villages used in the regression discontinuity estimates. \\

\end{figure}

\begin{table}[H] 
	\centering 
	\caption{Impact of rural roads on annual agricultural fires (count) and PM2.5 ($\mu g/m^{3}$) by agricultural mechanization at baseline}
	\label{tab:hetero_mech}
	\resizebox{0.75\linewidth}{!}{\begin{threeparttable}
			{
\def\sym#1{\ifmmode^{#1}\else\(^{#1}\)\fi}
\begin{tabular}{l*{4}{c}}
\hline\hline
                         &\multicolumn{2}{c}{High mechanization index}&\multicolumn{2}{c}{Low mechanization index}\\\cmidrule(lr){2-3}\cmidrule(lr){4-5}
                         &\multicolumn{1}{c}{(1)}&\multicolumn{1}{c}{(2)}&\multicolumn{1}{c}{(3)}&\multicolumn{1}{c}{(4)}\\
                         &\multicolumn{1}{c}{Fires}&\multicolumn{1}{c}{PM 2.5}&\multicolumn{1}{c}{Fires}&\multicolumn{1}{c}{PM 2.5}\\
\hline
Road built               &       1.007*  &       0.272   &       1.937   &       0.345   \\
                         &     (0.576)   &     (0.251)   &     (1.404)   &     (0.262)   \\
\hline
N                        &      57,552   &      57,552   &      59,388   &      59,388   \\
Control group mean       &        2.17   &       47.52   &        3.31   &       39.21   \\
\hline\hline
\end{tabular}
}
		\end{threeparttable}}
		\begin{minipage}{\linewidth} \footnotesize \smallskip 
			\scriptsize Notes: This table shows the effects of rural roads on village-level annual fires and PM 2.5 ($\mu g/m^{3}$). The sample consists of the panel of PMGSY villages from 2002 - 2013. High (low) mechanization index sample consists of districts which had above (below) sample median  agricultural mechanization index at baseline (2001). Mechanization index is measured using district-level data on mechanized input usage from the Agricultural Input Census (2001). The Agricultural Input Census provides district-level counts of the mechanized equipment used for various agricultural operations. We standardize the individual equipment counts into $z$-scores and construct the mechanization index as the sum across all $z$-scores.  All regressions include district-threshold fixed effects, year fixed effects and baseline control variables. Standard errors in parentheses are clustered at village level. Significance at 1\%, 5\% and 10\% are indicated by $^{***}$, $^{**}$ and $^{*}$, respectively.
		\end{minipage}
\end{table}

\begin{table}[H] 
	\centering 
	\caption{Impact of rural roads on agricultural fires (count) and PM2.5 ($\mu g/m^{3}$) by baseline agricultural mechanization in high rice acreage districts }
	\label{tab:hetero_mech_rice}
	\resizebox{0.75\linewidth}{!}{\begin{threeparttable}
			{
\def\sym#1{\ifmmode^{#1}\else\(^{#1}\)\fi}
\begin{tabular}{l*{4}{c}}
\hline\hline
                         &\multicolumn{2}{c}{High mechanization index}&\multicolumn{2}{c}{Low mechanization index}\\\cmidrule(lr){2-3}\cmidrule(lr){4-5}
                         &\multicolumn{1}{c}{(1)}&\multicolumn{1}{c}{(2)}&\multicolumn{1}{c}{(3)}&\multicolumn{1}{c}{(4)}\\
                         &\multicolumn{1}{c}{Fires}&\multicolumn{1}{c}{PM 2.5}&\multicolumn{1}{c}{Fires}&\multicolumn{1}{c}{PM 2.5}\\
\hline
Road built               &       2.915*  &       0.839   &       5.267   &       1.166***\\
                         &     (1.622)   &     (0.570)   &     (3.373)   &     (0.408)   \\
\hline
N                        &      29,700   &      29,700   &      28,452   &      28,452   \\
Control group mean       &        2.63   &       39.83   &        4.00   &       34.52   \\
\hline\hline
\end{tabular}
}
		\end{threeparttable}}
		\begin{minipage}{\linewidth} \footnotesize \smallskip 
			\scriptsize Notes: This table shows the effects of rural roads on village-level annual fire activity and PM 2.5 ($\mu g/m^{3}$). The sample consists of the panel of PMGSY villages from 2002 - 2013 within districts with high (above sample median) share of cropped area under rice. High (low) mechanization index sample consists of districts which had above (below) sample median level of agricultural mechanization index at baseline (2001). Mechanization index is measured using district-level data on mechanized input usage from the Agricultural Input Census (2001). All regressions include district-threshold fixed effects, year fixed effects and baseline control variables. Standard errors in parentheses are clustered at village level. Significance at 1\%, 5\% and 10\% are indicated by $^{***}$, $^{**}$ and $^{*}$, respectively.
		\end{minipage}
\end{table}

\begin{table}[H] 
	\centering 
	\caption{Impact of rural roads on agricultural fires (count) and PM2.5 ($\mu g/m^{3}$) by baseline  mechanization specific to rice harvesting in high rice acreage districts }
	\label{tab:hetero_mech_rice_v2}
	\resizebox{0.75\linewidth}{!}{\begin{threeparttable}
			{
\def\sym#1{\ifmmode^{#1}\else\(^{#1}\)\fi}
\begin{tabular}{l*{4}{c}}
\hline\hline
                         &\multicolumn{2}{c}{High mechanization index}&\multicolumn{2}{c}{Low mechanization index}\\\cmidrule(lr){2-3}\cmidrule(lr){4-5}
                         &\multicolumn{1}{c}{(1)}&\multicolumn{1}{c}{(2)}&\multicolumn{1}{c}{(3)}&\multicolumn{1}{c}{(4)}\\
                         &\multicolumn{1}{c}{Fires}&\multicolumn{1}{c}{PM 2.5}&\multicolumn{1}{c}{Fires}&\multicolumn{1}{c}{PM 2.5}\\
\hline
Road built               &       2.845   &       1.730   &       3.562*  &       0.753** \\
                         &     (3.396)   &     (1.413)   &     (2.075)   &     (0.294)   \\
\hline
N                        &      14,928   &      14,928   &      43,224   &      43,224   \\
Control group mean       &        2.98   &       44.00   &        3.42   &       34.91   \\
\hline\hline
\end{tabular}
}
		\end{threeparttable}}
		\begin{minipage}{\linewidth} \footnotesize \smallskip 
			\scriptsize Notes: This table shows the effects of rural roads on village-level annual fire activity and PM 2.5 ($\mu g/m^{3}$). The sample consists of the panel of PMGSY villages from 2002 - 2013 within districts with high (above sample median) share of cropped area under rice. High (low) mechanization index sample consists of districts which had above (below) sample median level of agricultural mechanization index at baseline (2001). Mechanization index is measured using district-level data on mechanized input usage from the Agricultural Input Census (2001). We standardize the equipment counts for rice harvesting equipment -- self-propelled combine harvesters  and tractor drawn combines -- into $z$-scores and construct the mechanization index as the sum across all $z$-scores. All regressions include district-threshold fixed effects, year fixed effects and baseline control variables. Standard errors in parentheses are clustered at village level. Significance at 1\%, 5\% and 10\% are indicated by $^{***}$, $^{**}$ and $^{*}$, respectively.
		\end{minipage}
\end{table}

\begin{table}[H] 
	\centering 
	\caption{Impact of rural roads on agricultural fires (count) activity by size of the expected opportunity cost and returns to education effects}
	\label{tab:hetero_educ}
	\resizebox{0.8\linewidth}{!}{\begin{threeparttable}
			{
\def\sym#1{\ifmmode^{#1}\else\(^{#1}\)\fi}
\begin{tabular}{l*{4}{c}}
\hline\hline
                         &\multicolumn{2}{c}{Opportunity cost effect}&\multicolumn{2}{c}{Returns to education effect}\\\cmidrule(lr){2-3}\cmidrule(lr){4-5}
                         &\multicolumn{1}{c}{(1)}&\multicolumn{1}{c}{(2)}&\multicolumn{1}{c}{(3)}&\multicolumn{1}{c}{(4)}\\
                         &\multicolumn{1}{c}{High}&\multicolumn{1}{c}{Low}&\multicolumn{1}{c}{High}&\multicolumn{1}{c}{Low}\\
\hline
Road built               &       4.537***&      -0.186   &       1.773** &       1.686   \\
                         &     (1.629)   &     (0.646)   &     (0.871)   &     (1.055)   \\
\hline
N                        &      65,172   &      68,616   &      67,620   &      64,596   \\
Control group mean       &        3.20   &        2.35   &        2.87   &        2.53   \\
\hline\hline
\end{tabular}
}
		\end{threeparttable}}
		\begin{minipage}{\linewidth} \footnotesize \smallskip 
			\scriptsize Notes: This table shows the effect of rural roads on village-level annual fire activity. The sample consists of the panel of villages from 2002 - 2013. Proxies for the opportunity cost and returns to education effects are estimated using pre-program survey data from the 1999 - 2000 NSSO survey data (Round 55). The size of the opportunity cost effect is proxied by the district-level mean low-skill urban wage minus the mean low-skill rural wage. The size of the returns to education effect is proxied by the difference between the urban and rural Mincerian returns to one additional year of education.  The Mincerian returns are estimated by regressing log wage on years of education, age, square of age, and the log of household land owned, separately for individuals in urban and rural locations in each district. For each measure, the sample is split into high/low based on whether the proxy is above/below the sample median.  Regressions include district-threshold fixed effects, year fixed effects and baseline control variables. Standard errors in parentheses are clustered at village level. Significance at 1\%, 5\% and 10\% are indicated by $^{***}$, $^{**}$ and $^{*}$, respectively.
		\end{minipage}
\end{table}

\newpage
\section{Replication of \cite{Asher2020}} \label{asher_novosad}
\setcounter{table}{0}
\renewcommand{\thetable}{\Alph{section}.\arabic{table}}
\setcounter{figure}{0}
\renewcommand{\thefigure}{\Alph{section}.\arabic{figure}}
\numberwithin{table}{section}

\begin{figure}[H]
	\begin{center}
		\caption{Distribution of running variable 	\label{fig:runningvar}}
		\subfigure[Histograph of Village Population]{
			\includegraphics[scale=0.4]{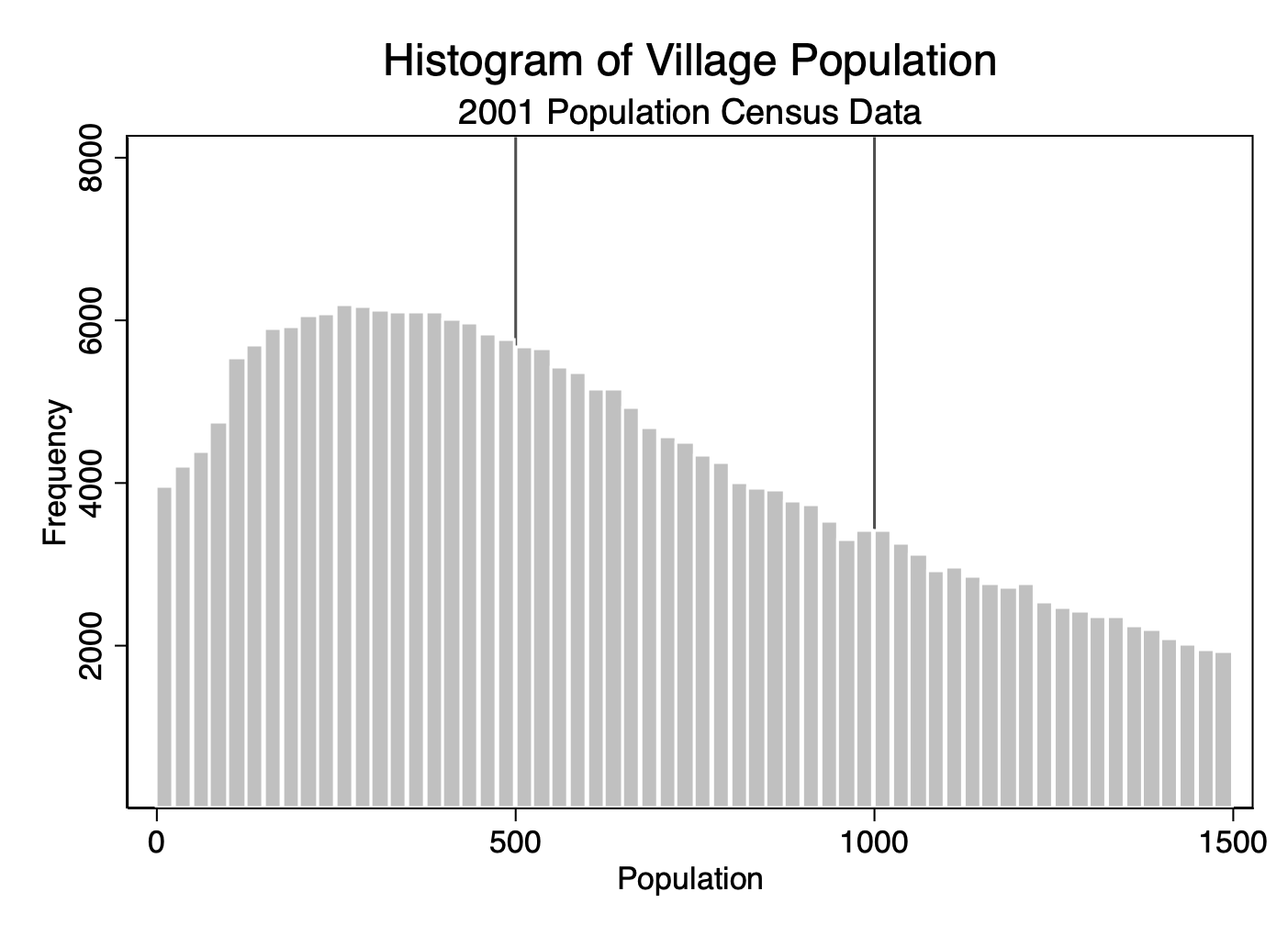}
		} 
		\subfigure[McCary Test]{
			\includegraphics[scale=0.4]{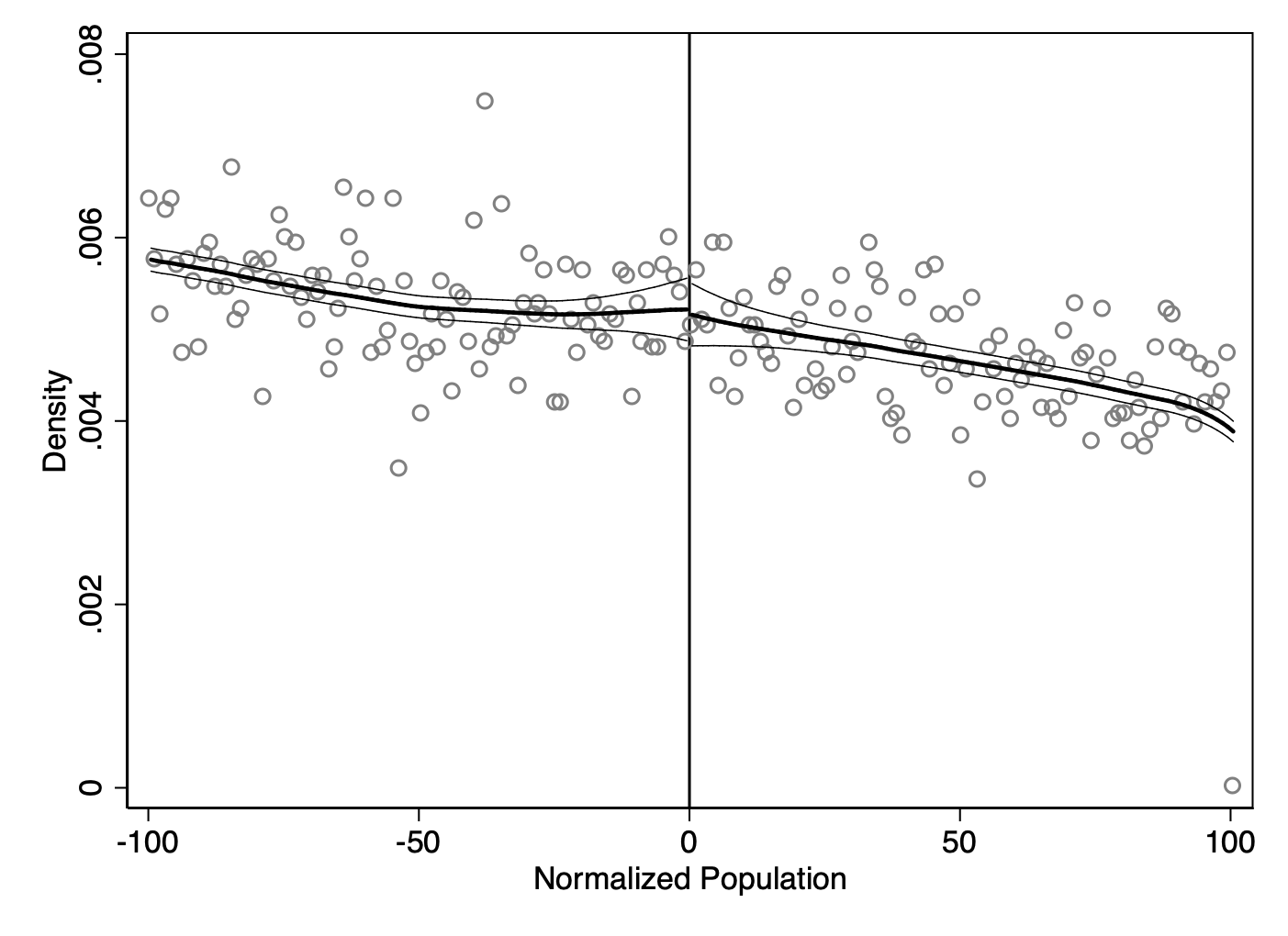} 
		}				
		
	\end{center}
	\scriptsize Notes: Panel (a) and (b) replicate results from Asher and Novosad (2020) showing the distribution of village population around the population thresholds. Panel (a) is a histogram of village population as recorded in the 2001 Population Census. The vertical lines show the program eligibility thresholds at 500 and 1,000. Panel (b) uses the normalized village population (reported population minus the threshold, either 500 or 1,000). It plots a non-parametric regression
	to each half of the distribution following \cite{McCrary2008}, testing for a discontinuity at zero. The point estimate for the discontinuity is  $-0.010$, with a standard error of $0.048$. \\
\end{figure}

\begin{table}[H] 
	\centering 
	\caption{Impact of rural roads on indices of major outcomes}
	\label{tab:indices}
	\resizebox{\linewidth}{!}{\begin{threeparttable}
			{
\def\sym#1{\ifmmode^{#1}\else\(^{#1}\)\fi}
\begin{tabular}{l*{5}{c}}
\hline\hline
                         &\multicolumn{1}{c}{(1)}&\multicolumn{1}{c}{(2)}&\multicolumn{1}{c}{(3)}&\multicolumn{1}{c}{(4)}&\multicolumn{1}{c}{(5)}\\
                         &\multicolumn{1}{c}{Transportation}&\multicolumn{1}{c}{Ag. occupation}&\multicolumn{1}{c}{Firms}&\multicolumn{1}{c}{Ag. production}&\multicolumn{1}{c}{Consumption}\\
\hline
Road built               &       0.432** &      -0.370** &       0.237   &       0.050   &       0.016   \\
                         &     (0.190)   &     (0.162)   &     (0.159)   &     (0.127)   &     (0.138)   \\
\hline
N                        &      11,151   &      11,151   &      10,405   &      11,151   &      11,151   \\
Control group mean       &       -0.02   &       -0.00   &        0.01   &        0.00   &       -0.00   \\
\hline\hline
\end{tabular}
}
	\end{threeparttable}}
	\begin{minipage}{\linewidth} \footnotesize \smallskip 
	\scriptsize Notes: This table replicates results from \cite{Asher2020} showing regression  discontinuity treatment estimates of the effect of new village roads on effect of a new road on indices of the major outcomes in each of the five families of outcomes: transportation, occupation, firms, agriculture, and welfare. Each index is generated following \cite{anderson2008multiple}. The transportation index consists of availability of public buses, private buses, vans, taxis and auto-rickshaws at the village level. The agricultural occupation index is a combination of the share of workers in agriculture and the inverse of the share of workers in manual labor. The firms index is the log of employment in all non-farm firms in the village. The agriculture production index is a combination of remotely-sensed Normalized Difference Vegetation Index (NDVI), share of households owning farm equipment, share of households owing irrigation equipment, share of households owning land, log total cultivated acres and presence of non-cereal or pulse crops among the primary three crops in the village. The consumption index consists of log per capita consumption, primary component asset index, log of night light luminosity and share of households with primary earner having a monthly earning of more than INR 5,000. All regressions include district-threshold fixed effects and baseline control variables.  Heteroskedasticity robust standard errors reported in parentheses. Significance at 1\%, 5\% and 10\% are indicated by $^{***}$, $^{**}$ and $^{*}$, respectively.
	\end{minipage}
\end{table}

\begin{table}[H] 
	\centering 
	\caption{Impact of rural roads on transportation}
	\label{tab:transport}
	\resizebox{0.8\linewidth}{!}{\begin{threeparttable}
			{
\def\sym#1{\ifmmode^{#1}\else\(^{#1}\)\fi}
\begin{tabular}{l*{5}{c}}
\hline\hline
                         &\multicolumn{1}{c}{(1)}&\multicolumn{1}{c}{(2)}&\multicolumn{1}{c}{(3)}&\multicolumn{1}{c}{(4)}&\multicolumn{1}{c}{(5)}\\
                         &\multicolumn{1}{c}{Govt. bus}&\multicolumn{1}{c}{Pvt. bus}&\multicolumn{1}{c}{Taxi}&\multicolumn{1}{c}{Van}&\multicolumn{1}{c}{Autorickshaw}\\
\hline
Road built               &       0.130** &       0.131*  &       0.012   &      -0.015   &       0.068   \\
                         &     (0.056)   &     (0.076)   &     (0.049)   &     (0.054)   &     (0.044)   \\
\hline
N                        &      11,151   &      11,151   &      11,151   &      11,151   &      11,151   \\
Control group mean       &        0.12   &        0.20   &        0.07   &        0.14   &        0.06   \\
$ R^{2} $                &        0.30   &        0.09   &        0.09   &        0.43   &        0.26   \\
\hline\hline
\end{tabular}
}
			\end{threeparttable}}
	\begin{minipage}{\linewidth} \footnotesize \smallskip 
	\scriptsize Notes: This table replicates results from \cite{Asher2020} showing regression  discontinuity treatment estimates of the effect of new village roads on availability of transportation services at the village level.  Columns (1) - (5) estimate the impact of new roads on five categories of transport services available at the village level from the Census in 2011.  Regressions include district-threshold fixed effects and baseline control variables.  Heteroskedasticity robust standard errors reported in parentheses. Significance at 1\%, 5\% and 10\% are indicated by $^{***}$, $^{**}$ and $^{*}$, respectively.
	\end{minipage}
\end{table}
	
\begin{table}[H] 
	\centering 
	\caption{Impact of rural roads occupation and income source}
	\label{tab:occup_inc_source}
	\resizebox{\linewidth}{!}{\begin{threeparttable}
			{
\def\sym#1{\ifmmode^{#1}\else\(^{#1}\)\fi}
\begin{tabular}{l*{6}{c}}
\hline\hline
                         &\multicolumn{4}{c}{Occupation}                                 &\multicolumn{2}{c}{Income source}\\\cmidrule(lr){2-5}\cmidrule(lr){6-7}
                         &\multicolumn{1}{c}{(1)}&\multicolumn{1}{c}{(2)}&\multicolumn{1}{c}{(3)}&\multicolumn{1}{c}{(4)}&\multicolumn{1}{c}{(5)}&\multicolumn{1}{c}{(6)}\\
                         &\multicolumn{1}{c}{Agriculture}&\multicolumn{1}{c}{Manual labor}&\multicolumn{1}{c}{Unemployed}&\multicolumn{1}{c}{Unclassifiable}&\multicolumn{1}{c}{Agriculture}&\multicolumn{1}{c}{Manual labor}\\
\hline
Road built               &      -0.100** &       0.086** &       0.011   &      -0.008   &      -0.045   &       0.003   \\
                         &     (0.044)   &     (0.044)   &     (0.025)   &     (0.010)   &     (0.045)   &     (0.044)   \\
\hline
N                        &      11,151   &      11,151   &      11,151   &      11,151   &      11,151   &      11,151   \\
Control group mean       &        0.48   &        0.45   &        0.43   &        0.02   &        0.41   &        0.51   \\
$ R^{2} $                &        0.29   &        0.26   &        0.30   &        0.17   &        0.31   &        0.28   \\
\hline\hline
\end{tabular}
}
		\end{threeparttable}}
	\begin{minipage}{\linewidth} \footnotesize \smallskip 
	\scriptsize Notes: This table replicates results from \cite{Asher2020} showing regression  discontinuity treatment estimates of the effect of new village roads on occupational choice among workers and households' primary income source based on data from the 2011 - 2012 Socioeconomic and Caste Census (SECC). Column (1) shows the effect on share of workers in agriculture, while column (2) shows the effect on share of workers in non-agriculture manual labor. These two categories make up 92\% of the workers, on average, in the sample villages.  To examine any potential changes in the workforce, columns (3) and (4) estimate the impact on the share of working age  adults (18-60) who are unemployed (including those who are studying, only involved in household work, etc.) and the share of working age adults whose work force participation is unclear in the SECC data, respectively. The results in columns (3) and (4) suggest that roads do not affect the share of adults in these categories, alleviating any concerns that the change in share of workers in agriculture or non-agricultural wage work is biased due to changes in the workforce. Additionally, results in Table~\ref{tab:pop_age_gender} show that rural roads had no effect on permanent out-migration. The absence of any impact on migration suggests that the observed sectoral reallocation of labor is the result of changes in occupational choice rather than compositional effects due to selective migration in the village population. Column (5) shows the impact on share of households in the village that report their main income source as cultivation. Column (6) shows the impact on share of households whose primary income source is manual labor. Regressions include district-threshold fixed effects and baseline control variables.  Heteroskedasticity robust standard errors reported in parentheses. Significance at 1\%, 5\% and 10\% are indicated by $^{***}$, $^{**}$ and $^{*}$, respectively.
	\end{minipage}
\end{table}
			
\begin{table}[H] 
	\centering 
	\caption{Impact of rural roads on employment in village non-farm firms}
	\label{tab:firms}
	\resizebox{0.9\linewidth}{!}{\begin{threeparttable}
			\begin{tabular}{l*{6}{c}} \hline
\hline \\ \multicolumn{7}{c}{\textit{Panel A: Log employment growth}} \\\\ [-1ex] \hline           &\multicolumn{1}{c}{(1)} & \multicolumn{1}{c}{(2)} &\multicolumn{1}{c}{(3)}          &\multicolumn{1}{c}{(4)} &\multicolumn{1}{c}{(5)} &\multicolumn{1}{c}{(6)} \\          &\multicolumn{1}{c}{Total} & \multicolumn{1}{c}{Livestock} &\multicolumn{1}{c}{Manufacturing}          &\multicolumn{1}{c}{Education} &\multicolumn{1}{c}{Retail} &\multicolumn{1}{c}{Forestry} \\ \hline
Road built               &       0.241   &       0.229   &       0.243   &       0.129   &       0.318** &      -0.107   \\
                         &     (0.162)   &     (0.190)   &     (0.196)   &     (0.145)   &     (0.157)   &     (0.110)   \\
\hline
N                        &      10,405   &      10,405   &      10,405   &      10,405   &      10,405   &      10,405   \\
\hline\hline \end{tabular}
	\end{threeparttable}}
	\begin{minipage}{\linewidth} \footnotesize \smallskip 
	\scriptsize Notes:  This table replicates results from \cite{Asher2020} showing regression  discontinuity treatment estimates of the effect of new village roads on employment in village level non-farm firms based on data from the 2013 Economic Census.  Column (1) shows the impact on total non-farm employment. Columns (2) - (6) show results for the five largest sectors in the sample which constitute 79\% of the non-farm firm employment (livestock, manufacturing, education, retail, and forestry). Panel A shows the impact on log employment in all non-farm firms and Panel B present same estimates using levels. The sample is limited to villages where total employment reported in the firm-level data is less than the total inhabitants in the village.  Regressions include district-threshold fixed effects and baseline control variables. Heteroskedasticity robust standard errors reported in parentheses. Significance at 1\%, 5\% and 10\% are indicated by $^{***}$, $^{**}$ and $^{*}$, respectively.
	\end{minipage}
\end{table}
				
\begin{table}[H] 
	\centering 
	\caption{Impact of rural roads on agricultural yields}
	\label{tab:agyields}
	\resizebox{0.9\linewidth}{!}{\begin{threeparttable}
		{
\def\sym#1{\ifmmode^{#1}\else\(^{#1}\)\fi}
\begin{tabular}{l*{6}{c}}
\hline\hline
                         &\multicolumn{3}{c}{NDVI}                       &\multicolumn{3}{c}{EVI}                        \\\cmidrule(lr){2-4}\cmidrule(lr){5-7}
                         &\multicolumn{1}{c}{(1)}&\multicolumn{1}{c}{(2)}&\multicolumn{1}{c}{(3)}&\multicolumn{1}{c}{(4)}&\multicolumn{1}{c}{(5)}&\multicolumn{1}{c}{(6)}\\
                         &\multicolumn{1}{c}{Max - June}&\multicolumn{1}{c}{Cumulative}&\multicolumn{1}{c}{Max}&\multicolumn{1}{c}{Max - June}&\multicolumn{1}{c}{Cumulative}&\multicolumn{1}{c}{Max}\\
\hline
Road built               &       0.028   &       0.002   &       0.015   &       0.043   &       0.000   &       0.024   \\
                         &     (0.027)   &     (0.013)   &     (0.014)   &     (0.034)   &     (0.016)   &     (0.019)   \\
\hline
N                        &      11,054   &      11,053   &      11,054   &      11,054   &      11,053   &      11,054   \\
Control group mean       &        8.24   &       10.52   &        8.81   &        7.96   &       10.17   &        8.48   \\
$ R^{2} $                &        0.70   &        0.88   &        0.80   &        0.71   &        0.85   &        0.69   \\
\hline\hline
\end{tabular}
}
	\end{threeparttable}}
	\begin{minipage}{\linewidth} \footnotesize \smallskip 
	\scriptsize Notes: This table replicates results from \cite{Asher2020} showing regression  discontinuity treatment estimates of the effect of new village roads on village-level measures of agricultural activity using three different remote-sensing NDVI-based proxies for agricultural yields. Regressions include district-threshold fixed effects and baseline control variables.  Heteroskedasticity robust standard errors reported in parentheses. Significance at 1\%, 5\% and 10\% are indicated by $^{***}$, $^{**}$ and $^{*}$, respectively.
	\end{minipage}
\end{table}

\begin{table}[H] 
	\centering 
	\caption{Impact of rural roads on agricultural inputs}
	\label{tab:aginputs}
	\resizebox{0.9\linewidth}{!}{\begin{threeparttable}
			{
\def\sym#1{\ifmmode^{#1}\else\(^{#1}\)\fi}
\begin{tabular}{l*{5}{c}}
\hline\hline
                         &\multicolumn{1}{c}{(1)}&\multicolumn{1}{c}{(2)}&\multicolumn{1}{c}{(3)}&\multicolumn{1}{c}{(4)}&\multicolumn{1}{c}{(5)}\\
                         &\multicolumn{1}{c}{Mech.}&\multicolumn{1}{c}{Irri.}&\multicolumn{1}{c}{Own ag. land}&\multicolumn{1}{c}{Non-cereal/pulse}&\multicolumn{1}{c}{Cult. land (log)}\\
\hline
Road built               &      -0.008   &      -0.009   &       0.005   &       0.002   &      -0.062   \\
                         &     (0.012)   &     (0.029)   &     (0.040)   &     (0.075)   &     (0.117)   \\
\hline
N                        &      11,150   &      11,151   &      11,151   &       8,001   &      10,885   \\
Control group mean       &        0.04   &        0.14   &        0.57   &        0.40   &        5.04   \\
$ R^{2} $                &        0.24   &        0.40   &        0.29   &        0.45   &        0.46   \\
\hline\hline
\end{tabular}
}
	\end{threeparttable}}
	\begin{minipage}{\linewidth} \footnotesize \smallskip 
	\scriptsize Notes: This table replicates results from \cite{Asher2020} showing regression  discontinuity treatment estimates of the effect of new village roads on the impact of roads on agricultural inputs based on data from the 2011-2012 Socioeconomic and Caste Census (Columns 1-3) and the 2011 Population Census (Columns 4-5). The outcome variables in Columns (1) and (2) are the share of households owning mechanized farm equipment and the share of households owning irrigation equipment, respectively. Column (3) shows the impact on the share of households owning agricultural land. The outcome in Column (4)  is a dummy variable taking the value 1 if a non-cereal and non-pulse crop is listed as one of its three major crops at the village-level census data. Column (5) examines the impact on  log total cultivated land in the village (villages that have no cultivable land are excluded from the sample). Regressions include district-threshold fixed effects and baseline control variables.  Heteroskedasticity robust standard errors reported in parentheses. Significance at 1\%, 5\% and 10\% are indicated by $^{***}$, $^{**}$ and $^{*}$, respectively.
	\end{minipage}		
\end{table}
	
\begin{table}[H] 
	\centering 
	\caption{Impact of rural roads on consumption and asset ownership}
	\label{tab:cons_assets}
	\resizebox{0.9\linewidth}{!}{\begin{threeparttable}
			\begin{tabular}{l*{4}{c}} \hline
\hline \\ \multicolumn{5}{c}{\textit{Panel A: Consumption indicators and asset index}} \\\\ [-1ex] \hline                           &\multicolumn{1}{c}{(1)} & \multicolumn{1}{c}{(2)} &\multicolumn{1}{c}{(3)}          &\multicolumn{1}{c}{(4)}  \\  &\multicolumn{1}{c}{Consumption per} & \multicolumn{1}{c}{Night lights (log)} &\multicolumn{1}{c}{Share of HH} &\multicolumn{1}{c}{Asset index} \\  &\multicolumn{1}{c}{capita (log)} &\multicolumn{1}{c}{} &\multicolumn{1}{c}{earning $\geq5k$} &\multicolumn{1}{c}{} \\ \hline
Road built               &       0.013   &       0.037   &      -0.008   &       0.120   \\
                         &     (0.039)   &     (0.169)   &     (0.032)   &     (0.135)   \\
\hline
N                        &      11,151   &      10,828   &      11,151   &      11,151   \\
Control group mean       &        9.56   &        1.58   &        0.15   &       -0.01   \\
$ R^{2} $                &        0.42   &        0.66   &        0.25   &        0.52   \\
\hline \\ \multicolumn{5}{c}{\textit{Panel B: Individual asset ownership}} \\\\ [-1ex] \hline                                   &\multicolumn{1}{c}{(1)} & \multicolumn{1}{c}{(2)} &\multicolumn{1}{c}{(3)}          &\multicolumn{1}{c}{(4)}  \\          &\multicolumn{1}{c}{Solid house} &\multicolumn{1}{c}{Refrigrator}          &\multicolumn{1}{c}{Any vehicle} &\multicolumn{1}{c}{Phone} \\ \hline
Road built               &       0.037   &       0.004   &      -0.007   &       0.020   \\
                         &     (0.030)   &     (0.013)   &     (0.024)   &     (0.041)   \\
\hline
N                        &      11,151   &      11,151   &      11,151   &      11,151   \\
Control group mean       &        0.22   &        0.04   &        0.14   &        0.44   \\
$ R^{2} $                &        0.66   &        0.26   &        0.38   &        0.48   \\
\hline\hline \end{tabular}
	\end{threeparttable}}
	\begin{minipage}{\linewidth} \footnotesize \smallskip 
	\scriptsize Notes: This table replicates results from \cite{Asher2020} showing regression  discontinuity treatment estimates of the effect of new village roads on indicators of consumption and asset ownership. Panel A, Column (1) shows the impact on imputed log consumption per capita (see \cite{Asher2020}, Data Appendix for details). Column (2) estimates the effect on log of mean total night light luminosity in 2011-13. Column (3) is the share of households whose highest earning member earns more than INR 5000 per month based on data from the 2011-12 Socioeconomic and Caste Census (SECC). Column (4) is village-level average of the primary component of indicator variables for all household assets included in the SECC. Panel B shows the impact on the village-level share of households owning major assets in the SECC. Regressions include district-threshold fixed effects and baseline control variables.  Heteroskedasticity robust standard errors reported in parentheses. Significance at 1\%, 5\% and 10\% are indicated by $^{***}$, $^{**}$ and $^{*}$, respectively.
	\end{minipage}		
\end{table}

\begin{table}[H] 
	\centering 
	\caption{Impact of rural roads on village-level population, age distribution, and gender ratios}
	\label{tab:pop_age_gender}
	\resizebox{0.7\linewidth}{!}{\begin{threeparttable}
			\begin{tabular}{l*{2}{c}} \hline
\hline \\ \multicolumn{3}{c}{\textit{Panel A: Population growth (2001 - 2011)}} \\\\ [-1ex] \hline                           &\multicolumn{1}{c}{(1)} & \multicolumn{1}{c}{(2)} \\                          &\multicolumn{1}{c}{Log} &\multicolumn{1}{c}{Level} \\ \hline
Road built               &      -0.023   &      -9.833   \\
                         &     (0.030)   &    (20.783)   \\
\hline
N                        &      11,151   &      11,151   \\
Control group mean       &        6.43   &      654.59   \\
$ R^{2} $                &        0.79   &        0.84   \\ \hline
\end{tabular}
\begin{tabular}{l*{6}{c}}
\\ \multicolumn{6}{c}{\textit{Panel B: Age group share}} \\\\ [-1ex] \hline            &\multicolumn{1}{c}{(1)} & \multicolumn{1}{c}{(2)} &\multicolumn{1}{c}{(3)}          &\multicolumn{1}{c}{(4)} &\multicolumn{1}{c}{(5)}  \\          &\multicolumn{1}{c}{11 - 20} &\multicolumn{1}{c}{21 - 30}          &\multicolumn{1}{c}{31 - 40} &\multicolumn{1}{c}{41 - 50}  &\multicolumn{1}{c}{51 - 60}  \\ \hline
Road built               &      -0.004   &      -0.004   &       0.002   &      -0.002   &       0.002   \\
                         &     (0.005)   &     (0.005)   &     (0.004)   &     (0.004)   &     (0.003)   \\
\hline
N                        &      11,151   &      11,151   &      11,151   &      11,151   &      11,151   \\
Control group mean       &        0.24   &        0.19   &        0.15   &        0.11   &        0.07   \\
$ R^{2} $                &        0.22   &        0.19   &        0.25   &        0.37   &        0.40   \\
\hline \\ \multicolumn{6}{c}{\textit{Panel C: Male share by age group}} \\\\ [-1ex] \hline            &\multicolumn{1}{c}{(1)} & \multicolumn{1}{c}{(2)} &\multicolumn{1}{c}{(3)}          &\multicolumn{1}{c}{(4)} &\multicolumn{1}{c}{(5)}  \\          &\multicolumn{1}{c}{11 - 20} &\multicolumn{1}{c}{21 - 30}          &\multicolumn{1}{c}{31 - 40} &\multicolumn{1}{c}{41 - 50}  &\multicolumn{1}{c}{51 - 60}  \\ \hline
Road built               &      -0.013   &       0.004   &       0.004   &      -0.007   &       0.016   \\
                         &     (0.009)   &     (0.008)   &     (0.008)   &     (0.010)   &     (0.013)   \\
\hline
N                        &      11,151   &      11,151   &      11,151   &      11,151   &      11,151   \\
Control group mean       &        0.52   &        0.52   &        0.51   &        0.52   &        0.51   \\
$ R^{2} $                &        0.12   &        0.19   &        0.10   &        0.07   &        0.06   \\
\hline\hline \end{tabular}
	\end{threeparttable}}
	\begin{minipage}{\linewidth} \footnotesize \smallskip 
	\scriptsize Notes: This table replicates results from \cite{Asher2020} showing regression  discontinuity treatment estimates of the effect of new village roads on population growth and demographic structure of the village population. Panel A shows the effect on village population from the Population Census 2011. Column (1) shows the outcome in log, and (2) shows the level effect. Panel B shows the effect on share of population in 10 year age groups, starting from 11 to 20 in Column (1) to 51 to 60 in Column (5). The outcome variables in Panel C are the share of men in the population within each 10-year age group. Outcomes in Panels B and C are based on 2011-12 Socioeconomic and Caste Census (SECC). Panel A shows that new roads do not affect population growth, suggesting no permanent migration impacts (short-term migrants and daily commuters are accounted for in the Census and SECC listings). The effect on population could be attenuated towards zero if there had been changes in fertility or mortality that offset any net migration. However, the null results in Panels B and C on age distribution and sex ratios by age cohort show changes in fertility and mortality were unlikely to be a contributing factor. Together, the results in this table suggest that roads did not lead to any substantial out-migration. Absent any impact on migration, the effects observed on labor reallocation (in Table~\ref{tab:occup_inc_source}) can be seen as a result of occupational choice rather than the result of selective migration.  Regressions include district-threshold fixed effects and baseline control variables.  Heteroskedasticity robust standard errors reported in parentheses. Significance at 1\%, 5\% and 10\% are indicated by $^{***}$, $^{**}$ and $^{*}$, respectively.
	\end{minipage}		
\end{table}

\newpage

\section{Harvest and Planting Dates} \label{dates}
\setcounter{table}{0}
\renewcommand{\thetable}{\Alph{section}.\arabic{table}}
\setcounter{figure}{0}
\renewcommand{\thefigure}{\Alph{section}.\arabic{figure}}
\numberwithin{table}{section}

In this section, we evaluate the effect of rural roads on harvest and planting dates. We procure satellite-based measures of harvest (end-)dates and planting dates for the kharif season, aggregated up to the village level from 250 m pixel data. The planting and harvest dates are estimated using Enhanced Vegetation Index (EVI) data from MODIS and were  validated using ground data.\footnote{We are grateful to Meha Jain, School for Environment and Sustainability, University of Michigan, for sharing these data.} Harvest (planting) date is measured as the median pixel value of the harvest (planting) dates within a 10 km buffer around the village. Unfortunately, these data are not available for states that followed population thresholds to determine rural road construction under PMGSY. Therefore, we are not able to use a regression discontinuity design. Instead, we estimate the following event study specification:

\begin{equation}\label{eq:ddreg2}
\Delta Date_{v,d,y} = \sum_{\tau, \tau \neq -1} \delta_{\tau}D_{t^{0} + \tau} + \lambda_{v} + \mu_{d,y} + \alpha_{y} X_{v} + \varepsilon_{v,d,y}
\end{equation}

where $\Delta Date_{v,d,y}$ is the change in harvest or sowing date for village $v$, located in district $d$ in year $y$ from the baseline (2002). $D_{t^{0} + \tau}$ are event time indicator variables that capture the average treatment effect, where $\tau$ indicates the year relative to when a village receives access to a rural road, with the year prior to treatment being the excluded category. $\lambda_{v}$ are village fixed effects and $\mu_{d,y}$ are district-by-year fixed effects. Village fixed effects control for time-invariant unobservables at the village level (e.g., soil type). District-by-year fixed effects control for time-varying district-specific confounders. For instance, the National Rural Employment Guarantee Scheme was rolled-out in a staggered manner across India between 2006 and 2008. Lastly, we include an interaction of baseline village characteristics $X_{v}$ with year fixed effects $\alpha_{y}$. Standard errors are clustered at the village level. 

The identifying assumption here is that there exist no village-specific time-varying confounders that are correlated with both access to rural roads as well as local agriculture. E.g., if rural roads are placed in villages where agricultural activities are changing, our estimates would be biased. While lack of pre-trends would bolster our confidence in said assumption, if change in local agriculture and rural road construction were to occur simultaneously, our estimates will still be biased.

\begin{figure} 
	\centering 
	\caption{Event study estimates: Effect of access to rural roads on monsoon (kharif) harvest date} 
	\label{fig:harvest}
	\includegraphics[width=\linewidth]{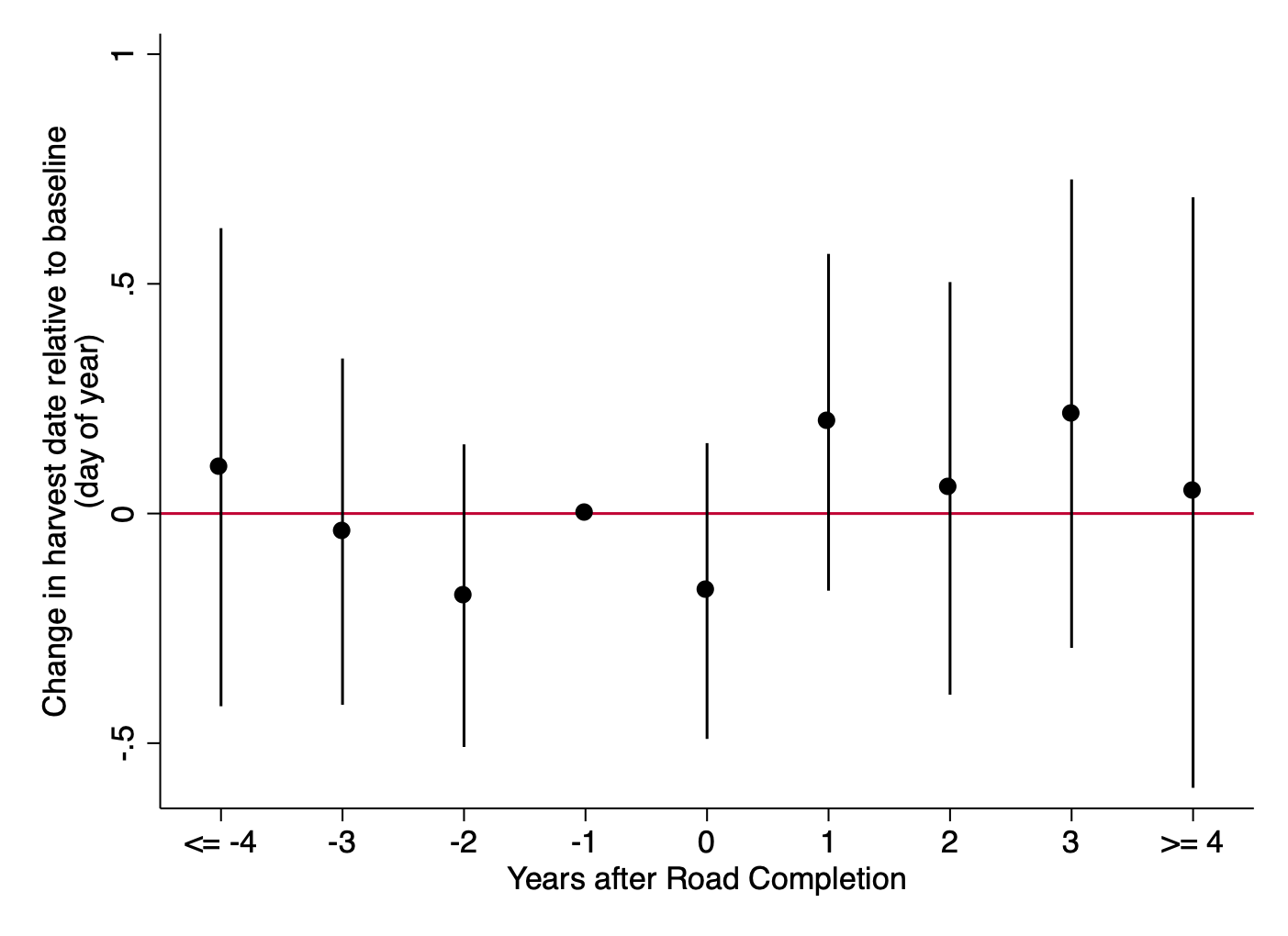}
	\begin{minipage}{\linewidth}\footnotesize 
		\scriptsize Notes: The graph shows event study estimates for the effect of new roads on monsoon harvest dates. Sample is limited to the states of Punjab, Haryana, Uttar Pradesh, and Bihar for which satellite-based sowing date measures are available. The sample period is 2003-2013. The dependent variable is the change in harvest completion day from the baseline (2002). The horizontal axis shows the event year relative to the year of road completion. Each point shows the coefficient and confidence interval on each event-time fixed effect relative to the omitted category which is the year before road completion ($ t = -1 $). All regressions include village FE, district $\times$ year fixed effects and the interactions of year fixed effects with baseline village characteristics and harvesting date/week in 2002. Standard errors in parentheses are clustered at village level.
	\end{minipage}
\end{figure}

\begin{figure} 
	\centering 
	\caption{Event study estimates: Effect of access to rural roads on monsoon (kharif) sowing date} 
	\label{fig:sowing}
	\includegraphics[width=\linewidth]{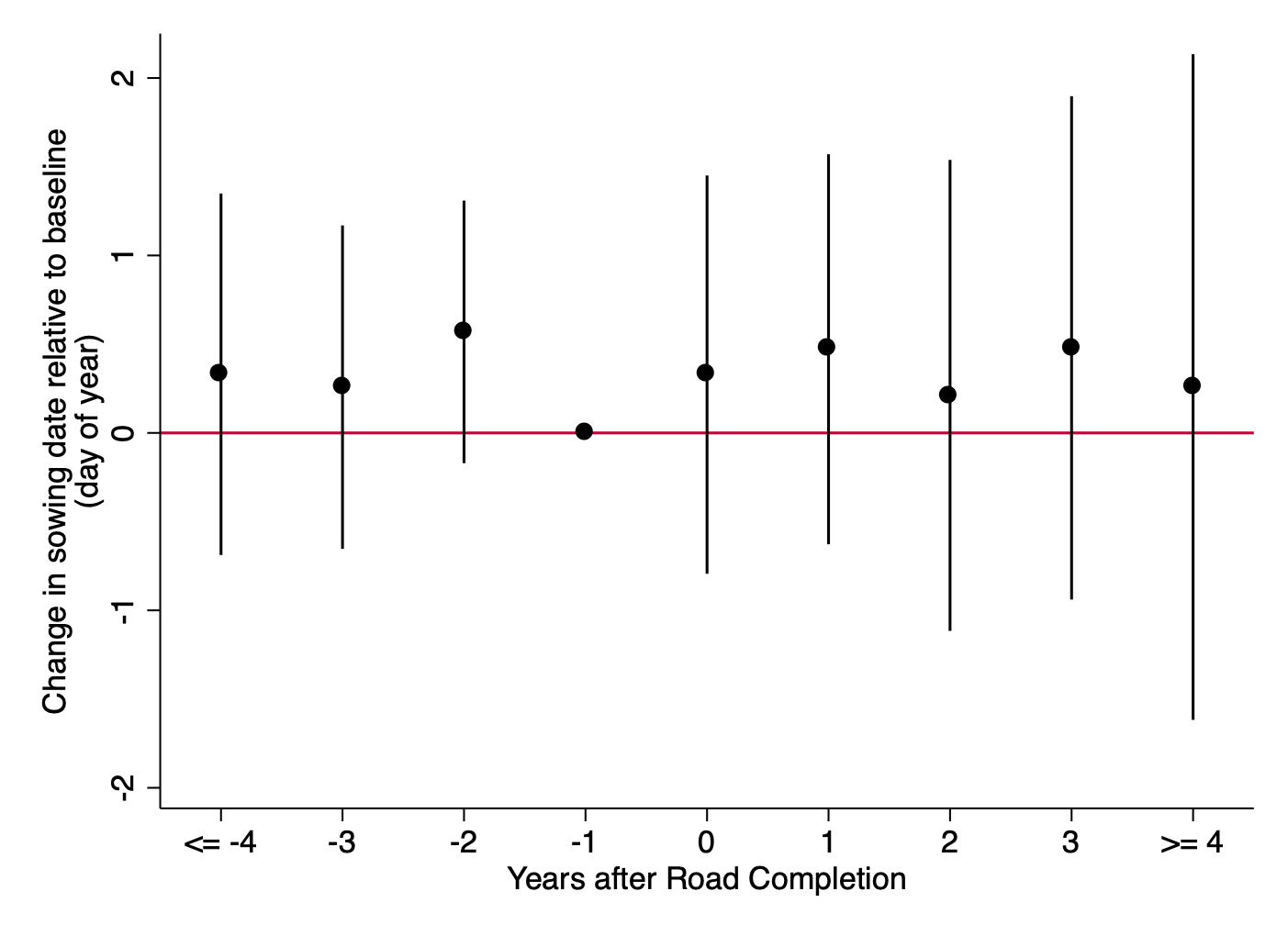}
	\begin{minipage}{\linewidth}\footnotesize 
		\scriptsize Notes: The graph shows event study estimates for the effect of new roads on monsoon sowing dates. Sample is limited to the states of Punjab, Haryana, Uttar Pradesh, and Bihar for which satellite-based sowing date measures are available. The sample period is 2003-2013. The dependent variable is the change in day of sowing from the baseline (2002). The horizontal axis shows the event year relative to the year of road completion. Each point shows the coefficient and confidence interval on each event-time fixed effect relative to the omitted category which is the year before road completion ($ t = -1 $). All regressions include village FE, district $\times$ year fixed effects and the interactions of year fixed effects with baseline village characteristics and sowing date/week in 2002. Standard errors in parentheses are clustered at village level.
	\end{minipage}
\end{figure}

Figures \ref{fig:sowing} and \ref{fig:harvest} present our results. First, we don't find evidence for any pre-trends. Second, and more importantly, we fail to find evidence that access to rural roads affects either harvest or planting dates. The point estimates are small and statistically insignificant.

\newpage

\section{Rural Economic and Demographic Survey \\(REDS)} \label{reds}
		\setcounter{table}{0}
		\renewcommand{\thetable}{\Alph{section}.\arabic{table}}
		\setcounter{figure}{0}
		\renewcommand{\thefigure}{\Alph{section}.\arabic{figure}}
		\numberwithin{table}{section}
					
We use village- and household-level surveys from the 1999 and 2006 rounds of the Rural Economic and Demographic Survey (REDS), administered by the National Council of Applied Economic Research (NCAER), to estimate the effect of rural roads on agricultural wages, local stock of combine harvesters, and use of hired mechanized agricultural equipment. REDS is a nationally representative survey of rural households in India spanning 221 villages across 100 districts in 17 major states. It includes a village survey that collects information on the prevailing  wage rates for agricultural labor as well as stock of agricultural machinery at the village-level . The household questionnaire provides detailed information on the use of agricultural inputs, including the use and cost of hired mechanized equipment.

Because the data includes few villages in states that followed population thresholds to determined rural road construction under PMGSY, we cannot use a regression discontinuity design. Instead, we estimate simple difference-in-difference specifications. `Treat' is an indicator variable that takes the value $1$ if a village receives a rural road between 1999 and 2006, $0$ otherwise. `Post' is an indicator variable that takes the value $1$ if the year is 2006, $0$ otherwise.   `TreatXPost' captures the effect of the construction of a rural road between 1999 and 2006. Since political patronage likely played a role in placement of roads as discussed in Section \ref{pmgsy}, the results discussed below should only be considered suggestive.

Table \ref{tab:reds_agwages} estimates the effect of rural roads on agricultural wages. We find access to rural roads increases agricultural wages for both men and women by 1.2\% and 2\%, respectively. This result is consistent with the key takeaway from  \cite{Asher2020} that rural roads induce movement of workers out of agriculture.

Table \ref{tab:reds_villagmach} estimates the effect of rural roads on the village-level stock of combine harvesters. Using a village-panel regression with village and year fixed effects, we fail to find an economically or statistically significant impact on the village-level stock agricultural machinery across a variety of agricultural implements either on the extensive (Panel A) or the intensive margin (Panel B). In particular, we see no effect on the presence or number of combine harvesters. Next, using a household-panel regression with household and year fixed effects, we fail to detect an impact on the use of hired mechanized agricultural equipment at the household level (Table \ref{tab:reds_hhagmach}).

\begin{table} [H]
	\centering 
	\caption{Impact of roads on agricultural wage rates - REDS village panel} 
	\label{tab:reds_agwages}
	\resizebox{0.8\linewidth}{!}{\begin{threeparttable}
		{
\def\sym#1{\ifmmode^{#1}\else\(^{#1}\)\fi}
\begin{tabular}{l*{6}{c}}
\hline\hline
                         &\multicolumn{3}{c}{Wage rate}                  &\multicolumn{3}{c}{Log wage}                   \\\cmidrule(lr){2-4}\cmidrule(lr){5-7}
                         &\multicolumn{1}{c}{(1)}&\multicolumn{1}{c}{(2)}&\multicolumn{1}{c}{(3)}&\multicolumn{1}{c}{(4)}&\multicolumn{1}{c}{(5)}&\multicolumn{1}{c}{(6)}\\
                         &\multicolumn{1}{c}{Male}&\multicolumn{1}{c}{Female}&\multicolumn{1}{c}{Average}&\multicolumn{1}{c}{Male}&\multicolumn{1}{c}{Female}&\multicolumn{1}{c}{Average}\\
\hline
Treat X Post             &       0.722*  &       0.918***&       0.856** &       0.012*  &       0.020** &       0.016** \\
                         &     (0.420)   &     (0.349)   &     (0.352)   &     (0.007)   &     (0.008)   &     (0.007)   \\
\hline
N                        &         442   &         442   &         442   &         442   &         442   &         442   \\
Sample mean              &       62.99   &       47.65   &       55.25   &       62.99   &       47.65   &       55.25   \\
$ R^{2} $                &        0.76   &        0.78   &        0.77   &        0.75   &        0.76   &        0.76   \\
\hline\hline
\end{tabular}
}
	\end{threeparttable}}	
	\begin{minipage}{\linewidth} \footnotesize \smallskip 
	\scriptsize Notes: Table reports regression estimates showing the impact of receiving a PMGSY road on  agricultural wage rates (in 2006 prices), in both levels and log wages. Sample used is village level panel from REDS 1999 - 2006. The coefficient ``Treat X Post'' takes the value 1 in the post (2006) year for villages that receive a road by 2006.  Regressions include village and year fixed effects. Standard errors in parentheses clustered at village level. Significance at 1\%, 5\% and 10\% are indicated by $^{***}$, $^{**}$ and $^{*}$, respectively.  
	\end{minipage}
\end{table}

\begin{table} [H]
	\centering 
	\caption{Impact of roads on village-level stock of agricultural machinery - REDS village panel } 
	\label{tab:reds_villagmach}
	\resizebox{0.65\linewidth}{!}{\begin{threeparttable}
		\begin{tabular}{l*{4}{c}} \hline
\hline \\ \multicolumn{5}{c}{\textit{Panel A: Present (Yes = 1)}} \\\\ [-1ex] \hline           &\multicolumn{1}{c}{Combines} & \multicolumn{1}{c}{Threshers} &\multicolumn{1}{c}{Tractors}          &\multicolumn{1}{c}{Power tillers} \\ \hline
Treat X Post             &       0.001         &       0.005         &       0.016\sym{*}  &       0.021         \\
                         &     (0.009)         &     (0.015)         &     (0.010)         &     (0.014)         \\
\hline
Sample mean              &        0.08         &        0.61         &        0.88         &        0.24         \\
$N$                      &         442         &         442         &         442         &         442         \\
$R^2$                    &        0.58         &        0.77         &        0.80         &        0.59         \\
\hline \\ \multicolumn{5}{c}{\textit{Panel B: Log count}} \\\\ [-1ex] \hline           &\multicolumn{1}{c}{Combines} & \multicolumn{1}{c}{Threshers} &\multicolumn{1}{c}{Tractors}          &\multicolumn{1}{c}{Power tillers} \\ \hline
Treat X Post             &       0.007         &       0.017         &       0.033         &       0.029         \\
                         &     (0.023)         &     (0.045)         &     (0.026)         &     (0.030)         \\
\hline
Sample mean              &        0.10         &        1.40         &        2.48         &        0.42         \\
$N$                      &         442         &         442         &         442         &         442         \\
$R^2$                    &        0.54         &        0.76         &        0.90         &        0.54         \\
\hline\hline \end{tabular}
	\end{threeparttable}}
	\begin{minipage}{\linewidth} \footnotesize \smallskip 
	\scriptsize Notes: Table reports regression estimates showing the impact of receiving a PMGSY road in agricultural machinery stock at the village level. Sample used is village level panel from REDS 1999 - 2006. The coefficient ``Treat X Post'' takes the value 1 in the post (2006) year for villages that receive a road by 2006. Regressions include village and year fixed effects. Standard errors in parentheses clustered at village level. Significance at 1\%, 5\% and 10\% are indicated by $^{***}$, $^{**}$ and $^{*}$, respectively.
	\end{minipage}
\end{table} 

\begin{table} [H]
	\centering 
	\caption{Impact of roads on household use of mechanized agricultural equipment - REDS household panel } 
	\label{tab:reds_hhagmach}
	\resizebox{0.8\linewidth}{!}{\begin{threeparttable}
		{
\def\sym#1{\ifmmode^{#1}\else\(^{#1}\)\fi}
\begin{tabular}{l*{4}{c}}
\hline\hline
                         &\multicolumn{1}{c}{(1)}&\multicolumn{1}{c}{(2)}&\multicolumn{1}{c}{(3)}&\multicolumn{1}{c}{(4)}\\
                         &\multicolumn{1}{c}{Used? (Yes=1)}&\multicolumn{1}{c}{Log cost}&\multicolumn{1}{c}{Log cost per acre}&\multicolumn{1}{c}{Share of total cost}\\
\hline
Treat X Post             &       0.002         &      -0.048         &      -0.046         &      -0.004         \\
                         &     (0.010)         &     (0.081)         &     (0.070)         &     (0.004)         \\
\hline
Sample mean              &        0.67         &        4.77         &        3.91         &        0.15         \\
$N$                      &        6090         &        5548         &        5548         &        5474         \\
$R^2$                    &        0.57         &        0.60         &        0.62         &        0.69         \\
\hline\hline
\end{tabular}
}
	\end{threeparttable}}	
	\begin{minipage}{\linewidth} \footnotesize \smallskip 
	\scriptsize Notes: Table reports regression estimates showing the impact of receiving a PMGSY road on household use of tractors, harvester, threshers or other mechanized equipment. Sample used is household panel from REDS 1999 and 2006. The coefficient ``Treat X Post'' takes the value 1 in the post (2006) year for households in villages that receive a road by 2006. Regressions include household and year fixed effects. Columns (1), (2) and (4) also control for cropped area. Standard errors in parentheses clustered at village level. Significance at 1\%, 5\% and 10\% are indicated by $^{***}$, $^{**}$ and $^{*}$, respectively.  
	\end{minipage}
\end{table}

\newpage
	
\section{Heterogeneity by Relative Agricultural Wage at Baseline} \label{heteroagwage}
		\setcounter{table}{0}
		\renewcommand{\thetable}{\Alph{section}.\arabic{table}}
		\setcounter{figure}{0}
		\renewcommand{\thefigure}{\Alph{section}.\arabic{figure}}
		\numberwithin{table}{section}

\begin{figure} [H]
	\begin{center}
		\caption{Heterogeneity by relative agricultural labor wages at baseline: Impact of rural roads on share of village labor in agriculture and non-agricultural sectors}	\label{fig:labshare_rdplots}
    		
    	\includegraphics[scale=0.6]{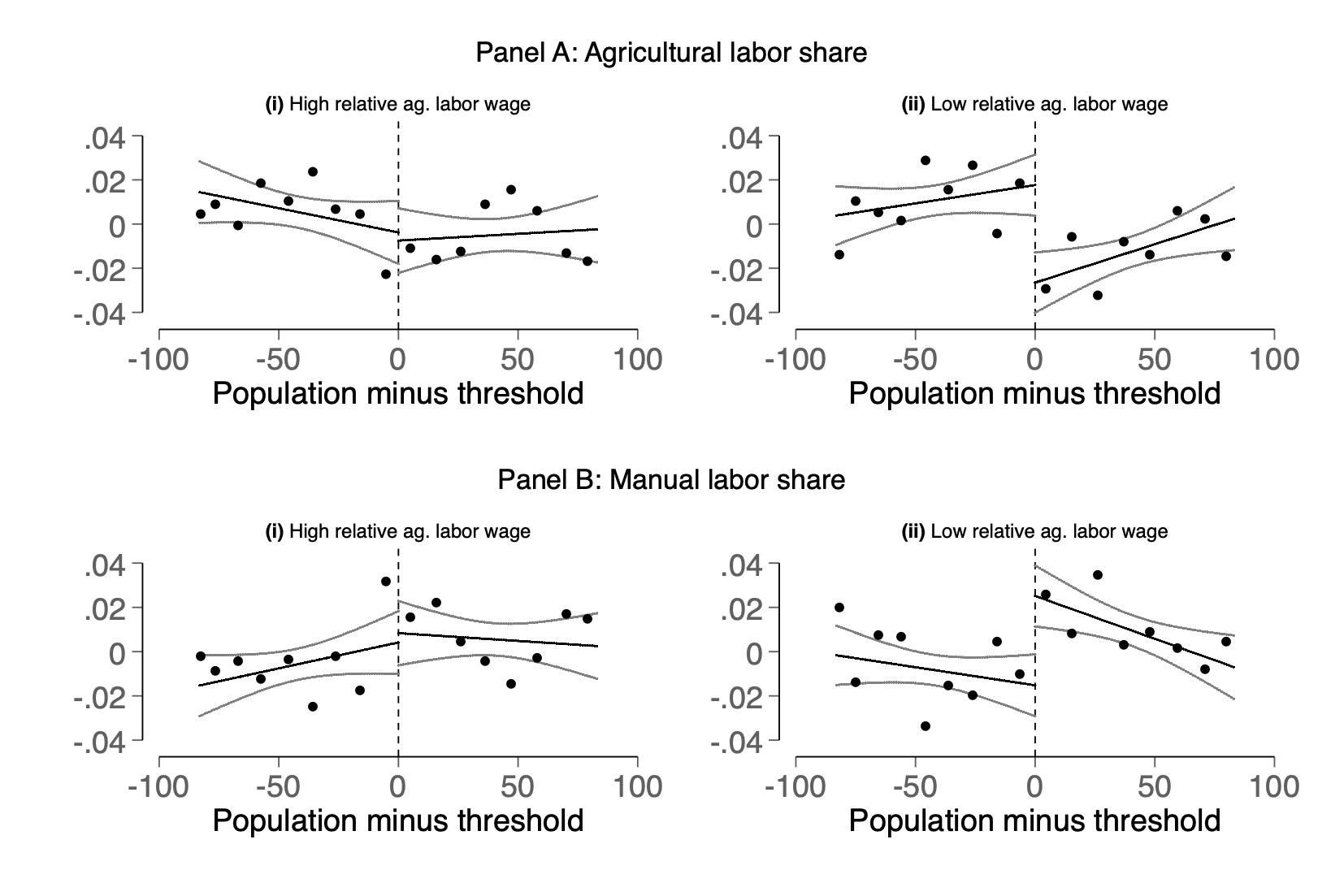} 	
		
	\end{center}
	\scriptsize Notes: All panels show regression discontinuity estimates by plotting the residualized values of village-level share of manual labor in agriculture (Panel A) and non-agricultural sectors (Panel B), after controlling for all variables in the main specification other than population, as a function of the normalized 2001 village population relative to the threshold. Each point represents the mean of all villages in a given population bin. Figure (i) of each panel plots the regression discontinuity  relationship  for  districts that had high (above sample median) agricultural labor wages relative to non-agricultural labor wage rates in rural areas at baseline. Figure (ii) of each panel plots the regression discontinuity  relationship for districts with below median relative agricultural wage rates. The outcome variables are based on the Socioeconomic and Caste Census 2011-2012 (see \citep{Asher2020} for details). Baseline rural agricultural and non-agricultural daily labor wage rates are based on the 1999 - 2000 NSSO survey data (Round 55).  Estimates in all panels control for district-threshold fixed effects, year fixed effects, and baseline village characteristics in 2001. Population is centered around the state-specific threshold used for road eligibility - either 500 or 1000, depending on the state. Standard errors clustered at the village level. \\
	
\end{figure}

\newpage 

\begin{figure} [H]
	\begin{center}
		\caption{Heterogeneity in impact of rural roads on annual agricultural fire activity by high versus low relative agricultural wage rates at baseline}
		\label{fig:heteroagwage_rdplots}
		\subfigure[Fires: High relative ag. labor wage]{
			\includegraphics[scale=0.24]{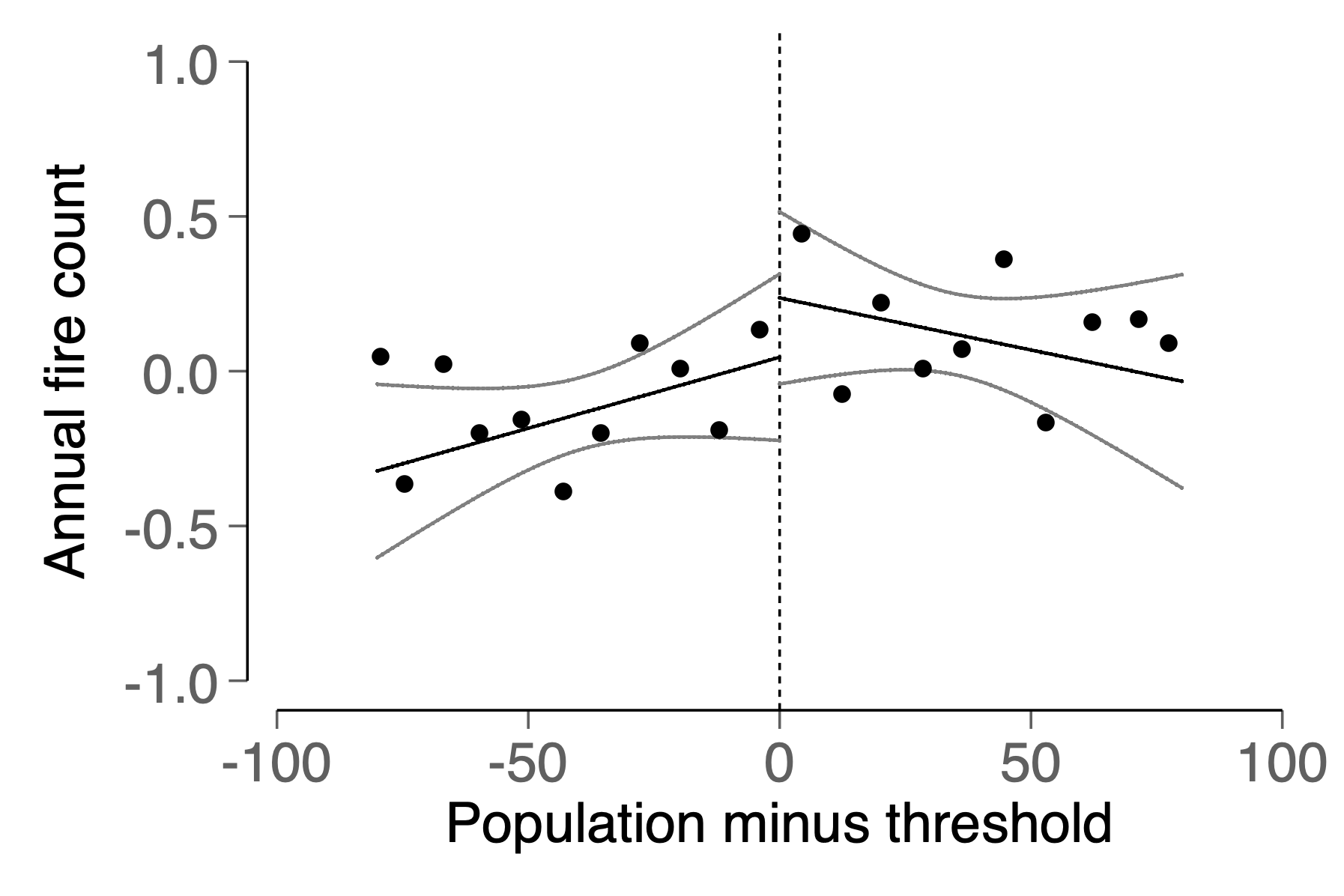}
		} 
		\subfigure[Fires: Low relative ag. labor wage]{
			\includegraphics[scale=0.24]{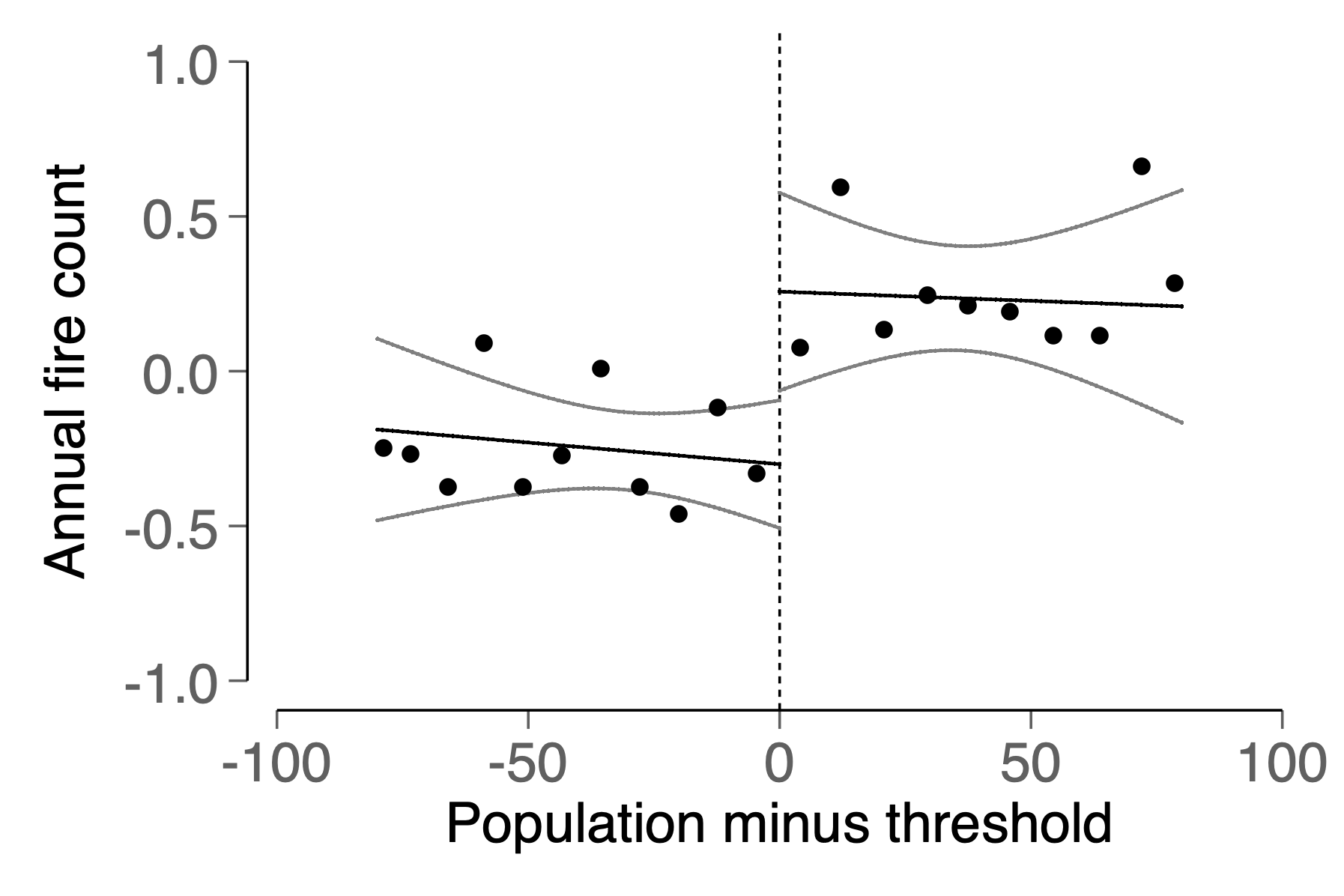} 
		}				
		\subfigure[PM2.5: High relative ag. labor wage]{
			\includegraphics[scale=0.24]{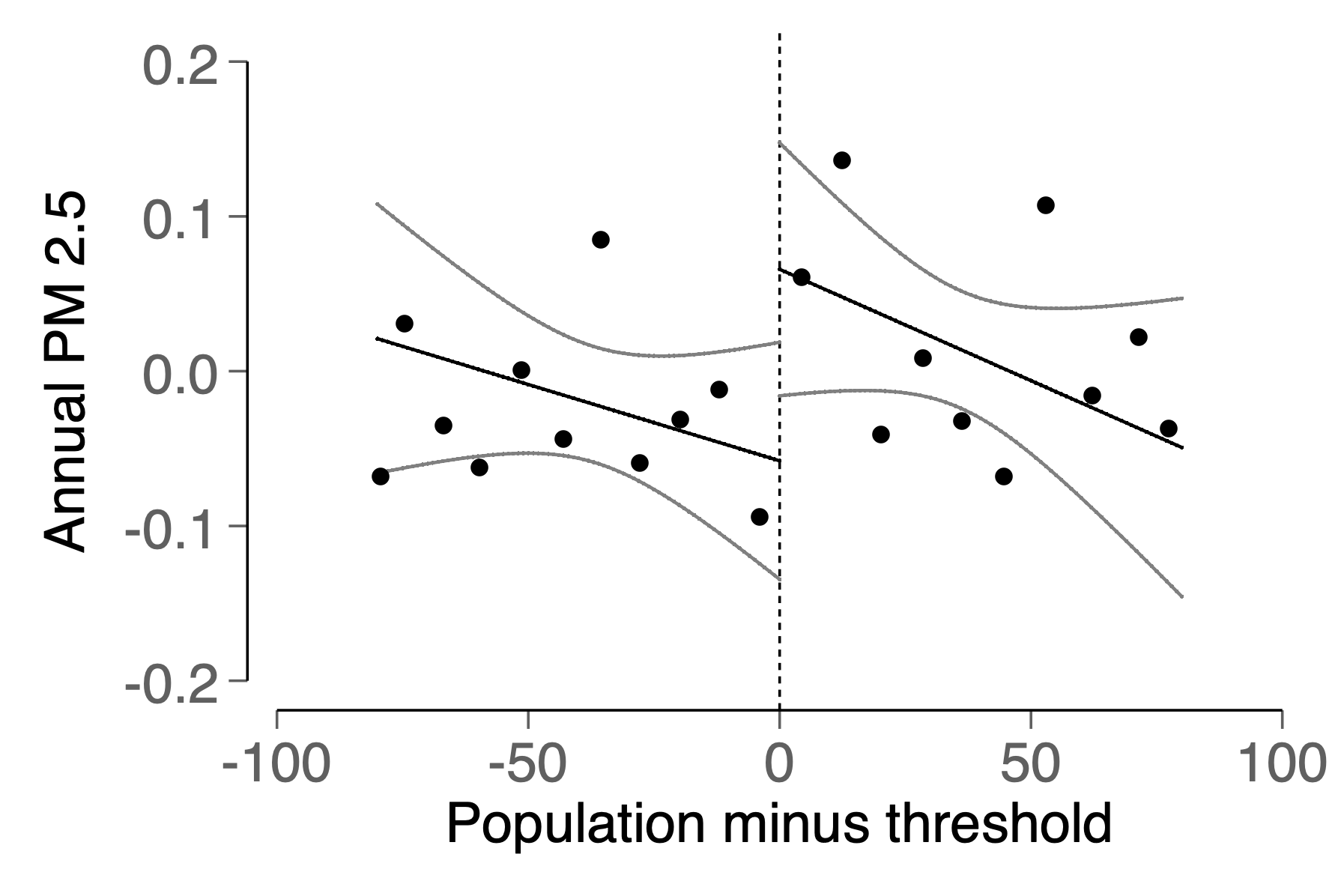}
		} 
		\subfigure[PM2.5: Low relative ag. labor wage]{
			\includegraphics[scale=0.24]{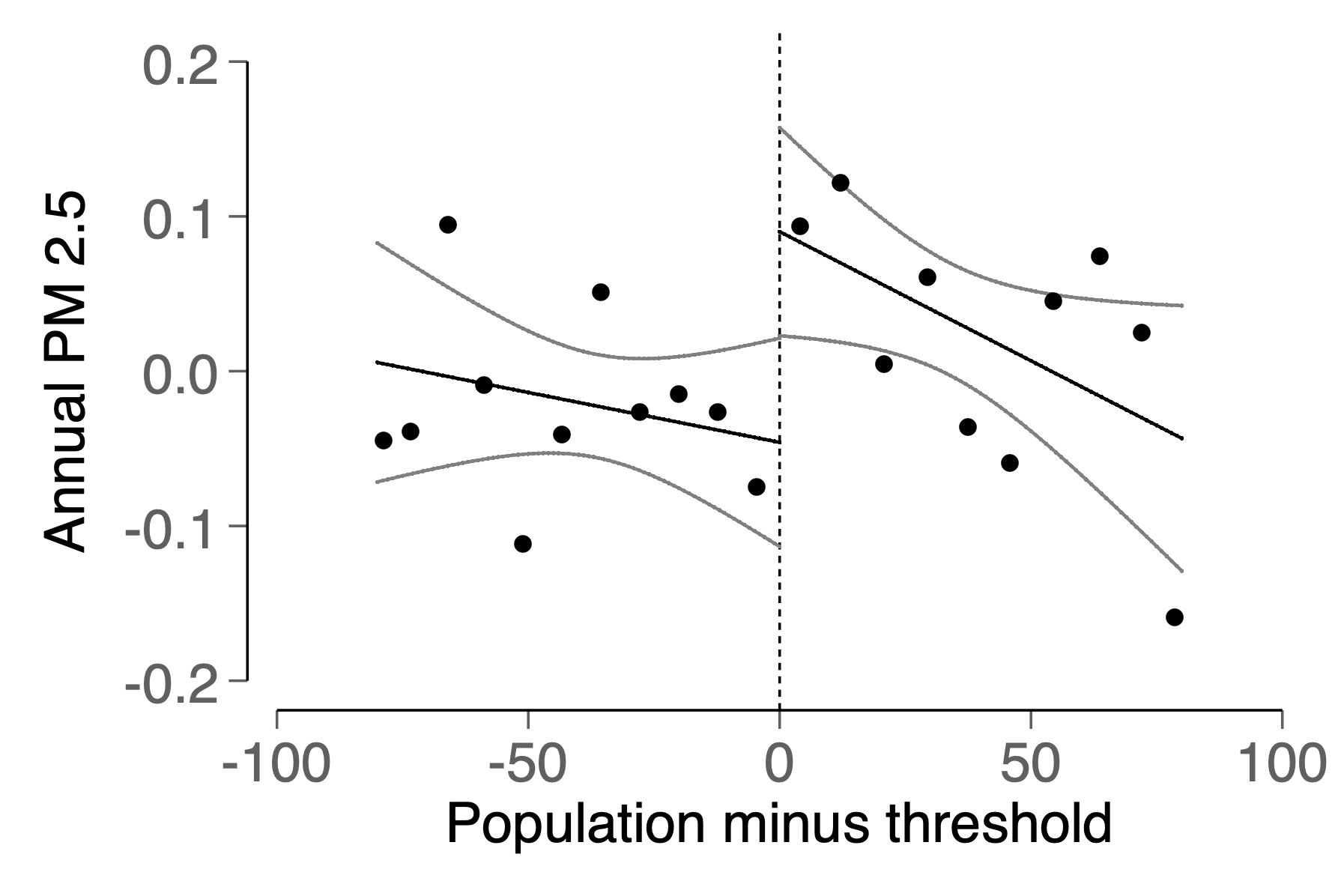} 
		}	
	\end{center}
	\scriptsize Notes: All panels show regression discontinuity estimates by plotting the residualized values of outcomes  (after controlling for all variables in the main specification other than population) as a function of the normalized 2001 village population relative to the threshold. Each point represents the mean of all villages in a given population bin. Panels (a) and (b) show results for the annual number of fires between 2002 - 2013 , while (c) and (d) show the same annual average PM 2.5 ($\mu g/ m^{3}$). Panels (a) and (c) plot the regression discontinuity relationship  for  districts which had high (above sample median) agricultural labor wages relative to non-agricultural labor wage rates in rural areas at baseline. Panels (b) and (d) plot the regression discontinuity relationship for districts with below median relative agricultural wage rates. Rural agricultural and non-agricultural daily labor wage rates are based on the 1999 - 2000 NSSO survey data (Round 55).  Estimates in both panels control for district-threshold fixed effects, year fixed effects, and baseline village characteristics in 2001. Population is centered around the state-specific threshold used for road eligibility - either 500 or 1000, depending on the state. Standard errors are clustered at the village level. \\
	
\end{figure}

\begin{table}[H] 
	\centering 
	\caption{Impact of rural roads on share of village labor in agriculture and non-agricultural sectors by relative agricultural labor wage rates}
	\label{tab:aglabexit_hetero}
	\resizebox{0.7\linewidth}{!}{\begin{threeparttable}
			{
\def\sym#1{\ifmmode^{#1}\else\(^{#1}\)\fi}
\begin{tabular}{l*{4}{c}}
\hline\hline
                         &\multicolumn{2}{c}{\shortstack{Share of labor\\in agriculture}}&\multicolumn{2}{c}{\shortstack{Share of non-agricultural\\manual labor}}\\\cmidrule(lr){2-3}\cmidrule(lr){4-5}
                         &\multicolumn{1}{c}{(1)}&\multicolumn{1}{c}{(2)}&\multicolumn{1}{c}{(3)}&\multicolumn{1}{c}{(4)}\\
                         &\multicolumn{1}{c}{\shortstack{High rel.\\ag. wage}}&\multicolumn{1}{c}{\shortstack{Low rel.\\ag. wage}}&\multicolumn{1}{c}{\shortstack{High rel.\\ag. wage}}&\multicolumn{1}{c}{\shortstack{Low rel.\\ag. wage}}\\
\hline
Road built               &      -0.029   &      -0.241***&       0.030   &       0.205** \\
                         &     (0.046)   &     (0.093)   &     (0.046)   &     (0.092)   \\
\hline
N                        &       5,403   &       5,484   &       5,403   &       5,484   \\
Control group mean       &        0.49   &        0.46   &        0.45   &        0.46   \\
\hline\hline
\end{tabular}
}
		\end{threeparttable}}
		\begin{minipage}{\linewidth} \footnotesize \smallskip 
			\scriptsize Notes: This table shows the effect of rural roads on share of manual labor at village-level in agriculture and non-agricultural sectors. The outcome variables are based on the Socioeconomic and Caste Census 2011-2012 (see \citep{Asher2020} for details). ``High rel. ag labor wage'' sample consists of districts which had high (above sample median) agricultural labor wages relative to non-agricultural labor wage rates, while ``Low rel. ag labor wage'' sample are districts with below median relative agricultural wage rates. Wage rates are based on the 1999 - 2000 NSSO survey data (Round 55). Regressions include district-threshold fixed effects, year and baseline control variables. Standard errors in parentheses are clustered at village level. Significance at 1\%, 5\% and 10\% are indicated by $^{***}$, $^{**}$ and $^{*}$, respectively.
		\end{minipage}
\end{table}

\begin{table}[H] 
	\centering 
	\caption{Impact of rural roads on annual agricultural fire activity and PM2.5 by high versus low relative agricultural wage rates at baseline}
	\label{tab:heteroagwage}
	\resizebox{0.7\linewidth}{!}{\begin{threeparttable}
			{
\def\sym#1{\ifmmode^{#1}\else\(^{#1}\)\fi}
\begin{tabular}{l*{4}{c}}
\hline\hline
                         &\multicolumn{2}{c}{High rel. ag. wage}&\multicolumn{2}{c}{Low rel. ag. wage}\\\cmidrule(lr){2-3}\cmidrule(lr){4-5}
                         &\multicolumn{1}{c}{(1)}&\multicolumn{1}{c}{(2)}&\multicolumn{1}{c}{(3)}&\multicolumn{1}{c}{(4)}\\
                         &\multicolumn{1}{c}{Fires}&\multicolumn{1}{c}{PM 2.5}&\multicolumn{1}{c}{Fires}&\multicolumn{1}{c}{PM 2.5}\\
\hline
Road built               &       0.803   &       0.367*  &       3.232** &       0.599*  \\
                         &     (0.702)   &     (0.215)   &     (1.389)   &     (0.364)   \\
\hline
N                        &      62,880   &      62,880   &      67,740   &      67,740   \\
Control group mean       &        2.68   &       45.38   &        2.88   &       40.09   \\
\hline\hline
\end{tabular}
}
		\end{threeparttable}}
		\begin{minipage}{\linewidth} \footnotesize \smallskip 
			\scriptsize Notes: This table shows the effect of rural roads on village-level annual fire activity and PM 2.5 ($\mu g/ m^{3}$). The sample consists of the panel of villages for the 5 year period from 2002 - 2013.  ``High rel. ag labor wage'' sample consists of districts which had high (above sample median) agricultural labor wages relative to non-agricultural labor wage rates, while ``Low rel. ag labor wage'' sample are districts with below median relative agricultural wage rates. Wage rates are based on the 1999 - 2000 NSSO survey data (Round 55). Regressions include district-threshold fixed effects, year fixed effects and baseline control variables. Standard errors in parentheses are clustered at village level. Significance at 1\%, 5\% and 10\% are indicated by $^{***}$, $^{**}$ and $^{*}$, respectively.
		\end{minipage}
\end{table}

\newpage

\section{Heterogeneity by Crops Grown} \label{heterocrop}
		\setcounter{table}{0}
		\renewcommand{\thetable}{\Alph{section}.\arabic{table}}
		\setcounter{figure}{0}
		\renewcommand{\thefigure}{\Alph{section}.\arabic{figure}}
		\numberwithin{table}{section}
		
\begin{figure} [H]
	\begin{center}
		\caption{Spatial distribution of average annual fire activity and baseline rice and sugarcane acreage shares}	\label{fig:fires_crops}
		\subfigure[Average annual fire counts across India]{
			\includegraphics[scale=0.09]{distfires_wstudysample_v2.png}
		} 
		\subfigure[Average annual fire counts in sample districts]{
			\includegraphics[scale=0.09]{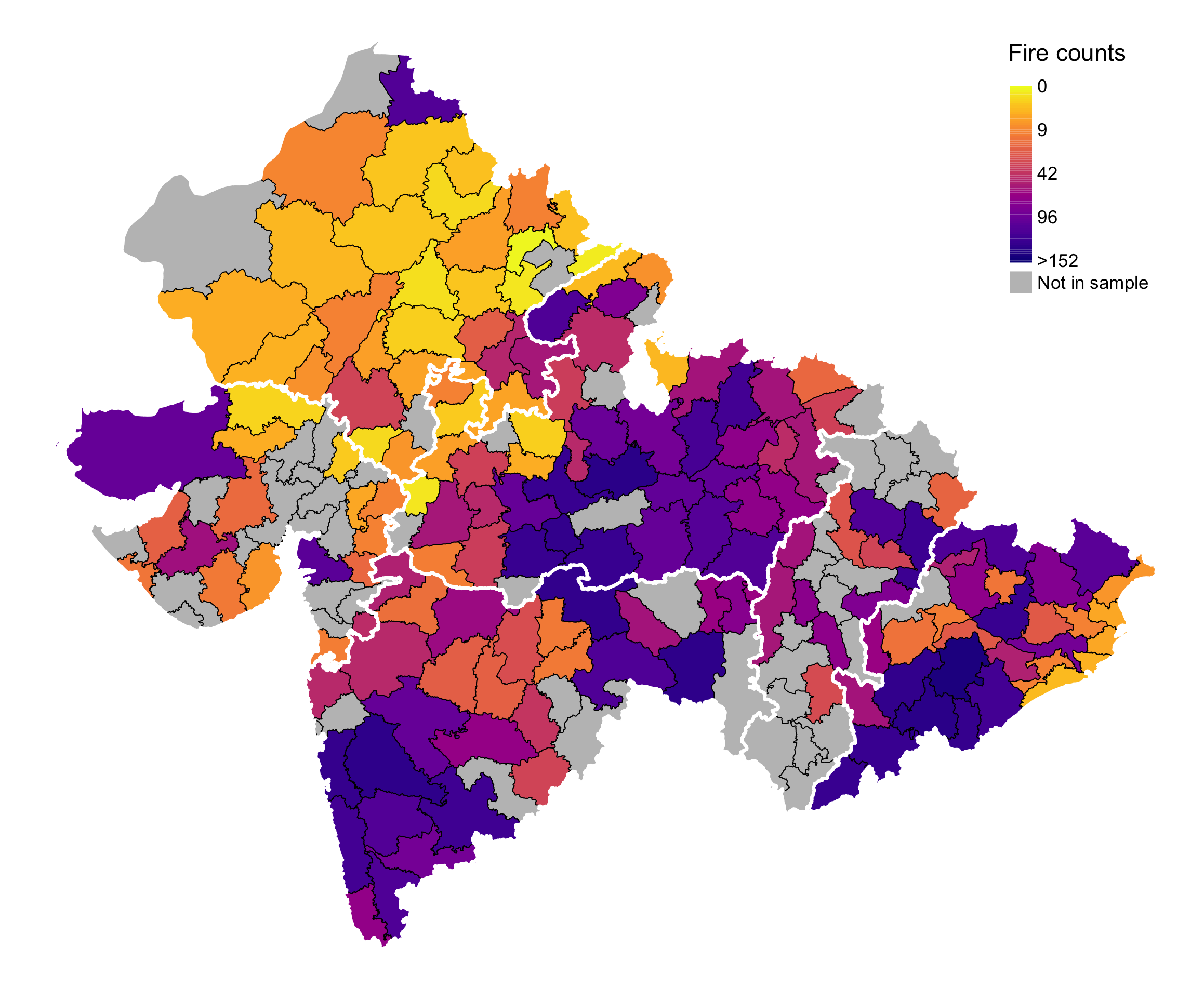} 
		}				
		\subfigure[Rice area share]{
			\includegraphics[scale=0.09]{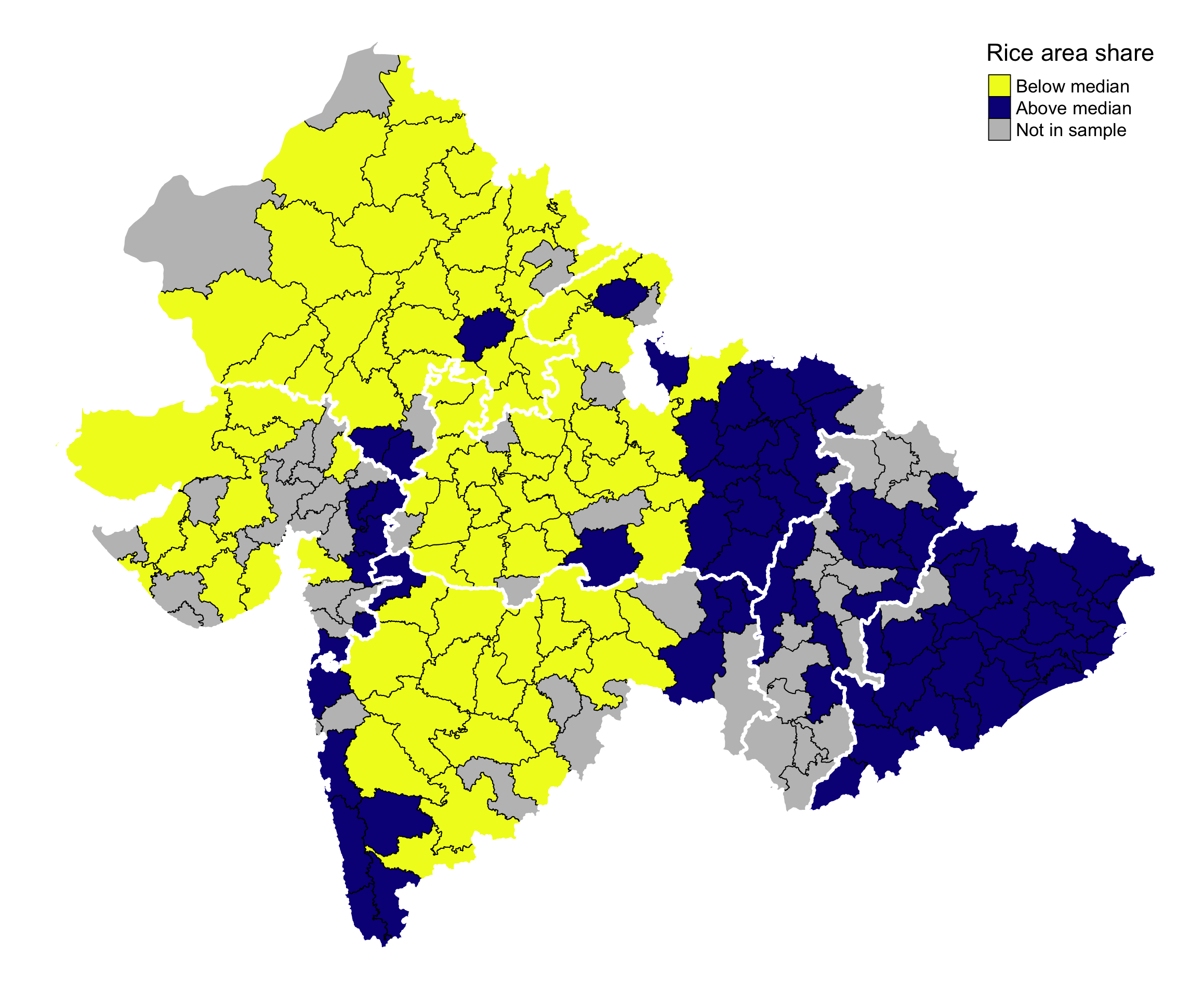}
		} 
		\subfigure[Sugarcane area share]{
			\includegraphics[scale=0.09]{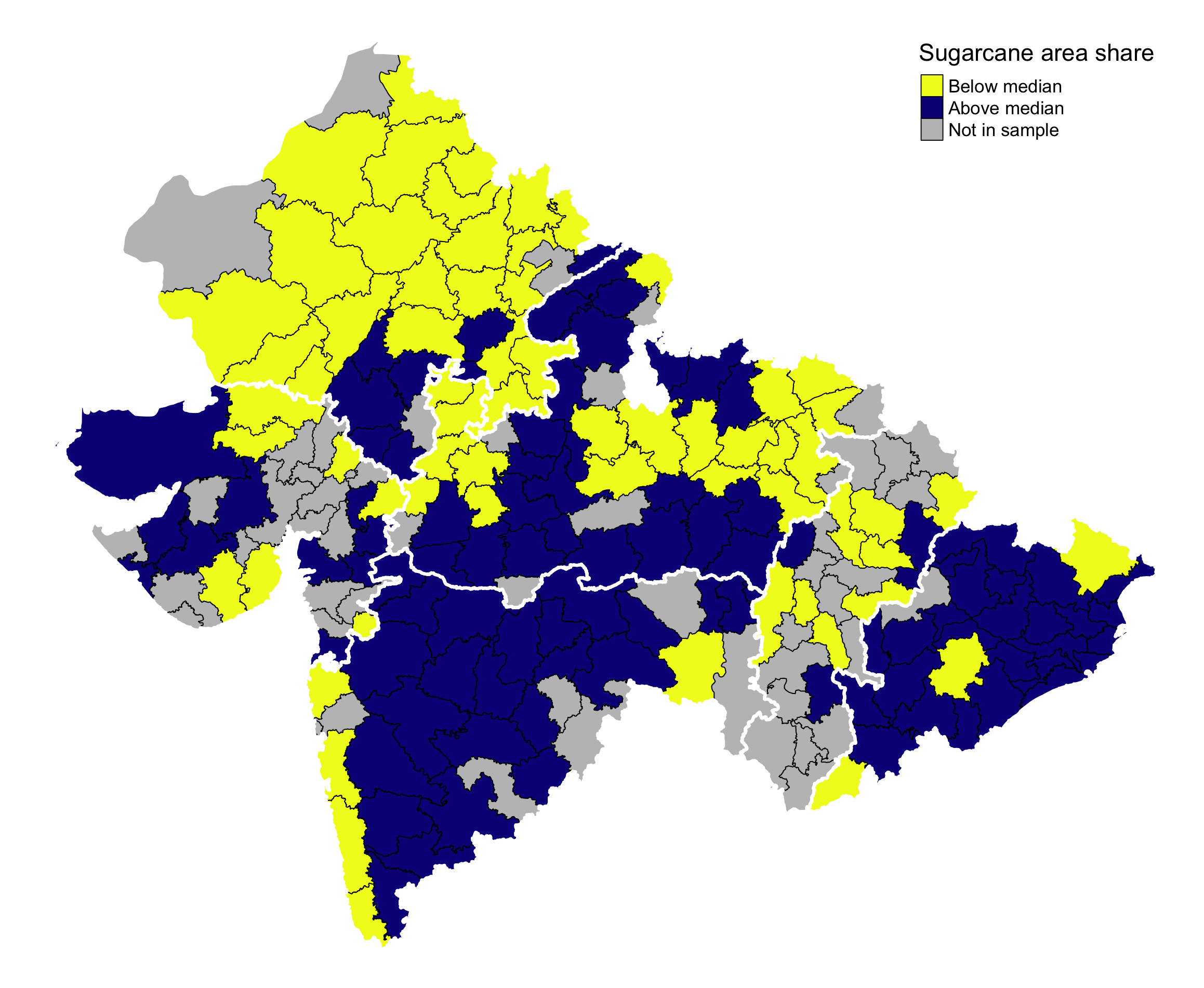} 
		}				

	\end{center}
	\scriptsize Notes: Panels (a) and (b) show the mean annual number of fire pixels detected in each district from MODIS satellite data for the period 2001 to 2013 for whole of India and within the sample districts, respectively. Panels (c) and (d) show districts with above/below sample median share of cropland under rice and sugarcane, respectively, at baseline (2001).  \\
	
\end{figure}

\begin{figure} [H]
	\begin{center}
		\caption{Heterogeneity in impact of rural roads on annual agricultural fire activity and PM2.5 by high rice \emph{or} high sugarcane districts versus low rice \emph{and} low sugar districts}
		\label{fig:rdricesugar}
		\subfigure[Fires: High rice \emph{or} high sugar]{
			\includegraphics[scale=0.25]{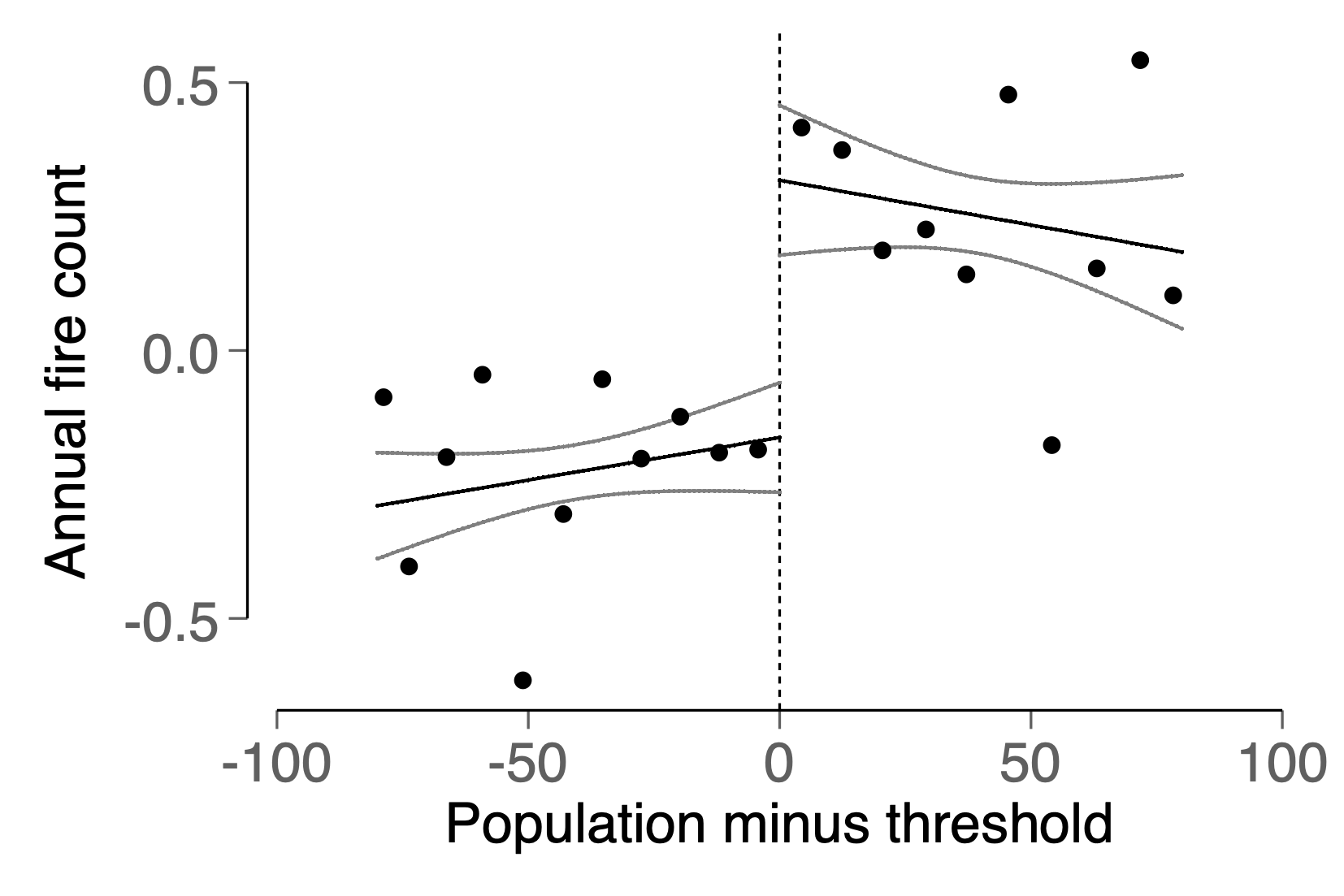}
		} 
		\subfigure[Fires: Low rice \emph{and} low sugar]{
			\includegraphics[scale=0.25]{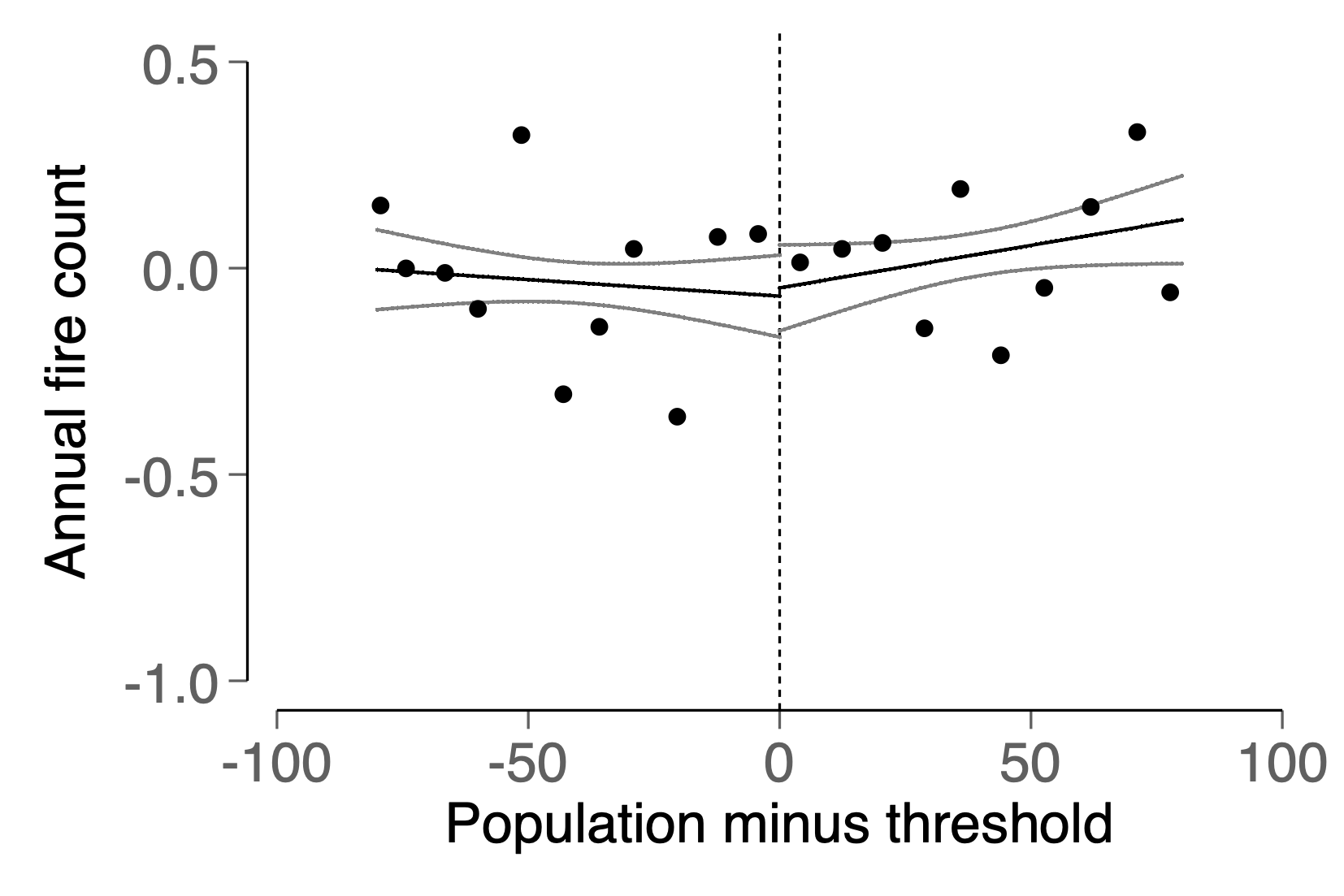} 
		}				
		\subfigure[PM 2.5: High rice \emph{or} high sugar]{
			\includegraphics[scale=0.25]{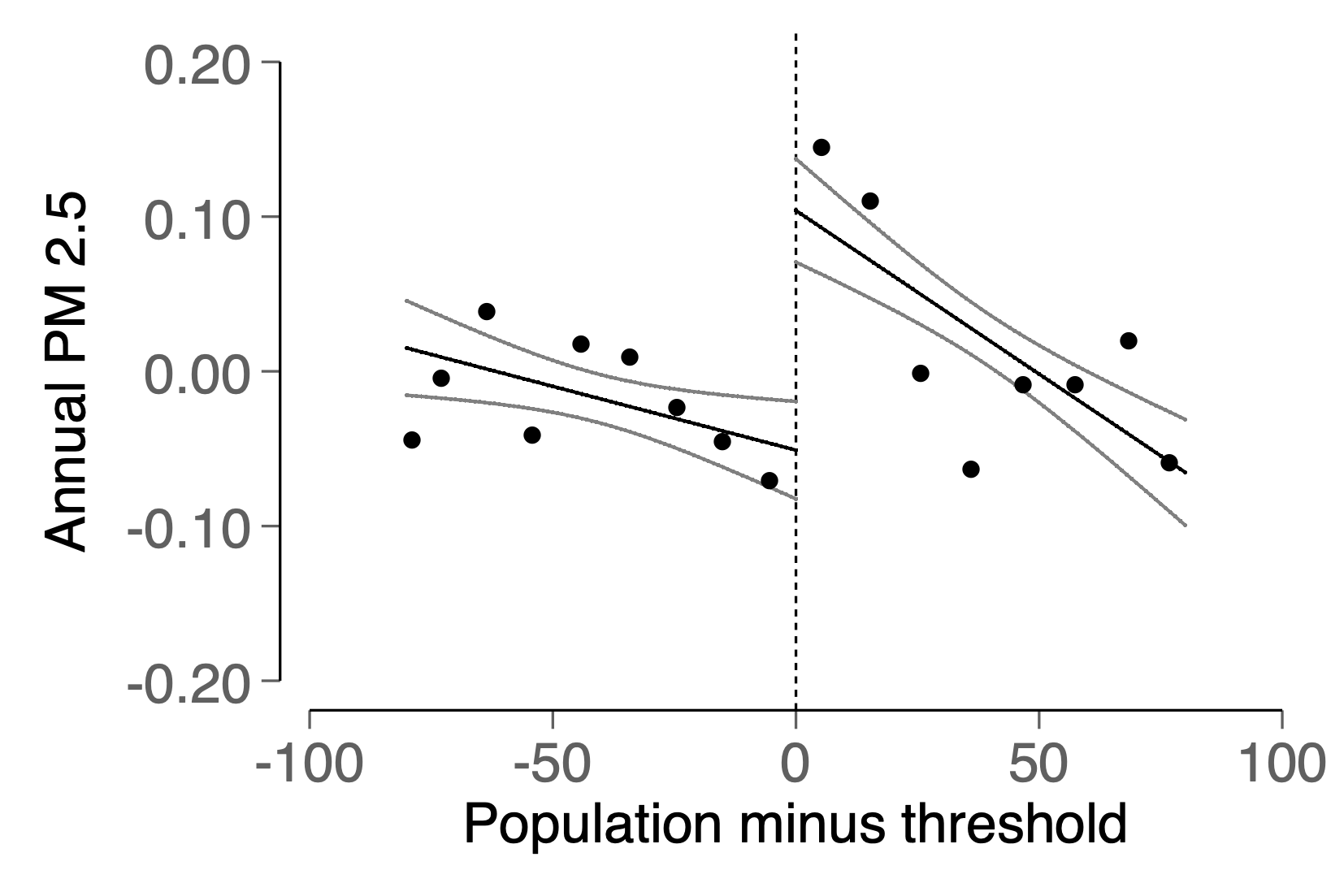}
		} 
		\subfigure[PM 2.5: Low rice \emph{and} low sugar]{
			\includegraphics[scale=0.25]{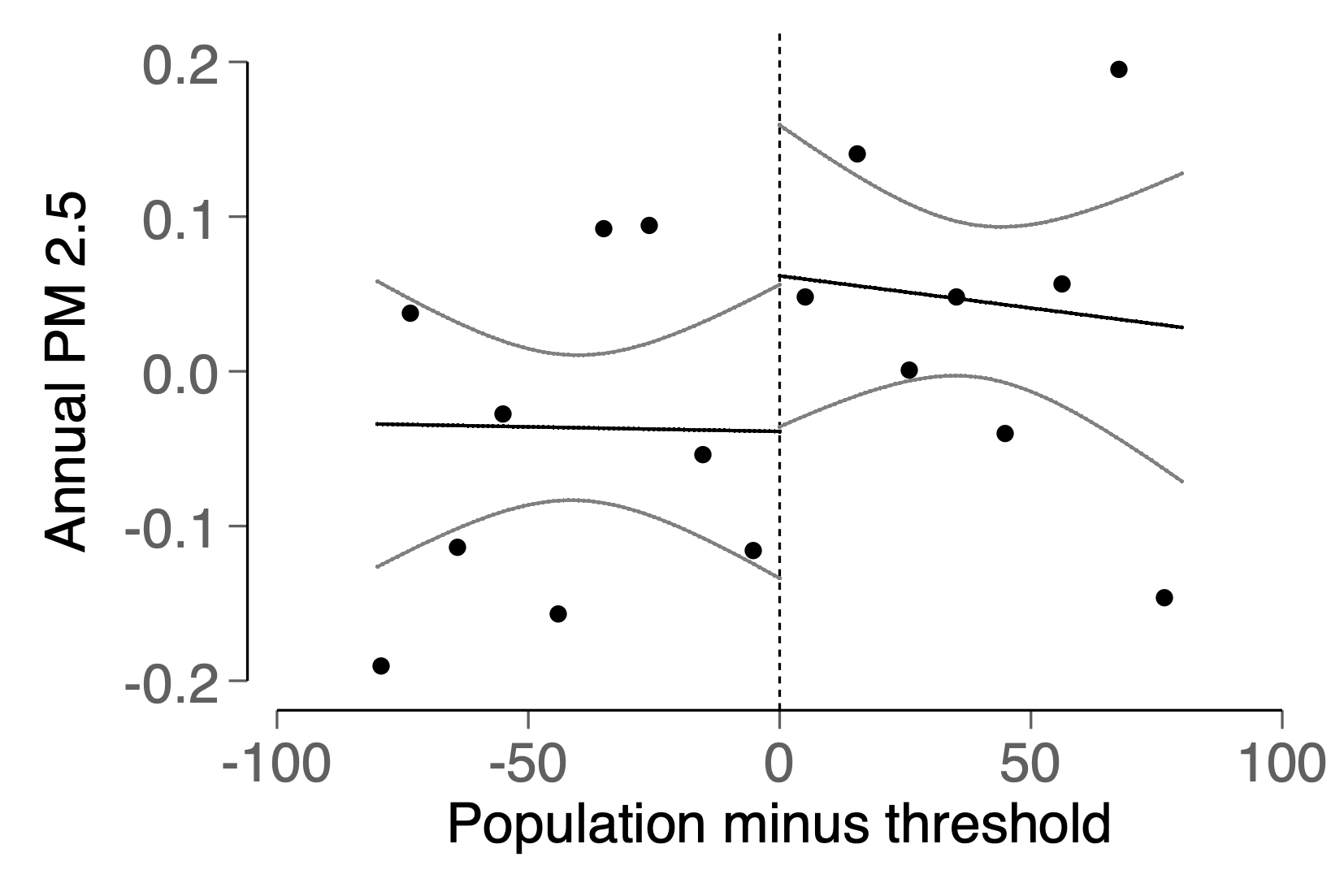} 
		}		
	\end{center}
	\scriptsize Notes: All panels show regression discontinuity estimates by plotting the residualized values of outcomes as a function of the normalized 2001 village population relative to the threshold (after controlling for fixed effects and all baseline variables in the main specification other than population). Each point represents the mean of all villages in a given population bin. Panels (a) and (b) plot the regression discontinuity relationship for  annual fire counts, while panels (c) and (d) plot the regression discontinuity relationship for annual  average PM 2.5 ($\mu g/ m^{3}$).  The sample used is the panel of villages for the period from 2002 - 2013. Panels (a) and (c) plot the regression discontinuity relationship for  districts with high (above sample median) rice \emph{or} sugarcane acreage share at baseline. Panels (b) and (d) plot the regression discontinuity relationship for districts with low (below sample median) rice \emph{and} sugarcane acreage share at baseline.  Estimates in all panels control for district-threshold fixed effects, year fixed effects, and baseline village characteristics in 2001. Population is centered around the state-specific threshold used for road eligibility - either 500 or 1000, depending on the state. Standard errors are clustered at the village level. \\
	
\end{figure}

\begin{table}[H] 
	\centering 
	\caption{Impact of rural roads on annual agricultural fire activity and PM2.5 by high rice \emph{or} high sugarcane districts versus low rice \emph{and} low sugar districts}
	\label{tab:ricesugar}
	\resizebox{0.7\linewidth}{!}{\begin{threeparttable}
			{
\def\sym#1{\ifmmode^{#1}\else\(^{#1}\)\fi}
\begin{tabular}{l*{4}{c}}
\hline\hline
                         &\multicolumn{2}{c}{High rice or high sugar}&\multicolumn{2}{c}{Low rice and low sugar}\\\cmidrule(lr){2-3}\cmidrule(lr){4-5}
                         &\multicolumn{1}{c}{(1)}&\multicolumn{1}{c}{(2)}&\multicolumn{1}{c}{(3)}&\multicolumn{1}{c}{(4)}\\
                         &\multicolumn{1}{c}{Fires}&\multicolumn{1}{c}{PM 2.5}&\multicolumn{1}{c}{Fires}&\multicolumn{1}{c}{PM 2.5}\\
\hline
Road built               &       3.191** &       1.045***&       0.183   &       0.176   \\
                         &     (1.291)   &     (0.308)   &     (0.382)   &     (0.273)   \\
\hline
N                        &      92,772   &      92,772   &      31,920   &      31,920   \\
Control group mean       &        3.30   &       38.86   &        1.43   &       52.93   \\
\hline\hline
\end{tabular}
}
		\end{threeparttable}}
		\begin{minipage}{\linewidth} \footnotesize \smallskip 
			\scriptsize Notes: This table shows the effect of rural roads on village-level annual fire activity and annual average PM 2.5 ($\mu g/ m^{3}$). The sample consists of the panel of villages from 2002 - 2013. ``High rice or high sugar'' sample consists of districts with high (above sample median) rice \emph{or} sugarcane acreage share at baseline (2001).  ``Low rice and low sugar'' sample consists of districts with below median acreage share of rice and below median sugarcane acreage share at baseline.  Regressions include district-threshold fixed effects, year fixed effects and baseline control variables. Standard errors in parentheses are clustered at village level. Significance at 1\%, 5\% and 10\% are indicated by $^{***}$, $^{**}$ and $^{*}$, respectively.
		\end{minipage}
\end{table}

\begin{table}[H] 
	\centering 
	\caption{Impact of rural roads on annual agricultural fire activity and PM2.5 in high rice versus high sugarcane districts}
	\label{tab:ricehi_sugarhi}
	\resizebox{0.6\linewidth}{!}{\begin{threeparttable}
			{
\def\sym#1{\ifmmode^{#1}\else\(^{#1}\)\fi}
\begin{tabular}{l*{4}{c}}
\hline\hline
                         &\multicolumn{2}{c}{High rice}  &\multicolumn{2}{c}{High sugar} \\\cmidrule(lr){2-3}\cmidrule(lr){4-5}
                         &\multicolumn{1}{c}{(1)}&\multicolumn{1}{c}{(2)}&\multicolumn{1}{c}{(3)}&\multicolumn{1}{c}{(4)}\\
                         &\multicolumn{1}{c}{Fires}&\multicolumn{1}{c}{PM 2.5}&\multicolumn{1}{c}{Fires}&\multicolumn{1}{c}{PM 2.5}\\
\hline
Road built               &       3.560*  &       1.392***&       3.245** &       0.705** \\
                         &     (2.022)   &     (0.453)   &     (1.315)   &     (0.305)   \\
\hline
N                        &      62,400   &      62,400   &      65,712   &      65,712   \\
Control group mean       &        3.46   &       37.05   &        3.09   &       38.54   \\
\hline\hline
\end{tabular}
}
		\end{threeparttable}}
		\begin{minipage}{\linewidth} \footnotesize \smallskip 
			\scriptsize Notes: This table shows the effect of rural roads on village-level annual fire activity and annual average PM 2.5 ($\mu g/ m^{3}$) from 2002-2013. The sample for columns (1) and (2) consists of villages in districts with high (above sample median) rice acreage share at baseline (2001).  The sample in columns (3) and (4) consists of villages in districts with high (above sample median) sugarcane acreage share at baseline (2001).  Regressions include district-threshold fixed effects, year fixed effects and baseline control variables. Standard errors in parentheses are clustered at village level. Significance at 1\%, 5\% and 10\% are indicated by $^{***}$, $^{**}$ and $^{*}$, respectively.
		\end{minipage}
\end{table}

\begin{table}[H] 
	\centering 
	\caption{Impact of rural roads on annual agricultural fire activity and PM2.5 in high rice and low sugarcane versus high sugarcane and low rice districts}
	\label{tab:ricehi_sugarhi_only}
	\resizebox{0.7\linewidth}{!}{\begin{threeparttable}
			{
\def\sym#1{\ifmmode^{#1}\else\(^{#1}\)\fi}
\begin{tabular}{l*{4}{c}}
\hline\hline
                         &\multicolumn{2}{c}{High rice \& low sugar}&\multicolumn{2}{c}{High sugar \& low rice}\\\cmidrule(lr){2-3}\cmidrule(lr){4-5}
                         &\multicolumn{1}{c}{(1)}&\multicolumn{1}{c}{(2)}&\multicolumn{1}{c}{(3)}&\multicolumn{1}{c}{(4)}\\
                         &\multicolumn{1}{c}{Fires}&\multicolumn{1}{c}{PM 2.5}&\multicolumn{1}{c}{Fires}&\multicolumn{1}{c}{PM 2.5}\\
\hline
Road built               &       3.169   &       2.223** &       2.641*  &       0.479   \\
                         &     (3.951)   &     (1.133)   &     (1.495)   &     (0.423)   \\
\hline
N                        &      27,060   &      27,060   &      30,372   &      30,372   \\
Control group mean       &        3.80   &       39.61   &        2.96   &       42.76   \\
\hline\hline
\end{tabular}
}
		\end{threeparttable}}
		\begin{minipage}{\linewidth} \footnotesize \smallskip 
			\scriptsize Notes: This table shows the effect of rural roads on village-level annual fire activity and annual average PM 2.5 ($\mu g/ m^{3}$) from 2002-2013. The sample for columns (1) and (2) consists of villages in districts with high (above sample median) rice acreage share but low (below median) sugarcane share at baseline (2001).  The sample in columns (3) and (4) consists of villages in districts with high (above sample median) sugarcane acreage but low rice share at baseline (2001).  Regressions include district-threshold fixed effects, year fixed effects and baseline control variables. Standard errors in parentheses are clustered at village level. Significance at 1\%, 5\% and 10\% are indicated by $^{***}$, $^{**}$ and $^{*}$, respectively.
		\end{minipage}
\end{table}

\newpage

\section{Fire Activity in Cropland versus Non-Cropland Areas} \label{landcover}
		\setcounter{table}{0}
		\renewcommand{\thetable}{\Alph{section}.\arabic{table}}
		\setcounter{figure}{0}
		\renewcommand{\thefigure}{\Alph{section}.\arabic{figure}}
		\numberwithin{table}{section}

We use remotely-sensed land cover classification data from 1995 to categorize fires into those occurring on cropland versus fires on non-cropland land cover classes around the PMGSY sample villages \citep{roy2016}. The land cover data assigns 100 m X 100 m grid cells to 19 land use categories based on the International Geosphere-Biosphere Programme (IGBP) classification scheme. These categories include cropland, various type of forests, built-up land, and other categories. We overlay the land cover data on 1-km grid cells around each fire pixel (this reflects the resolution of the MODIS data we use to measure fires). Fire pixels that have most of their 1-km grid cell falling in the cropland land class are categorized as cropland fires. We find the increase in fire activity in response to rural roads is concentrated in cropland areas, with small and statistically insignificant effects in non-cropland areas (Table~\ref{tab:fires_cropland}).

\paragraph{Measurement error.}
It is important to note that classifying remotely-sensed fires by type of vegetation using land cover data as we do here is prone to measurement error. The first source of error arises due to the coarse spatial resolution of the fire data.  The MODIS fire data we use only tells us the coordinates at the center of a 1 km grid cell that contains a fire -- but cannot place the exact location of the fire within that grid cell. As a result,  a fire occurring on cropland at the edge of a 1 X 1 km grid cell, with the majority of that grid cell consisting of non-cropland land classes, would be mislabelled as a non-cropland fire.  Second, the accuracy of land cover data itself can be poor, particularly in regions with poor ground-level validation data and in fragmented landscapes such as those at forest edges \citep{zubkova2019changes}. Finally, the land cover data we use reflects average land usage over the year in 1995. It will not capture seasonal changes in land use or the changes in land use that occurred in the years leading to the start of the PMGSY program in 2002. Studies in the remote sensing literature have noted that relying on land cover data to classify fires by type of vegetation (crops, forests, or other) is not validated and can result in errors of omission and commission \citep{roy2017multi}. Instead, a more accurate approach would rely on extensive labeled field-level data to serve as a training and validation data set \citep{hall2021validation}. However, such ground truth data for agricultural fires in India, especially in the time period for our analysis, is unavailable. Given these limitations, we focus our primary analysis on all types of vegetative fires that are detectable from satellite data around the sample villages.

\begin{table}[H] 
	\centering 
	\caption{ IV estimates of impact of rural roads on fires (count) in cropland areas versus non-cropland areas}
	\label{tab:fires_cropland}
	\resizebox{0.6\linewidth}{!}{\begin{threeparttable}
			{
\def\sym#1{\ifmmode^{#1}\else\(^{#1}\)\fi}
\begin{tabular}{l*{2}{c}}
\hline\hline
                         &\multicolumn{2}{c}{Annual fire count}\\\cmidrule(lr){2-3}
                         &\multicolumn{1}{c}{(1)}&\multicolumn{1}{c}{(2)}\\
                         &\multicolumn{1}{c}{Cropland}&\multicolumn{1}{c}{Non-cropland}\\
\hline
Road built               &       1.061*  &       0.248   \\
                         &     (0.558)   &     (0.387)   \\
\hline
N                        &     133,788   &     133,788   \\
Control group mean       &        1.71   &        1.27   \\
\hline\hline
\end{tabular}
}
		\end{threeparttable}}
		\begin{minipage}{\linewidth} \footnotesize \smallskip 
			\scriptsize Notes: This table shows  the IV estimates for the effects of rural roads on the annual count of fires in cropland (column (1)) versus non-cropland (column (2)) land cover categories. Classification of cropland versus non-cropland areas is based on pre-program (1995) land cover classification data from \cite{roy2016}. The sample used is the panel of villages for the period from 2002 - 2013. All regressions include district-threshold fixed effects, year fixed effects, and baseline control variables. Standard errors in parentheses are clustered at village level. Significance at 1\%, 5\% and 10\% are indicated by $^{***}$, $^{**}$ and $^{*}$, respectively.
		\end{minipage}
\end{table}

\clearpage

\section{Infant Mortality} \label{imr}
		\setcounter{table}{0}
		\renewcommand{\thetable}{\Alph{section}.\arabic{table}}
		\setcounter{figure}{0}
		\renewcommand{\thefigure}{\Alph{section}.\arabic{figure}}
		\numberwithin{table}{section}
		\setcounter{equation}{0}
		\renewcommand{\theequation}{\Alph{section}.\arabic{figure}}
		\numberwithin{equation}{section}

\begin{table}[H] 
	\centering 
	\caption{Differences between PMGSY villages that are matched vs. not matched with NFHS-IV villages within 50 km, and balance and falsification tests for the matched sample}
	\label{tab:rdbaldown}
	\resizebox{\linewidth}{!}
		{\begin{threeparttable}
			\setlength{\linewidth}{.1cm} \begin{center} 
{\begin{tabular}{l c c c c c c c}
\hline\hline 
& \multicolumn{1}{c}{(1)} & \multicolumn{1}{c}{(2)} & \multicolumn{1}{c}{(3)} & \multicolumn{1}{c}{(4)} &
\multicolumn{1}{c}{(5)} & \multicolumn{1}{c}{(6)} & \multicolumn{1}{c}{(7)} \\
Variable  & Full    & NFHS-IV matched    & No NFHS-IV match    & Difference    & p-value on & RD estimate  & p-value on \\ 
          & sample  & sample    & sample      & of means     & difference    & (matched sample)  & RD estimate \\ 
\hline                                                               
Primary school                     & 0.959           & 0.958           & 0.969           & 0.011                      & 0.07    & -0.029   & 0.42    \\
Medical center                     & 0.166               & 0.164               & 0.180               & 0.015                             & 0.09 & -0.090 & 0.20               \\
Electrified                        & 0.430                    & 0.435                    & 0.367                    & -0.067                                      & 0.67 & 0.018 & 0.84               \\
Distance from nearest town (km)    & 26.490                    & 26.514                    & 26.190                    & -0.324                                     & 0.78     & -3.146 & 0.43               \\
Land irrigated (share)             & 0.281                & 0.282                & 0.266                & -0.016                               & 0.81 & -0.011 & 0.83                \\
\hline 
\end{tabular} }  
\end{center}
		\end{threeparttable}}
		\begin{minipage}{\linewidth} \footnotesize  
			\scriptsize Notes: This table presents mean values for village characteristics, measured in the baseline period. The first eight variables are from the 2001 Population Census, the next three (below the first line) are from the 2002 BPL Census, while the final six variables are our outcome variables measured at baseline (2001). Columns 1-3 show the unconditional means for all villages, villages matched to NFHS-IV survey locations within 50 km, and villages with no NFHS-IV matches, respectively. Column 4 shows the difference of means across matched versus unmatched samples, and Column 5 shows the p-value for the difference of means between (2) and (3). Column 6 shows the regression discontinuity estimate for matched PMGSY villages, following the main estimating equation, of the effect of being above the treatment threshold on the baseline variable (with the outcome variable omitted from the set of controls), and Column 7 is the p-value for this estimate, using heteroskedasticity robust standard errors.  
		\end{minipage}
\end{table}

\begin{table}[H] 
	\centering 
	\caption{First stage and IV estimates of impact of rural roads on agricultural fires (count) and PM2.5 ($\mu g/ m^{3}$) for PMGSY villages matched to NFHS-IV villages}
	\label{tab:fsiv_nfhsmatched}
	\resizebox{0.9\linewidth}{!}{\begin{threeparttable}
			\begin{tabular}{lccc}
   \toprule
                        & Road built    & Annual fire activity & Annual average PM 2.5\\
                        & (1)           & (2)                  & (3)\\
    & $1^{st}$ stage & \multicolumn{2}{c}{IV} \\ \cmidrule(lr){2-2} \cmidrule(lr){3-4}
   \midrule 
   Above threshold pop. & 0.224$^{***}$ &                      &   \\
                        & (0.018)       &                      &   \\
   Road built           &               & 2.182$^{***}$        & 0.724$^{***}$\\
                        &               & (0.740)              & (0.226)\\
   \midrule 
   Control group mean   & 0.242         & 2.947                & 43.105\\
   Observations         & 129,379       & 129,379              & 129,379\\
   \bottomrule
\end{tabular}
		\end{threeparttable}}
		\begin{minipage}{\linewidth} \footnotesize \smallskip 
			\scriptsize Notes: This table shows  the first stage estimates (probability of receiving a rural road) as well as the effects of rural roads on count of agricultural fires and PM 2.5. The sample consists of the panel of PMGSY villages matched to NFHS-IV villages from 2002 - 2013. ``Above threshold pop." is an indicator for the village population being above the treatment threshold. Column (1) shows the first stage, with the dependent variable (``Road built") taking the value 1 if the village received a new road during 2002-2013, 0 otherwise. Columns (2) and (3) present the IV estimates of the treatment effects of new roads on annual fire counts and annual average PM 2.5  ($\mu g/m^{3}$), respectively. All regressions include district-threshold fixed effects, year fixed effects, and baseline control variables. Standard errors in parentheses are clustered at village level. Significance at 1\%, 5\% and 10\% are indicated by $^{***}$, $^{**}$ and $^{*}$, respectively.
		\end{minipage}
\end{table}

\end{document}